\documentclass[12pt,a4paper]{article}

\usepackage{jheppub}

\usepackage{etoolbox}
    \makeatletter
    \patchcmd{\maketitle}{\@fpheader}{}{}{}
    \makeatother

\usepackage[utf8]{inputenc}
\usepackage{lmodern}
\usepackage{exscale}

\usepackage{amsmath}
\usepackage{amssymb}
\usepackage{mathtools}

\usepackage{mathrsfs}

\usepackage{relsize,exscale}
\usepackage{tensor}

\usepackage[low-sup]{subdepth}

\newcommand{\scr}{\scriptscriptstyle}

\usepackage{graphicx}

\newcommand{\longsim}{\scalebox{1.8}[1]{$\sim$}}

\newcommand{\ch}{ {\rm ch} }
\newcommand{\sh}{ {\rm sh} }

\usepackage{ulem}
\usepackage{cancel}

\usepackage[dvipsnames]{xcolor}
\hypersetup{
    colorlinks,
    linkcolor={black},
    citecolor={black},
    urlcolor={black}
}

\makeatletter
\newcommand{\dalembertian}{\mathop{\mathpalette\dalembertian@\relax}}
\newcommand{\dalembertian@}[2]{%
  \begingroup
  \sbox\z@{$\m@th#1\square$}%
  \dimen0=\fontdimen8
    \ifx#1\displaystyle\textfont\else
    \ifx#1\textstyle\textfont\else
    \ifx#1\scriptstyle\scriptfont\else
    \scriptscriptfont\fi\fi\fi3
  \makebox[\wd\z@]{%
    \hbox to \ht\z@{%
      \vrule width \dimen0
      \kern-\dimen0
      \vbox to \ht\z@{
        \hrule height \dimen0 width \ht\z@
        \vss
        \hrule height 2\dimen0
      }%
      \kern-2.5\dimen0
      \vrule width 2.5\dimen0
    }%
  }%
  \endgroup
}
\makeatother

\title{Photon propagator for inflation \\   in the
 general covariant gauge}

\author[]{Silvije Domazet,}
\emailAdd{sdomazet@mail.com}
\author[a]{Dra\v{z}en Glavan,}
\emailAdd{glavan@fzu.cz}
\author[b]{Tomislav Prokopec}
\emailAdd{t.prokopec@uu.nl}

\affiliation[a]{CEICO, FZU --- Institute of Physics of the Czech Academy of Sciences,
	\\
	Na Slovance 1999/2, 182 21 Prague 8, Czech Republic}
\affiliation[b]{Institute for Theoretical Physics, Spinoza Institute \& EMME$\Phi$,
	Utrecht University, 
	\\
	Princetonplein 5, 3584 CC Utrecht, The Netherlands}

\abstract{
Photon propagator for power-law inflation is considered in the general covariant gauges within the canonical quantization formalism. Photon mode functions in covariant gauges are considerably more complicated than their scalar counterparts, except for the special choice of the gauge-fixing parameter we call the simple covariant gauge. We explicitly construct the position space photon propagator in the simple covariant gauge, and find the result considerably more complicated than its scalar counterpart. This is because of the need for explicitly inverting the Laplace operator acting on the scalar propagator, which results in Appell’s fourth function. Our propagator correctly reproduces the de Sitter and flat space limits. We use this propagator to compute two simple observables: the off-coincident field strength-field strength correlator and the energy-momentum tensor, both of which yield consistent results. As a spinoff of our computation we also give the exact expression for the Coulomb gauge propagator in power-law inflation in arbitrary dimensions.
}

\notoc

\begin{document}

\maketitle

\titlepage

\section{Introduction}
\label{sec: Introduction}

The rapid expansion of primordial inflation quickly dilutes the Universe.
Therefore, for most of inflation, it is the coupling of fields to the expanding background that 
controls the dynamics of their fluctuations. 
Massless vector fields, such as the electromagnetic field, couple conformally to gravity
and thus do not directly sense the expansion of the Universe. Only when their conformality 
is broken can the effects 
of the expansion be communicated to them. This happens in instances when
the vector field couples to another background field (condensate). Such is
the case of symmetry-breaking scalar electrodynamics, where the scalar condensate 
generates a possibly time-dependent effective mass for the gauge vector.
Some of the notable early Universe models that exhibit 
gauge field amplification due to this mechanism are Higgs inflation
(see e.g.~\cite{Bezrukov:2007ep,George:2012xs,Bezrukov:2013fka,Bezrukov:2014ipa,George:2013iia,George:2015nza,Ema:2016dny,Adshead:2023mvt,Sfakianakis:2018lzf,Adshead:2017xll,Obata:2014qba})
and axion inflation (see e.g.~\cite{Adshead:2015pva,Figueroa:2023oxc,Caravano:2021bfn,Gorbar:2021rlt,Domcke:2018eki,Adshead:2016iae,Pajer:2013fsa}).
The Ratra model~\cite{Ratra:1991bn} is another case of conformality breaking,
where a time-dependent scalar condensate couples to the vector kinetic term, 
thus amplifying vector perturbations in inflation.
However, arguably more intriguing are examples where the breaking of conformality is
due to loop effects from interactions with non-confomal fluctuations.

Due to the effect of gravitational particle 
production~\cite{Parker:1968mv,Parker:1969au,Parker:1971pt},
infrared fluctuations of non-conformally coupled 
fields, such as scalars~\cite{Mukhanov:1981xt} and gravitons~\cite{Starobinsky:1979ty}, 
are copiously produced by the almost exponential rapid expansion of
the primordial inflationary Universe. Coupling of vectors to such
fluctuations can lead to large secular loop corrections for vectors
in inflation. 
This is true in inflationary scalar electrodynamics~\cite{Kahya:2005kj,Kahya:2006ui,Prokopec:2006ue,FabianGonzalez:2016vgo,Popov:2017xut,Glavan:2019uni,Chen:2016uwp,Chen:2016hrz}, where the photon can develop a mass gap~\cite{Prokopec:2002jn,Prokopec:2002uw,Prokopec:2003bx,Prokopec:2003tm} due to large infrared fluctuations
of the charged scalar, which in turn exponentially damps its spatial correlations.
A nonperturbative analysis~\cite{Prokopec:2007ak,Prokopec:2008gw},
based on Starobinsky's stochastic formalism~\cite{Starobinsky:1986fx,Starobinsky:1994bd},
shows the generated photon mass to be noneperturbatively large. 
Coupling to the charged scalar fluctuations
can also induce generation of magnetic fields during inflation~\cite{Turner:1987bw,Calzetta:1997ku,Giovannini:2000dj,Prokopec:2001nc,Davis:2000zp}
which could be observable at late times (see e.g.~\cite{Durrer:2013pga} for a review).
For electromagnetism interacting with quantum gravity, loops of inflationary 
gravitons~\cite{Hu:1997iu,Dalvit:2000ay,Leonard:2012si,Leonard:2013xsa,Glavan:2013jca,Wang:2014tza,Glavan:2015ura,Wang:2015eaa,Glavan:2016bvp,Miao:2018bol,Glavan:2023tet} engender secular corrections to the Coulomb potential, and
the electric field of the propagating photon.

Thus far most of the loop computations have been performed in 
rigid de Sitter space where the Universe expands exactly exponentially.
This idealization is very often a good approximation for inflationary spacetime,
and it typically
leads to considerable computational simplifications due to enhanced symmetry.
However, de Sitter space is not always the most appropriate description for
the inflating spacetime. In fact, observations~\cite{Planck:2018jri} are good enough
to decisively point at a small deviation from the exponential expansion of de Sitter 
space, characterized by slow-roll parameters which are a measure of the rate of decrease 
in the expansion rate. These slow-roll parameters are some of the key inflationary 
observables that discriminate between the models. We are 
interested in corrections that slow-roll parameters can engender in inflationary 
loop corrections, especially in the infrared. This is generally not a mathematically 
tractable problem,
but given that slow-roll parameters are small and evolve slowly,
deviations from de Sitter expansion can often be treated adiabatically. 
Here we consider power-law inflation~\cite{Lucchin:1984yf,La:1989za}
as an analytically tractable 
model.~\footnote{
Fermions also couple conformally to gravity, and experience the 
expansion only by coupling to non-conformally coupled fields. The effects 
of quantum corrections in power-law
inflation with fermions running in the loops can be studies using the propagator
worked out in~\cite{Koksma:2009tc}.}

Power-law inflation is characterized by a 
constant non-vanishing principal
slow-roll parameter. Even though a constant slow-roll parameter is excluded 
by data~\cite{Planck:2018jri}, 
 it is closer to the realistic slow-roll inflation
than a rigid de Sitter space, and allows to capture the effects of non-vanishing principal 
slow-roll parameter. This should also be a good approximation when the
evolution of the slow-roll parameter can be considered adiabatic,
which indeed is the case in realistic inflationary models.
On a technical side power-law inflation is considered almost as tractable as de 
Sitter space, because of the experience with scalar fields whose mode functions retain
the same functional form as in de Sitter space, and scalar propagators that also
retain the same functional form as in de Sitter, being expressible in terms of 
the hypergeometric function. Some of the notable works dealing with
questions of quantum loop corrections in inflating spacetimes with a
non-vanishing slow-roll parameter are~\cite{Bilandzic:2007nb,Herranen:2013raa,Frob:2018tdd,Lima:2020prb,Miao:2015oba,Katuwal:2021kry,Miao:2019bnq,Kyriazis:2019xgj,Markkanen:2017edu,Glavan:2023lvw,Janssen:2009pb}.

In order to perform dimensionally regulated loop computations involving vector fields
in power-law inflation we need to have the necessary two-point functions (propagators)
in~$D$ dimensions. While the propagator for the massive vector in the Abelian Higgs 
model has been worked out in the unitary gauge~\cite{Glavan:2020zne}, 
the massless vector propagator
has not been worked out. That is why in this 
paper we compute the propagator for the massless vector field in power-law inflation. 
Massless vector field~$A_\mu$, that we henceforth refer to as the photon, is a~$U(1)$ 
gauge field, and working out its two-point functions requires fixing a gauge. 
Perhaps the most natural gauge choice is the general covariant gauge,
characterized by adding the covariant gauge-fixing term to the action,
\begin{equation}
S_{\rm gf}[A_\mu] = \int\! d^{D\!}x \, \sqrt{-g} \,
	\biggl[
	- \frac{1}{2\xi} \bigl( \nabla^\mu A_\mu \bigr)^{\!2}
	\biggr]
	\, ,
\label{gauge}
\end{equation}
that contains one free gauge-fixing parameter~$\xi$, where~$\nabla^\mu A_\mu$
is the covariant derivative of the vector potential with respect to the space time 
metric~$g_{\mu\nu}$. This gauge in general preserves all of the spacetime isometries.
However, even in de Sitter space where this is expected to be quite beneficial, 
the covariant gauges cannot preserve all de Sitter 
symmetries~\cite{Glavan:2022dwb,Glavan:2022nrd} of the two-point functions, except in 
the exact transverse gauge~\cite{Tsamis:2006gj}.
Even though de Sitter invariant solutions to propagator equations of motion do 
exist~\cite{Allen:1985wd,Frob:2013qsa}, they predict spurious behaviour
in the infrared~\cite{Youssef:2010dw,Rendell:2018qid} due to not respecting 
the necessary Ward-Takahashi identity~\cite{Glavan:2022pmk}.
Nonetheless  the covariant gauge~(\ref{gauge}) still seems as 
a reasonable gauge choice to consider in power-law inflation,
and we construct the photon propagator in this gauge. 
Our computations yield the propagator not as practical as expected.
While in de Sitter space the covariant gauge two-point functions are generally expressible 
in terms of the corresponding scalar two-point functions, in power-law inflation
this is no longer the case, which complicates matter significantly.

Determining the photon propagator in momentum space amounts 
to computing the photon mode functions. We compute them 
for an arbitrary gauge-fixing parameter~$\xi$ in~(\ref{gauge}).  Their solutions
in power-law inflation is considerably more complicated than the corresponding
solutions in de Sitter space. However, we identify a particular choice for the
gauge-fixing parameter, that we refer to as the simple covariant gauge, 
for which the mode functions take on a simple form and are expressible in terms
of scalar mode functions and their derivatives. We then use these mode functions
to compute the position space photon propagator in the simple covariant gauge.
Despite the relative simplicity of the mode functions, this propagator in power-law inflation
is still much more complicated that its de Sitter counterpart, and cannot be expressed 
in terms of a finite number of derivatives of a scalar propagator. Rather, the
propagator involves complicated Appell's fourth functions which are two-variable
generalizations of the hypergeometric function. This complication appears
because of the necessity to explicitly invert the Laplace operator acting on the
scalar propagator. This inversion is also necessary to obtain
the Coulomb gauge propagator, the result for which we report as well.

\medskip

Following the introductory section that is now concluding, Sec.~\ref{sec: Preliminaries}
collects some of the definitions and results for the scalar 
mode functions in power-law inflation, and the corresponding two-point functions, 
that are used throughout.
Section~\ref{sec: Photon in FLRW} summarizes the canonical quantization procedure in the
general covariant gauge, giving equations of motion for the field operators and the
constraints. The photon mode functions are given in the general covariant gauge in
Sec.~\ref{sec: Field operator dynamics},
and in Sec.~\ref{sec: Two-point function} the photon two-point function is computed
in the simple covariant gauge. In Sec.~\ref{sec: Various limits} the flat space and the
de Sitter space limits are worked out and compared with the literature. 
Section~\ref{sec: Simple observables} gives results for two simplest
observables. The concluding section summarizes and discusses the main results.
A considerable amount of mathematical details, alternative derivations, 
and checks are collected in three sizable 
appendices~\ref{sec: Particular mode functions}--\ref{app: Checks for two-point function}.

\section{Preliminaries}
\label{sec: Preliminaries}

Here we collect definitions of the background power-law inflation spacetime, 
and the results for quantum scalar fields in power-law inflation that are used 
throughout.

\subsection{Power-law inflation}
\label{subsec: Power-law inflation}

The geometry of the~$D$-dimensional spatially flat Friedman-Lema\^itre-Robertson-Walker (FLRW) 
spacetime in conformal coordinates is described by the line element,
\begin{equation}
ds^2 = a^2(\eta) \bigl( -d\eta^2  +d\vec{x}^{\,2} \bigr) \, ,
\end{equation}
that defines the conformally flat 
metric~$g_{\mu\nu}\!=\! a^2(\eta) \, {\rm diag} (-1 , 1 , \dots , 1)$,
where~$a$ is the scale factor that encodes the dynamics of the expansion.
The rate of expansion is conveniently captured by the conformal Hubble rate,
\begin{equation}
\mathcal{H} = \frac{1}{a} \frac{d a}{d \eta} \, ,
\end{equation}
which is related to the physical Hubble rate~$H \!=\! \mathcal{H}/a$.
The acceleration of the expansion is captured by the principal slow-roll parameter,
\begin{equation}
\epsilon 
	= - \frac{1}{H^2} \frac{dH}{dt}
	= 1 - \frac{1}{ \mathcal{H}^2 } \frac{d\mathcal{H}}{ d\eta }
	\, ,
\end{equation}
that is connected to the deceleration parameter~$q \!=\! \epsilon \!-\! 1$.
In our conventions the Riemann tensor is defined 
as~${R^\alpha}_{\mu\beta\nu} \!=\! \partial_\beta \Gamma^\alpha_{\mu\nu}
	\!-\! \partial_\nu \Gamma^\alpha_{\mu\beta}
	\!+\! \Gamma^\rho_{\mu\nu} \Gamma^\alpha_{\beta\rho}
	\!-\! \Gamma^\rho_{\mu\beta} \Gamma^{\alpha}_{\nu\rho} $, 
where the Christoffel symbol 
is~$\Gamma^\alpha_{\mu\nu} \!=\! \frac{1}{2} g^{\alpha\beta} ( \partial_\mu g_{\nu\beta}
	\!+\! \partial_\nu g_{\mu\beta} \!-\! \partial_\beta g_{\mu\nu} )$.
Thus, the curvature tensors in FLRW spacetime are,
\begin{align}
R_{\mu\nu\rho\sigma} ={}&
	2 H^2 g_{\mu[\rho} g_{\sigma]\nu}
	+ 4 \epsilon H^2 \bigl( a^2 \delta^0_{[\mu} g_{\nu] [ \sigma} \delta^0_{\rho]} \bigr)
	\, ,
\\
R_{\mu\nu} \equiv {R^\alpha}_{\mu\alpha\nu} ={}&
	(D\!-\!1\!-\!\epsilon) H^2 g_{\mu\nu}
	+ (D\!-\!2) \epsilon H^2 \bigl( a^2 \delta_\mu^0 \delta_\nu^0 \bigr)
	\, ,
\\
R \equiv g^{\mu\nu} R_{\mu\nu}={}&
	(D\!-\!1)(D\!-\!2\epsilon) H^2 
	\, .
\end{align}

If the principal slow-roll parameter is smaller than one,~$\epsilon\!<\!1$, 
the expansion is accelerating,
which is the case in primordial sow-roll inflation where~$0 \!<\! \epsilon \! \ll\! 1$. 
The special case of inflation is the power-law inflation~\cite{Lucchin:1984yf,La:1989za}
characterized by a constant principal slow-roll parameter,
\begin{equation}
\epsilon = {\tt const.} \ll 1
\quad \Longrightarrow \quad
\mathcal{H} = \frac{H_0}{1 \!-\! (1\!-\!\epsilon)H_0(\eta\!-\!\eta_0)}
	= H_0 a^{1-\epsilon}
	\, ,
\quad 
a(\eta) = \Bigl( \frac{\mathcal{H}}{H_0} \Bigr)^{ \! \frac{1}{1-\epsilon}} \, ,
\end{equation}
where~$\eta_0$ is the initial time at which~$a(\eta_0) \!=\! 1$ 
and~$\mathcal{H}(\eta_0)\!=\! H(\eta_0) \!=\! H_0$.
The conformal time coordinate then ranges on an 
interval~$\eta\!\in\!(-\infty,\overline{\eta})$, where~$\overline{\eta}\!=\!\eta_0 \!+\! 1/[(1\!-\!\epsilon)H_0]$,
while spatial coordinates cover~$(D\!-\!1)$-dimensional Euclidean space. 
The special case of power-law inflation for~$\epsilon\!=\!0$ is the de Sitter space,
that has a constant physical Hubble rate~$H\!=\!H_0$.

When expressing bi-local quantities, dependent on two spacetime 
points~$x$ and~$x'$, such as the photon two-point functions,
it is usefult to employ bi-local variables that respect cosmological symmetries.
The convenient choice for these variables is,
\begin{align}
y(x;x') = (1\!-\!\epsilon)^2 \mathcal{H} \mathcal{H}' \Delta x^2 \, ,
\quad
u(x;x') = (1\!-\!\epsilon) \ln(aa') \, ,
\quad
v(x;x') = (1\!-\!\epsilon) \ln(a/a')
\, ,
\label{bilocal variables}
\end{align}
where henceforth primed quantities always refer to the primed coordinate, and
where $\Delta x^2 \!=\! \| \vec{x} \!-\! \vec{x}^{\,\prime} \|^2 \!-\! (\eta \!-\! \eta')^2 $
is the Lorentz invariant distance between points. In order to represent the two-point
functions as distributional limits of analytic bi-local functions, the bi-local 
variables above will acquire infinitesimal imaginary parts. Different prescriptions for
these imaginary parts are defined in the two following subsections.

\subsection{Scalar mode functions}
\label{subsec: Scalar mode functions}

The equation of motion for scalar modes is ubiquitous in cosmology, 
\begin{equation}
\biggl[ \partial_0^2 + k^2 - \Bigl( \lambda^2 \!-\! \frac{1}{4} \Bigr)(1\!-\!\epsilon)^2 \mathcal{H}^2 \biggr]
	\mathscr{U}_{\lambda}(\eta,\vec{k} ) 
	= 0
	\, ,
\label{mode eq}
\end{equation}
where~$\mathscr{U}_\lambda$ is the suitably rescaled mode function of a non-minimally
coupled scalar field, and
where~$\lambda$ is a constant that can be related to the non-minimal coupling.
The general solution in power-law inflation, where~$\epsilon\!=\!{\tt const.}$, is a linear combination,
\begin{equation}
\mathscr{U}(\eta, \vec{k}) 
	= \alpha(\vec{k}) \, U_\lambda(\eta,k)
	+ \beta(\vec{k}) \, U_\lambda^*(\eta,k) \, ,
\label{general mode function}
\end{equation}
of the positive-frequency Chernikov-Tagirov-Bunch-Davies (CTBD) mode 
function~\cite{Chernikov:1968zm,Bunch:1978yq},
\begin{equation}
U_\lambda(\eta,k) = 
	e^{\frac{i\pi}{4} (2\lambda+1) } e^{ \frac{- i k}{ (1-\epsilon) H_0 } } 
	\sqrt{ \frac{ \pi }{ 4(1\!-\!\epsilon) \mathcal{H} } } \,
	H_\lambda^{\scr (1)}\biggl( \frac{ k }{ (1\!-\!\epsilon) \mathcal{H} } \biggr)
	\, ,
\label{mode function}
\end{equation}
and its complex conjugate,
where~$H_\lambda^{\scr (1)}$ is the Hankel function of the first kind.
The Wronskian for these two linearly independent solutions is,
\begin{equation}
{\rm Im} \Bigl[ U_\lambda(\eta,k) \partial_0 U_\lambda^*(\eta,k) \Bigr]
	= \frac{1}{2} 
	\, .
\label{wrons}
\end{equation}
In flat space the CTBD mode function reduces to,
\begin{align}
\MoveEqLeft[7]
U_\lambda(\eta,k)
	\ \overset{H_0\to0}{\longsim} \
	\frac{ e^{- i k (\eta-\eta_0) } }{ \sqrt{2k} }
	\biggl\{
	1 + \frac{i}{2} \Bigl( \lambda^2 \!-\! \frac{1}{4} \Bigr) \frac{ (1\!-\!\epsilon) H_0 }{ k }
\nonumber \\
&
	- \frac{1}{8} \Bigl( \lambda^2 \!-\! \frac{1}{4} \Bigr) 
	\Bigl[ \lambda^2 \!-\! \frac{9}{4} - 4 i k (\eta\!-\!\eta_0) \Bigr]
		\frac{ (1\!-\!\epsilon)^2 H_0^2 }{ k^2 }
		+ \dots
	\biggr\}
	\, ,
\end{align}
which is closely related to the UV limit~$k/\mathcal{H}\!\gg\!1$.
On the other hand the IR expansion, i.e.
the small momentum 
expansion~$k/\mathcal{H}\!\ll\!1$ of the CTBD mode function is
\begin{align}
U_\lambda(\eta,k) 
	\ \overset{k/\mathcal{H} \ll 1}{\longsim} \
	{}&
	e^{\frac{i\pi}{4} (2\lambda+1) } e^{ \frac{- i k}{ (1-\epsilon) H_0 } } \,
	\frac{ \Gamma(\lambda) \, \Gamma(1\!-\!\lambda) }{ \sqrt{ 4 \pi(1\!-\!\epsilon) \mathcal{H} } } \,
	\Biggl\{
	\sum_{n=0}^{\infty} \frac{ i (-1)^{n+1} }{ n! \, \Gamma(n \!+\! 1 \!-\! \lambda) } \Bigl[ \frac{k}{2(1\!-\!\epsilon) \mathcal{H}} \Bigr]^{2n-\lambda}
\nonumber \\
&	\hspace{2.5cm}
	- 
	e^{-i\lambda \pi}
	\sum_{n=0}^{\infty} \frac{ i (-1)^{n+1} }{ n! \, \Gamma(n \!+\! 1 \!+\! \lambda) } \Bigl[ \frac{k}{2(1\!-\!\epsilon) \mathcal{H}} \Bigr]^{2n+\lambda}
	\Biggr\}
	\, .
\label{more IR}
\end{align}
We make frequent use of recurrence relations between contiguous scalar mode functions
(scalar mode functions whose indices differ by one),
\begin{equation}
\biggl[ \partial_0 + \Bigl( \frac{1}{2} \!+\! \lambda \Bigr) (1\!-\!\epsilon) \mathcal{H} \biggr] U_\lambda
	= -ik U_{\lambda+1} \, ,
\qquad
\biggl[ \partial_0 + \Bigl( \frac{1}{2} \!-\! \lambda \Bigr) (1\!-\!\epsilon) \mathcal{H} \biggr] U_{\lambda}
	= -ik U_{\lambda-1} \, ,
\label{mode recurrence}
\end{equation}
which follow from recurrence relations for Hankel functions~(c.f.~10.6.2 
in~\cite{Olver:2010,Olver:web}).
These allow to express the Wronskian~(\ref{wrons}) as,
\begin{equation}
{\rm Re} \Bigl[ U_\lambda(\eta,k) U_{\lambda+1}^*(\eta,k) \Bigr] = \frac{1}{2k} \, . 
\label{Wronskian}
\end{equation}
We will also use the following identity,
\begin{align}
&
\biggl[ \partial_0^2 + k^2 
	- \Bigl( \lambda^2 \!-\! \frac{1}{4} \Bigr) 
		(1\!-\!\epsilon)^2 \mathcal{H}^2 \biggr] \Bigl( \frac{a^2}{\mathcal{H}} U_{\lambda+1} \Bigr)
\nonumber \\
&	\hspace{4.5cm}
		=
		2 a^2 \mathcal{H} \biggl[
		2 \bigl[ \lambda(1\!-\!\epsilon) \!+\! 1 \bigr] U_{\lambda+1}
		- (1\!+\!\epsilon) \frac{ik}{\mathcal{H}} U_\lambda
		\biggr]
		\, ,
\label{mode id1}
\end{align}
 that follows from the equation of motion~(\ref{mode eq})
and the recurrence relations in~(\ref{mode recurrence}).

\subsection{Scalar two-point functions}
\label{subsec: Scalar two-point functions}

The positive-frequency Wightman two-point function for the scalar field 
in power-law inflation
can be constructed as a sum-over-modes over its mode functions
that satisfy the equation of motion~(\ref{mode eq}),
\begin{equation}
i \bigl[ \tensor*[^{\scr \! - \!}]{\Delta}{^{\scr \!+ \! } } \bigr]_\lambda(x;x')
	=
(aa')^{-\frac{D-2}{2}} \!\!
\int\! \frac{ d^{D-1}k }{ (2\pi)^{D-1 } } \, e^{i\vec{k} \cdot (\vec{x} - \vec{x}^{\, \prime}) } \,
	\mathscr{U}_\lambda(\eta,k) \mathscr{U}_\lambda^*(\eta',k)
	\, ,
\label{int over modes}
\end{equation}
where different states are captured by different Bogolyubov coefficients in
mode functions~(\ref{general mode function}). 
Implicit in this construction is the analytic continuation~$\eta \!\to\! \eta \!-\! i \delta/2$
and~$\eta' \!\to\! \eta' \!+\! i \delta/2$ (applied after the complex conjugation
of the mode function), that preserves the property under complex 
conjugation~$\bigl\{ i \bigl[ \tensor*[^{\scr \! - \!}]{\Delta}{^{\scr \! + \! }} \bigr]_\lambda(x;x') \bigr\}^{\!*} \!\!=\! i \bigl[ \tensor*[^{\scr \! - \!}]{\Delta}{^{\scr \! + \! }} \bigr]_\lambda(x';x)$. 
Nonequilibrium quantum field theory~\cite{Schwinger:1960qe,Mahanthappa:1962ex,Bakshi:1962dv,Bakshi:1963bn,Keldysh:1964ud,Jordan:1986ug,Calzetta:1986ey}
requires the use of three other two-point functions~(see~\cite{Berges:2004yj,NoneqLectures} for an introduction).
The complex conjugate of~(\ref{int over modes}) is the negative-frequency
Wightman function,~$i \bigl[ \tensor*[^{\scr \! + \!}]{\Delta}{^{\scr \! - \!}} \bigr]_\lambda (x;x') 	
	\!=\! \bigl\{ i \bigl[ \tensor*[^{\scr \! - \!}]{\Delta}{^{\scr \! + \!}} \bigr]_\lambda (x;x') \bigr\}^{*}$. The Feynman propagator is then constructed from the two
Wightman functions,
\begin{equation}
i \bigl[ \tensor*[^{\scr \! + \!}]{\Delta}{^{\scr \! + \!}} \bigr]_\lambda(x;x')
	=
	\theta(\eta \!-\! \eta') \, 
	i \bigl[ \tensor*[^{\scr \! - \!}]{\Delta}{^{\scr \!+ \! } } \bigr]_\lambda(x;x')
	+
	\theta(\eta' \!-\! \eta)
	i \bigl[ \tensor*[^{\scr \! + \!}]{\Delta}{^{\scr \!- \! } } \bigr]_\lambda(x;x')
	\, ,
\label{Feynman definition}
\end{equation}
and its complex conjugate is the Dyson propagator,~$i \bigl[ \tensor*[^{\scr \! - \!}]{\Delta}{^{\scr \! - \!}} \bigr]_\lambda (x;x') 	
	\!=\! \bigl\{ i \bigl[ \tensor*[^{\scr \! + \!}]{\Delta}{^{\scr \! + \!}} \bigr]_\lambda (x;x') \bigr\}^{*}$.
It then follows, from the definitions and from the mode equation~(\ref{mode eq}),
that the two-point functions satisfy the following equation of motion,
\begin{align}
\biggl[ \, \dalembertian - (1\!-\!\epsilon)^2 H^2
	\biggl( \Bigl[ \frac{D\!-\!1\!-\!\epsilon}{2(1\!-\!\epsilon)} \Bigr]^2 
		\!\! -\! \lambda^2 \biggr) \biggr] 
	i \bigl[ \tensor*[^{\tt a \!}]{\Delta}{^{\tt \! b}} \bigr]_\lambda (x;x') 
	= {\tt S}^{\tt ab} \, \frac{ i \delta^D(x\!-\!x') }{ \sqrt{-g}} \, ,
\label{scalar EOM}
\end{align}
where the scalar d'Alembertian is given 
by~$\dalembertian \!=\! \nabla^\mu \nabla_\mu\!=\!
	 - a^{-2} \bigl[ \partial_0^2 \!+\! (D\!-\!2) \mathcal{H} \partial_0 \!-\! \nabla^2 \bigr]$,
and~$\nabla^2\!=\!\partial_i \partial_i$ is the Laplacian.
Henceforth, the typewriter font indices denote the Schwinger-Keldysh
polarity indices~$\tt a,b \!=\! \pm$, and we introduce an accompanying sign 
symbol~${\tt S}^{\tt ab}$ defined by,
\begin{equation}
{\tt S}^{\scr ++} \!=\! - {\tt S}^{\scr --} \!=\! 1 \, ,
\qquad \qquad
{\tt S}^{\scr -+} \!=\! {\tt S}^{\scr +-} \!=\! 0 \, .
\label{SK sign}
\end{equation}
 Note that, in addition to the equation of motion~(\ref{scalar EOM}) with respect 
to the unprimed coordinate~$x$, we have another independent equation with respect 
to the primed coordinate~$x'$,
%
\begin{equation}
\biggl[ \, {\dalembertian}' - (1\!-\!\epsilon)^2 H'^2
	\biggl( \Bigl[ \frac{D\!-\!1\!-\!\epsilon}{2(1\!-\!\epsilon)} \Bigr]^2 
		\!\! -\! \lambda^2 \biggr) \biggr] 
	i \bigl[ \tensor*[^{\tt a \!}]{\Delta}{^{\tt \! b}} \bigr]_\lambda (x;x') 
	= {\tt S}^{\tt ab} \, \frac{ i \delta^D(x\!-\!x') }{ \sqrt{-g}} \, .
\label{scalar EOM primed}
\end{equation}

In general the scalar two-point functions that respect cosmological symmetries 
of spatial homogeneity and isotropy can be expressed in terms of
bi-local variables~(\ref{bilocal variables}) 
with appropriate infinitesimal complex parts~$\propto \! i \delta$ appended.
This allows the two-point functions to be expressed as distributional limits
of analytic functions.
The complex parts appended to bi-local variables depend on the type of the two-point function, 
and different prescriptions are labeled by the
Schwinger-Keldysh polarity indices. For the first 
variable,
\begin{equation}
y_{\tt ab} \!=\! (1\!-\!\epsilon)^2 \mathcal{H} \mathcal{H}' \Delta x^2_{\tt ab} \, ,
\label{yab}
\end{equation}
the~$i\delta$ is in the Lorentz invariant distance,~\footnote{The time
dependent factor multiplying the Lorentz invariant distance in~(\ref{yab}) strictly speaking 
also has complexified parts as does the~$u$ variable in~(\ref{u complex}), though
it is irrelevant in all the expressions we consider and we thus omit it.}
\begin{equation}
\Delta x^2_{\scr -+} = \| \Delta \vec{x} \|^2 - \bigl( \Delta \eta - i \delta \bigr)^2 \, ,
\qquad \quad
\Delta x^2_{\scr ++} = \| \Delta \vec{x} \|^2 - \bigl( |\Delta \eta| - i \delta \bigr)^2 \, ,
\label{Delta x2 prescription}
\end{equation}
and~$\Delta x_{\scr + - }^2 \!=\! \bigl(\Delta x_{\scr - + }^2 \bigr)^{\!*}$,
and~$\Delta x_{\scr -- }^2 \!=\! \bigl(\Delta x_{\scr + + }^2 \bigr)^{\!*}$.
The complexified second variable is given by,
\begin{equation}
u_{\scr -+} = u - \frac{1}{2} (1\!-\!\epsilon) ( \mathcal{H} \!-\! \mathcal{H}' )
	i \delta
	\, ,
\qquad \quad
u_{\scr ++} = u - \frac{1}{2} (1\!-\!\epsilon)
	| \mathcal{H} \!-\! \mathcal{H}' | i \delta
	\, ,
\label{u complex}
\end{equation}
and~$u_{\scr +-} \!=\! ( u_{\scr -+} )^*$,~$u_{\scr --} \!=\! ( u_{\scr ++} )^*$,
while the third is given by,
\begin{align}
v_{\scr - +} 
	=
	v
	-
	\frac{1}{2} (1\!-\!\epsilon) (\mathcal{H} \!+\! \mathcal{H}' ) i \delta
\qquad \quad
v_{\scr + +} = 
	v 
	-
	\frac{1}{2}(1\!-\!\epsilon) (\mathcal{H} \!+\! \mathcal{H}' ) 
		{\rm sgn}(\Delta\eta) i \delta
	\, ,
\label{v prescription}
\end{align}
and~$v_{\scr +-} \!=\! (v_{\scr -+} )^*$ and~$v_{\scr --} \!=\! (v_{\scr ++} )^*$.
For the sake of the flat space limit it is useful to define the~$i \delta$-prescription
for the time difference as well,
\begin{equation}
\Delta \eta_{\scr -+} = \Delta\eta - i \delta \, ,
\qquad \qquad
\Delta \eta_{\scr ++} = | \Delta\eta | - i \delta \, ,
\end{equation}
and the accompanying complex 
conjugates,~$\Delta \eta_{\scr +-} \!=\! (\Delta \eta_{\scr -+})^*$
and~$\Delta \eta_{\scr --} \!=\! (\Delta \eta_{\scr ++})^*$.
All of the prescriptions essentially follow from the one for
the Wightman function~(\ref{int over modes}).

\medskip

The scalar two-point functions in power-law inflation can differ qualitatively
depending on the index~$\lambda$ that they inherit from the mode
equation~(\ref{mode eq}). The preferred mode function is the CTBD
one~(\ref{mode function}) which corresponds to the state that minimizes
energy mode-per-mode in the asymptotic past. While this mode function is
defined for any~$\lambda$, it is clear from its IR behaviour~(\ref{more IR})
that for~$\lambda\!\geq \!(D\!-\!1)/2$ this leads to an IR divergent 
sum-over-modes~(\ref{int over modes})~\cite{Ford:1977in,Allen:1985ux,Allen:1987tz}, and that the Bogolyubov
coefficients in the mode function~(\ref{general mode function}) 
need to be chosen differently~\cite{Ford:1977in,Janssen:2009nz}.
We consider the two cases separately in the two following subsections.

\subsubsection{IR finite scalar two-point functions}

When the mode function index~$\lambda\!<\! (D\!-\!1)/2$ there is no obstruction to
choosing the mode function in the sum-over-modes~(\ref{int over modes}) 
to be the CTBD one~(\ref{mode function}),
\begin{equation}
i \bigl[ \tensor*[^{\scr \! - \!}]{\Delta}{^{\scr \!+ \! } } \bigr]_\lambda(x;x')
	=
(aa')^{-\frac{D-2}{2}} \!\!
\int\! \frac{ d^{D-1}k }{ (2\pi)^{D-1 } } \, e^{i\vec{k} \cdot (\vec{x} - \vec{x}^{\, \prime}) }
	U_\lambda(\eta,k) U_\lambda^*(\eta',k)
	\, .
\label{int over modes finite}
\end{equation}
This integral representation evaluates to~\cite{Janssen:2007ht,Janssen:2008px},
%
%
%
\begin{equation}
i \bigl[ \tensor*[^{\scr \! - \!}]{\Delta}{^{\scr \! + \! } } \bigr]_\lambda(x;x')
	=
	i \Delta_\lambda(y_{\scr -+} , u )
	=
	e^{-\frac{(D-2)\epsilon}{2(1-\epsilon)} u } \mathcal{F}_\lambda(y_{\scr -+}) 
	\, ,
\label{scalar prop solution}
\end{equation}
where~$y_{\scr -+}$ is the~$i\delta$-regulated distance 
function appropriate for 
the positive-frequency Wightman function,
as introduced in~(\ref{Delta x2 prescription}) together with~$u$,
and the rescaled propagator function is
expressed in terms of the hypergeometric function,
\begin{align}
\mathcal{F}_\lambda(y)
	={}&
	\frac{\bigl[ (1\!-\!\epsilon) H_0 \bigr]^{D-2} }{ (4\pi)^{ \frac{D}{2} } }
	\frac{ \Gamma\bigl( \frac{D-1}{2} \!+\! \lambda \bigr) \, 
		\Gamma\bigl( \frac{D-1}{2} \!-\! \lambda \bigr) }{ \Gamma\bigl( \frac{D}{2} \bigr) }
\nonumber \\
&	\hspace{2cm}
	\times
	{}_2F_1\biggl( \Bigl\{ \frac{D\!-\!1}{2} \!+\! \lambda , \frac{D\!-\!1}{2} \!-\! \lambda \Bigr\} , 
		\Bigl\{ \frac{D}{2} \Bigr\} , 1 \!-\! \frac{y}{4} \biggr)
		\, ,
\label{F def}
\end{align}
that satisfies the hypergeometric equation in disguise,
\begin{equation}
\biggl[
	(4y \!-\! y^2) \frac{\partial^2}{\partial y^2}
	+ D (2\!-\!y) \frac{ \partial }{ \partial y }
	+ \lambda^2 - \Bigl( \frac{ D\!-\!1 }{2} \Bigr)^{\!2}
	\biggr]
	\mathcal{F}_\lambda(y) 
	=
	0
	\, .
\label{F equation}
\end{equation}
The function in~(\ref{F def}) admits a power series representation around~$y\!=\!0$,
\begin{align}
\MoveEqLeft[1]
\mathcal{F}_\lambda(y) 
	=
	\frac{\bigl[ (1\!-\!\epsilon) H_0 \bigr]^{D-2} \, \Gamma\bigl( \frac{D-2}{2} \bigr) }{ (4\pi)^{ \frac{D}{2} } } 
	\biggl\{
	\Bigl( \frac{y}{4} \Bigr)^{\! -\frac{D-2}{2} }
	+ \frac{ \Gamma\bigl( \frac{4-D}{2} \bigr) }{ \Gamma\bigl( \frac{1}{2} \!+\! \lambda \bigr) \, \Gamma\bigl( \frac{1}{2} \!-\! \lambda \bigr) } 
		\sum_{n=0}^{\infty}
\label{power series}
\\
&
	\times \!
	\biggl[
	\frac{ \Gamma\bigl( \frac{3}{2} \!+\! \lambda \!+\! n \bigr) \, \Gamma\bigl( \frac{3}{2} \!-\! \lambda \!+\! n \bigr) }
		{ \Gamma\bigl( \frac{6-D}{2} \!+\! n \bigr) \, (n\!+\!1)! } \Bigl( \frac{y}{4} \Bigr)^{\!n - \frac{D-4}{2}}
	-
	\frac{ \Gamma\bigl( \frac{D-1}{2} \!+\! \lambda \!+\! n \bigr) \, \Gamma\bigl( \frac{D-1}{2} \!-\! \lambda \!+\! n \bigr) }
		{ \Gamma\bigl( \frac{D}{2} \!+\! n \bigr) \, n! }
	\Bigl( \frac{y}{4} \Bigr)^{\!n }
	\biggr]
	\biggr\} 
	\, ,
\nonumber 
\end{align}
that we will make use of.  We will also use recurrence relations between 
propagator functions with contiguous indices,
\begin{align}
2 \frac{\partial \mathcal{F}_\lambda}{\partial y}
	={}&
	(2\!-\!y) \frac{\partial \mathcal{F}_{\lambda+1} }{ \partial y }
	- \Bigl( \frac{D\!-\!3}{2} \!-\! \lambda \Bigr) \mathcal{F}_{\lambda+1}
	\, ,
\label{contiguousF 1}
\\
2 \frac{\partial \mathcal{F}_{\lambda}}{\partial y}
	={}&
	(2\!-\!y) \frac{\partial \mathcal{F}_{\lambda-1} }{ \partial y }
	- \Bigl( \frac{D\!-\!3}{2} \!+\! \lambda \Bigr) \mathcal{F}_{\lambda-1}
	\, .
\label{contiguousF 2}
\end{align}
that follow from Gauss' recursion relations
for hypergeometric functions~(see~(9.137) in~\cite{Gradshteyn:2007}). 

\medskip

The solution for the scalar Feynman propagator, that is expressed in terms of the
Wightman function as in~(\ref{Feynman definition}) is then given by,
%
%
%
\begin{equation}
i \bigl[ \tensor*[^{\scr \! + \!}]{\Delta}{^{\scr \! + \! } } \bigr]_\lambda(x;x')
	=
	i \Delta_\lambda(y_{\scr ++} , u )
	=
	e^{-\frac{(D-2)\epsilon}{2(1-\epsilon)} u } \mathcal{F}_\lambda(y_{\scr ++}) 
	\, ,
\end{equation}
where~$y_{\scr ++}$ is complexified according to the prescription in~(\ref{Delta x2 prescription}). The remaining two-point functions are obtained from the definitions,
which in the end amounts to simply changing the polarity indices on the~$y$ variable,
denoting the appropriate prescription.
Henceforth we suppress denoting explicitly the Schwinger-Keldysh polarity indices 
on variables, as they are either defined by the corresponding indices of the object
being computed, or otherwise should be clear from the context.
The expressions without any indices are meant to be valid in general.

\subsubsection{IR divergent scalar two-point functions}
\label{subsubsec: IR divergent scalar two-point functions}

When the index of the CTBD mode function is~$\lambda\!\geq \! (D\!-\!1)/2$ the integral 
in~(\ref{int over modes finite}) is divergent in the IR~\cite{Ford:1977in,Janssen:2009nz}. 
This implies that~(\ref{mode function}) is not a legitimate 
choice for the mode functions in the IR sector.
Rather, one should resort to choosing Bogolyubov coefficients 
in~(\ref{general mode function}) such that 
the sum-over-modes representation~(\ref{int over modes}) of the two-point function
is well defined and convergent in the IR.
This is guaranteed by fixing initial conditions at some time~$\eta_0$,
\begin{equation}
\mathscr{U}_\lambda(\eta_0,k) 
	= 
	\overline{U}(k)
	\, ,
\qquad \qquad
\partial_0\mathscr{U}_\lambda(\eta_0,k) 
	= 
	\partial_0\overline{U}(k)
	\, ,
\end{equation}
such that~$\overline{U}(k) \!\overset{ k\to0 }{\propto} \! k^p $, 
where~$p\!\geq\! - (D\!-\!1)/2$.
This in turn determines the Bogolyubov coefficients in~(\ref{general mode function}),
\begin{subequations}
\begin{align}
\alpha(k) 
	={}&
	k U_{\lambda+1}^*(\eta_0,k) \overline{U} (k)
	+ 
	i U_\lambda^*(\eta_0,k) \biggl[ \partial_0 \overline{U}(k)
		+ \Bigl( \lambda \!+\! \frac{1}{2} \Bigr) (1\!-\!\epsilon) H_0 \overline{U}(k) \biggr]
	\, ,
\\
\beta(k)
	={}&
	k U_{\lambda+1}(\eta_0,k) \overline{U} (k)
	- 
	i U_\lambda(\eta_0,k) \biggl[ \partial_0 \overline{U}(k)
		+ \Bigl( \lambda \!+\! \frac{1}{2} \Bigr) (1\!-\!\epsilon) H_0 \overline{U}(k) \biggr]
	\, .
\end{align}
\label{Bogolyubov}%
\end{subequations}
The evolution cannot generate IR divergences in~(\ref{int over modes}) 
once they are absent at 
the initial time~\cite{Ford:1977in}. 
There are many possible choices for defining an IR finite state, but since we are interested in
studying loop corrections due to interactions, the appropriate choice for Bogolyubov 
coefficients in~(\ref{general mode function})
is the one that minimizes the sensitivity to the details of regulating the IR.
This we accomplish by assuming the full mode function to be the CTBD one all the way down
to some deep IR scale~$k_0\!\ll\!H_0$,
\begin{equation}
k > k_0 \qquad : \qquad
\overline{U}(k) = U_\lambda(\eta_0,k)  \, ,
\quad {\tt and} \quad
\partial_0 \overline{U}(k) = \partial_0 U_\lambda(\eta_0,k)  \, ,
\end{equation}
while the suppression for long wavelength modes~$k\!<\!k_0$ is implemented via 
Bogolyubov coefficients~(\ref{Bogolyubov}) in some reasonable way.
But given that~$k_0/H_0 \!\ll\! 1$ we can neglect all positive powers of~$k_0$ that
appear in the two-point function, and the modes below~$k_0$ can generate only such 
contributions. 
In accelerating spaces this approximation only becomes better at later times, as the
infrared scale always appears as a decaying ratio~$k_0/\mathcal{H}$.
Thus, keeping only the negative powers of the deep IR scale~$k_0$,
the two-point function is very well approximated by effectively
introducing an IR cutoff~$k_0$ into the integral~(\ref{int over modes finite}),
\begin{equation}
i \bigl[ \tensor*[^{\scr \! - \!}]{\Delta}{^{\scr \!+ \! } } \bigr]_\lambda(x;x')
	=
(aa')^{-\frac{D-2}{2}} \!\!
\int\! \frac{ d^{D-1}k }{ (2\pi)^{D-1 } } \, e^{i\vec{k} \cdot (\vec{x} - \vec{x}^{\, \prime}) } \,
	\theta(k \!-\! k_0)
	U_\lambda(\eta,k) U_\lambda^*(\eta',k)
	\, .
\label{scalar 2pt cutoff def}
\end{equation}
The two-point function can now be computed analytically by splitting it into two 
parts, according to the split of the step 
function~$\theta(k \!-\! k_0) \!=\! 1 - \theta(k_0 \!-\! k)$,
and compute each part separately. However, since now each of the two parts will
be given by an integral with zero as the lower limit of integration, both integrals
will be IR divergent on their own. Despite this, their sum is finite because they come from splitting
a manifestly finite quantity~(\ref{scalar 2pt cutoff def}). This is why we are allowed 
to compute each individual part by regulating the IR by dimensional regularization,
and after adding them together we are guaranteed to obtain a result that is 
correct as the two errors cancel each other out.~\footnote{Dimensional 
regularization automatically subtracts any power-law divergences, whether UV or IR. 
While for the UV this is a convenient property, as it reduces the labor needed to renormalize 
a theory, it should not be employed to subtract divergences in the IR, as they are 
not universal but rather are symptomatic of unphysical properties of a chosen state,
and need to
be approached with due care~\cite{Janssen:2009nz}.}
We denote the split as,
\begin{equation}
i \bigl[ \tensor*[^{\scr \!-\! } ]{\Delta}{^{\scr \!+\!} } \bigr]_{\lambda}(x;x')
	=
	e^{-\frac{(D-2)\epsilon}{2(1-\epsilon)} u}
		\Bigl[
		\mathcal{F}_\lambda(y) + \mathcal{W}_\lambda(y,u,v)
		\Bigr]
	=
	i \Delta_\lambda(y, u , v)
		\, .
\label{IR breaking scalar 2pt}
\end{equation}
where the first (bulk) part corresponds to the rescaled propagator function~(\ref{F def}),
and the IR part is given by,
\begin{equation}
\mathcal{W}_\lambda(y,u,v) = 
	- (aa')^{-\frac{(D-2)(1-\epsilon)}{2}} \!\!
\int\! \frac{ d^{D-1}k }{ (2\pi)^{D-1 } } \, e^{i\vec{k} \cdot (\vec{x} - \vec{x}^{\, \prime}) } \,
	\theta(k_0\!-\!k)
	U_\lambda(\eta,k) U_\lambda^*(\eta',k)
	\, .
\end{equation}
We evaluate this integral by only keeping terms that can potentially contribute
negative powers of the effective IR cutoff~$k_0$.

To evaluate the IR part we first integrate over the angular coordinates.
This is accomplished as in~\cite{Janssen:2008px}, by first writing out the volume element
in angular coordinates,~$d^{D-1}k \!=\! k^{D-2} dk \, d \Omega_{D-2}$,
where the surface element of the~$(D\!-\!2)$-sphere,
$d\Omega_{D-2} \!=\! \sin^{D-3}(\theta_{D-3}) d\theta_{D-3} \sin^{D-4}(\theta_{D-4}) d\theta_{D-4} \dots d\varphi$,
is  parametrized by $(D\!-\!2)$ angles 
$\theta_{D-3}, \theta_{D-4}, \dots, \varphi$.
Then using the expression for the surface of the~($D\!-\!3$)-sphere,
and the representation of the Bessel function 8.411.7 from~\cite{Gradshteyn:2007},
\begin{equation}
\int\! d \Omega_{D-3} = \frac{2 \pi^{\frac{D-2}{2}} }{ \Gamma\bigl( \frac{D-2}{2} \bigr) } \, ,
\qquad
J_\lambda(z) = \frac{ \bigl( \frac{z}{2} \bigr)^{\lambda} }{ \sqrt{\pi} \, \Gamma\bigl( \frac{1}{2} \!+\! \lambda \bigr) }
	\int_{0}^{\pi}\! d \theta \, e^{ \pm i z \cos(\theta) } \sin^{2\lambda}(\theta) \, ,
\label{some integrals}
\end{equation}
we arrive at the one-dimensional integral expression,
\begin{equation}
\mathcal{W}_\lambda
	=
	-
	\frac{ (aa')^{-\frac{(D-2)(1-\epsilon)}{2}} }{ (2\pi)^{\frac{D-1}{2}} \| \Delta\vec{x} \|^{ \frac{D-3}{2} }  } 
\int_{0}^{k_0}\!\! dk \, k^{ \frac{D-1}{2} }
	J_{ \frac{D-3}{2} } \bigl( k \| \Delta\vec{x} \| \bigr)
	U_\lambda(\eta,k) U_\lambda^*(\eta',k)
	\, .
\end{equation}
Next we expand the mode functions in the small argument limit, keeping only
terms potentially generating negative powers of~$k_0$,
and collecting the same powers of~$k$,
\begin{align}
\mathcal{W}_\lambda
	={}&
	- \frac{ (aa')^{-\frac{(D-2)(1-\epsilon)}{2}} \, 
		2^{2\lambda-1} \, \Gamma^2(\lambda) \, \Gamma^2(1\!-\!\lambda) }
		{  (2\pi)^{\frac{D+1}{2}} (1\!-\!\epsilon) \sqrt{\mathcal{H} \mathcal{H}' } \| \Delta\vec{x} \|^{ \frac{D-3}{2} }  } 
\nonumber \\
&
	\times
	\sum_{N=0}^{\infty} \sum_{n=0}^{N} 
	\frac{ \bigl( -\frac{1}{4} \bigr)^{N} }{ n! ( N \!-\! n )! \, \Gamma(n \!+\! 1 \!-\! \lambda) \, \Gamma(N \!-\! n \!+\! 1 \!-\! \lambda) } 
	\Bigl( \frac{ \mathcal{H} }{\mathcal{H}' } \Bigr)^{\!N-2n }
\nonumber \\
&	\hspace{1cm}
	\times
	\int_{0}^{k_0}\!\! dk \, k^{ \frac{D-1}{2} }
	J_{ \frac{D-3}{2} } \bigl( k \| \Delta\vec{x} \| \bigr)
	\Bigl[ \frac{k^2}{ (1\!-\!\epsilon)^2 \mathcal{H} \mathcal{H}' } \Bigr]^{N -\lambda}
	\, .
\end{align}
The integral over a single Bessel function and a power
can be found in 1.8.1.1.~of~\cite{Prudnikov2},
\begin{equation}
\int_{0}^{z} \!\! dz' \, z'^\rho J_{\sigma}(z')
	\! = \!
	\frac{ z^{\rho+\sigma+1} }{ 2^\sigma (\rho \!+\! \sigma \!+\! 1) \, \Gamma(\sigma\!+\!1) }
	\, {}_1F_2\biggl( \Bigl\{ \frac{\rho + \sigma + 1}{2} \Bigr\} , 
		\Bigl\{ \frac{\rho + \sigma + 3}{2} , \sigma \!+\! 1 \Bigr\} , - \frac{z^2}{4} \biggr)
		\, ,
\label{IR int ref}
\end{equation}
so that after some rearrangements, and after
applying the Legendre duplication formula for the gamma function in the overall factor,
the IR series takes the form,
\begin{align}
\MoveEqLeft[2]
\mathcal{W}_\lambda
	=
	-
	\frac{ \bigl[ (1 \!-\! \epsilon) H_0 \bigr]^{D-2 }  }
		{ (4\pi)^{\frac{D}{2}} } 
	\frac{ \Gamma(\lambda) \, \Gamma(2\lambda)  }
		{ \Gamma\bigl( \frac{1}{2} \!+\! \lambda \bigr) \, \Gamma\bigl( \frac{D-1}{2} \bigr)  } 
	\sum_{N=0}^{\infty}
	\sum_{n=0}^{N} 
	\frac{ 1 }{  \bigl( \frac{D - 1}{2} \!+\! N \!-\! \lambda \bigr) }
\nonumber \\
&
	\times
	\frac{ \bigl( -\frac{1}{4} \bigr)^{N} \, \Gamma^2(1\!-\!\lambda) }
		{ n! ( N \!-\! n )! \, \Gamma(n \!+\! 1 \!-\! \lambda) \, \Gamma(N \!-\! n \!+\! 1 \!-\! \lambda) } 
	\Bigl[ \frac{ k_0^2 }{ (1\!-\!\epsilon)^2 \mathcal{H} \mathcal{H}' } 
		\Bigr]^{ \frac{D-1}{2} + N -\lambda}
	\Bigl( \frac{ \mathcal{H} }{\mathcal{H}' } \Bigr)^{\!N-2n }
\nonumber \\
&	\hspace{1cm}
	\times
	{}_1F_2\biggl( \Bigl\{ \frac{ D \!-\! 1 }{2} \!+\! N \!-\! \lambda \Bigr\} , 
		\Bigl\{ \frac{ D \!+\! 1 }{2} \!+\! N \!-\! \lambda , \frac{D \!-\! 1}{2} \Bigr\} , 
		- \frac{ (k_0 \| \Delta\vec{x} \| ) ^2}{4} \biggr)
	\, .
\end{align}
Finally, restricting to the domain where~$k_0 \| \Delta\vec{x} \| \!\ll\! 1$,
expanding the hypergeometric function above  in a power series, and rearranging the 
triple series gives,
\begin{align}
\MoveEqLeft[2]
\mathcal{W}_\lambda(y,u,v)
	=
	-
	\frac{ \bigl[ (1 \!-\! \epsilon) H_0 \bigr]^{D-2 }  }
		{ (4\pi)^{\frac{D}{2}} } 
	\frac{ \Gamma(\lambda) \, \Gamma(2\lambda)  }
		{ \Gamma\bigl( \frac{1}{2} \!+\! \lambda \bigr) \, \Gamma\bigl( \frac{D-1}{2} \bigr)  } 
	\sum_{N= 0}^{ \lfloor \lambda - \frac{D-1}{2} \rfloor }
	\sum_{n=0}^{N}
	\sum_{\ell=0}^{N-n} 
	\frac{ c_{Nn\ell} }{ \bigl( \frac{D-1}{2} \!+\! N \!-\! \lambda \bigr) }
\nonumber  \\
&
	\times
	\Bigl[ \frac{ k_0^2 e^{-u} }{ (1\!-\!\epsilon)^2 H_0^2 } 
		\Bigr]^{ \frac{D-1}{2} -\lambda + N}
	\Bigl[ y + 4 \, {\rm sh}^2\Bigl( \frac{v}{2} \Bigr) \Bigr]^n
	\,
	\ch \bigl[ ( N \!-\! n \!-\! 2\ell ) v  \bigr]
	\, ,
\label{W result}
\end{align}
where the coefficients are,
\begin{equation}
c_{Nn\ell}
	=
	\frac{ \bigl( - \frac{1}{4} \bigr)^{\!N}  }{ n! \, \ell! \, (N \!-\! n \!-\! \ell) ! }
	\frac{ \Gamma^2(1\!-\!\lambda) \, \Gamma\bigl( \frac{D-1}{2} \bigr) }
		{ \Gamma\bigl( \frac{D-1}{2} \!+\! n \bigr) \, \Gamma(\ell \!+\! 1 \!-\! \lambda) \, \Gamma( N \!-\! n \!-\! \ell \!+\! 1 \!-\! \lambda) } 
	\, ,
\label{c coeff def}
\end{equation}
and where the sum over~$N$
is cut by the floor function in the upper limit to~$N\!\le\!\lambda \!-\! (D\!-\!1)/2$, which guarantees
that only the negative powers of~$k_0$ are kept, while positive ones are neglected.
This corresponds to the result obtained in~\cite{Janssen:2008px}, where the infrared regulation
was implemented by considering the spatial sections of the spacetime to be flat torii with a 
large radius. Here, however, it is advantageous to consider the IR regulation to be implemented by
the Bogolyubov coefficients defining the initial state, in order to avoid having to deal with
linearization instability~\cite{Moncrief:1976un,Arms:1979au,Higuchi:1991tk,Higuchi:1991tm,Tsamis:1992xa,Marolf:1995cn,Marolf:2008it,Marolf:2008hg,Miao:2009hb,Gibbons:2014zya,Altas:2021htf} 
that appears in gauge theories on compact manifolds. 
The fact that the two expressions for the two-point function
agree regardless of the difference in the method for regulating the IR is due to the expressions in actuality being
valid only for spatial separations~$ k_0 \| \Delta \vec{x} \| \!\ll\! 1$.
However, since $k_0\ll H_0$ is chosen to be smaller than any physical scale of interest, that limitation is of no real significance.

\bigskip

Writing out the equation of motion~(\ref{scalar EOM}) off-coincidence
in terms of the bi-local variabls reads,
\begin{align}
\MoveEqLeft[0.5]
\Biggl[
\bigl( 4y \!-\! y^2 \bigr) \frac{\partial^2}{\partial y^2}
	+ D(2 \!-\! y) \frac{\partial}{\partial y}
	+ 2 \bigl( 2 \!-\! y \!-\! 2 e^{-v} \bigr)
		\biggl( \frac{\partial}{\partial u} \!+\! \frac{\partial}{\partial v}
			\!+\! \frac{ (D\!-\!2)\epsilon }{ 2 (1\!-\!\epsilon) } \biggr) \frac{\partial}{\partial y}
\nonumber \\
&
	- \biggl( \frac{\partial}{\partial u} \!+\! \frac{\partial}{\partial v} 
		\!+\! \frac{ D \!-\! 1 \!-\! \epsilon }{ 1 \!-\! \epsilon } \biggr) 
		\Bigl( \frac{\partial}{\partial u} \!+\! \frac{\partial}{\partial v}  \Bigr)
	+
	\lambda^2 \!-\! \Bigl( \frac{ D \!-\! 1 \!-\! \epsilon }{ 2(1\!-\!\epsilon) } \Bigr)^{\!2}
	\Biggr]
	i \Delta_\lambda(y,u,v)
	=
	0
	\, .
\label{Delta lambda EOM}
\end{align}
This is a generalization of Eq.~(\ref{F equation}) to the case when the CTBD mode function
cannot be used due to an IR divergence. In this case there is another independent equation
descending from the primed equation of motion~(\ref{scalar EOM primed}), that is obtained by
substituting~$v \!\to\! -v$ everywhere inside the square brackets in~(\ref{Delta lambda EOM}). The two equations can be combined into another two equations of well 
defined parity that have to be satisfied independently. Namely, the even equation is,
\begin{align}
&
\Biggl[
	\bigl( 4y \!-\! y^2 \bigr) \frac{\partial^2}{\partial y^2}
	+ D(2 \!-\! y) \frac{\partial}{\partial y}
	- 2 \Bigl[ y \!+\! 4 \, \sh^2 \Bigl( \frac{v}{2} \Bigr) \Bigr]
		\biggl( \frac{\partial}{\partial u} 
			\!+\! \frac{ (D\!-\!2)\epsilon }{ 2 (1\!-\!\epsilon) } \biggr) \frac{\partial}{\partial y}
\label{scalar eom even}
\\
&
	- \biggl( \frac{\partial}{\partial u}
		\!+\! \frac{ D \!-\! 1 \!-\! \epsilon }{ 1 \!-\! \epsilon } \biggr) 
		\frac{\partial}{\partial u}
	+ \biggl( 4 \, \sh(v) \frac{\partial}{\partial y}
		\!-\! \frac{\partial}{\partial v} \biggr) \frac{\partial}{\partial v} 
	+
	\lambda^2 \!-\! \Bigl( \frac{ D \!-\! 1 \!-\! \epsilon }{ 2(1\!-\!\epsilon) } \Bigr)^{\!2}
	\Biggr]
	i \Delta_\lambda(y,u,v)
	=
	0
	\, ,
\nonumber 
\end{align}
and the odd equation reads,
\begin{align}
&
\Biggl[
	\biggl( \Bigl[ y + 4 \, \sh^2\Bigl( \frac{v}{2} \Bigr) \Bigr]
		\frac{\partial}{\partial y}
	+
	\frac{\partial}{\partial u} 
	+ 
	\frac{ D \!-\! 1 \!-\! \epsilon }{ 2(1\!-\!\epsilon) }
		\biggr) \frac{\partial}{\partial v} 
\nonumber \\
&	\hspace{4cm}
	- 2 \, \sh(v)  \biggl( \frac{\partial}{\partial u} + \frac{(D\!-\!2)\epsilon}{2 (1\!-\!\epsilon)} \biggr) 
		 \frac{\partial}{\partial y}
	\Biggr]
	i \Delta_\lambda(y,u,v)
	=
	0
	\, .
\label{scalar eom odd}
\end{align}
We can also derive generalizations of recurrence relations
for rescaled propagator functions~(\ref{contiguousF 1}) 
and~(\ref{contiguousF 2}) that apply for the 
full scalar two-point functions. These are derived from
recurrence relations between mode functions~(\ref{mode recurrence})
that imply the following two reflection identities for time derivatives,
\begin{subequations}%
\begin{align}
\MoveEqLeft[10]
\biggl[ \partial_0 - \biggl( \lambda \!-\! \frac{D \!-\! 3 \!+\! \epsilon}{2(1 \!-\! \epsilon)} \biggr) (1\!-\!\epsilon) \mathcal{H} \biggr]
	i \bigl[ \tensor*[^{\tt a \!}]{\Delta}{^{\! \tt b}} \bigr]_{\lambda+1}(x;x')
\nonumber \\
	={}&
	- \biggl[ \partial_0' 
		+ \biggl( \lambda \!+\! 1 \!+\! \frac{D \!-\! 3 \!+\! \epsilon}{2(1 \!-\! \epsilon)} \biggr) (1\!-\!\epsilon) \mathcal{H}' \biggr]
		i \bigl[ \tensor*[^{\tt a \!}]{\Delta}{^{\! \tt b}} \bigr]_{\lambda}(x;x')
		\, ,
\\
\MoveEqLeft[10]
\biggl[ \partial_0' - \biggl( \lambda \!-\! \frac{D \!-\! 3 \!+\! \epsilon}{2(1\!-\!\epsilon)} \biggr) (1\!-\!\epsilon) \mathcal{H}' 
		\biggr] i \bigl[ \tensor*[^{\tt a \!}]{\Delta}{^{\! \tt b}} \bigr]_{\lambda+1}(x;x')
\nonumber \\
	={}&
	- \biggl[ \partial_0
		+ \biggl( \lambda \!+\! 1 \!+\! \frac{D \!-\! 3 \!+\! \epsilon}{2(1\!-\!\epsilon)} \biggr) (1\!-\!\epsilon) \mathcal{H} 
		\biggr] i \bigl[ \tensor*[^{\tt a \!}]{\Delta}{^{\! \tt b}} \bigr]_{\lambda}(x;x')
		\, .
\end{align}
\end{subequations}
Then writing these out in terms of bi-local variables 
yields the desired generalized recurrence relations,
\begin{subequations}
\begin{align}
\MoveEqLeft[4]
	\biggl[ 
	2 \frac{\partial}{\partial y}
	-
	\frac{ 1 }{ \sh(v) } 
		\frac{\partial}{\partial v}
		\biggr] 
	i \Delta_\lambda(y,u,v)
\nonumber \\
={}&
	\biggl[ 
	( 2 \!-\! y ) \frac{\partial}{\partial y}
	-
	\Bigl( \frac{D \!-\! 3 \!+\! \epsilon }{2(1 \!-\! \epsilon) } \!-\! \lambda \Bigr) 
	-
	\frac{\partial}{\partial u}
	-
	\frac{\ch(v)}{\sh(v)}
		\frac{\partial}{\partial v}
	 \biggr]
	i \Delta_{\lambda+1}(y,u,v)
	\, ,
\label{generalized recurrence 2}
\\
\MoveEqLeft[4]
	\biggl[ 
	2 \frac{\partial}{\partial y}
	-
	\frac{ 1 }{ \sh(v) } 
		\frac{\partial}{\partial v}
	 \biggr]
	i \Delta_{\lambda}(y,u,v)
\nonumber \\
={}&
	\biggl[ 
	( 2 \!-\! y ) \frac{\partial}{\partial y}
	-
	\Bigl( \frac{D \!-\! 3 \!+\! \epsilon }{2(1 \!-\! \epsilon) } \!+\! \lambda \Bigr)
	-
	 \frac{\partial}{\partial u}
	-
	\frac{\ch(v)}{\sh(v)}
		\frac{\partial}{\partial v}
		\biggr] 
	i \Delta_{\lambda-1}(y,u,v) 
	\, ,
\label{generalized recurrence 1}
\end{align}
\label{generalized recurrence}%
\end{subequations}
that we make frequent use of in the remainder of the paper.

\section{Photon in FLRW}
\label{sec: Photon in FLRW}

This section is devoted to a brief recap of 
the procedure for imposing the multiplier gauge for electromagnetism
that corresponds to the general covariant gauge-fixing term~(\ref{gauge}), 
and to the canonical quantization
of the resulting gauge-fixed theory. 
The details of this procedure, applicable to more general 
gauges and general FLRW spacetimes, are given in~\cite{Glavan:2022pmk}.

\subsection{Classical photon in FLRW}
\label{subsec: Classical photon in FLRW}

Electromagnetism in~$D$-dimensional curved spacetime is given by the covariantized Maxwell action,
\begin{equation}
S[A_\mu] = \int\! d^{D\!}x \, \sqrt{-g} \, \biggl[ - \frac{1}{4} g^{\mu\rho} g^{\nu\sigma} F_{\mu\nu} F_{\rho\sigma} \biggr] \, ,
\label{invariant action}
\end{equation}
that is invariant under~$U(1)$ gauge 
transformations,~$A_\mu(x) \!\to\! A_\mu(x) \!+\! \partial_\mu \Lambda(x)$.
In FLRW cosmological spaces the corresponding canonical formulation of electromagnetism 
is given by the canonical action~\cite{Glavan:2022pmk},
\begin{equation}
\mathscr{S} \bigl[ A_0, \Pi_0, A_i, \Pi_i, \ell \, \bigr]
	= \int\! d^{D\!}x \, \Bigl[
		\Pi_0 \partial A_0 + \Pi_i \partial_0 A_i - \mathscr{H} - \ell \, \Psi_1
		\Bigr] 
		\, ,
\label{invariant canonical action}
\end{equation}
where~$(A_0, \Pi_0)$ and~$(A_i,\Pi_j)$ are canonical pairs of vector potentials and 
their conjugate momenta, whose Poisson brackets are determined by the symplectic 
part of the action, with the non-vanishing ones being,
\begin{equation}
\bigl\{ A_0(\eta,\vec{x}) , \Pi_0(\eta, \vec{x}^{\,\prime}) \bigr\} = \delta^{D-1}(\vec{x} \!-\! \vec{x}^{\,\prime})
\, ,
\qquad
\bigl\{ A_i(\eta,\vec{x}) , \Pi_j(\eta, \vec{x}^{\,\prime}) \bigr\} = \delta_{ij} \delta^{D-1}(\vec{x} \!-\! \vec{x}^{\,\prime})
\, .
\label{Poisson brackets}
\end{equation}
The canonical Hamiltonian is given by,
\begin{equation}
\mathscr{H} = \frac{ a^{4-D} }{2} \Pi_i \Pi_i
	- A_0 \partial_i \Pi_i 
	+ \frac{a^{D-4}}{4} F_{ij} F_{ij} \, ,
\label{canonical Hamiltonian}
\end{equation}
where repeated lower spatial indices are summed over,
and~$\ell$ is a Lagrange multiplier that generates the primary constraint,
\begin{equation}
\Psi_1 = \Pi_0 \approx 0 \, ,
\label{primary constraint}
\end{equation}
that vanishes on-shell, and whose conservation in turn
generates a secondary constraint,
\begin{equation}
0 \approx \partial_0 \Psi_1 \approx \Psi_2
	= \partial_i \Pi_i \, .
\label{secondary constraint}
\end{equation}
that also vanishes on-shell.
The conservation of the secondary constraint generates no further constraints,
and thus the two form a complete set of first-class 
constraints,~$\bigl\{ \Psi_1 , \Psi_2 \bigr\} \!=\! 0$.
The fact that these two formulations --- configuration space one given by~(\ref{invariant action})
and the canonical one given by~(\ref{invariant canonical action}) --- are equivalent can be 
confirmed by checking that the two generate equivalent sets of equations of motion.

\medskip

The multiplier gauge is imposed by promoting the Lagrange multiplier in the
canonical action~(\ref{invariant canonical action}) to a function of canonical 
variables,~$\ell \!\to\! \overline{\ell}(A_0,\Pi_0,A_i,\Pi_i)$.
This way of fixing the gauge does not allow to reduce the phase space 
(such as the Coulomb gauge would), but rather we work with all the components 
of the vector potential.
On the other hand, multiplier gauges are much more conducive to preserving the 
symmetries of the gauge-invariant action. 
The particular choice for the multiplier that we consider here,
\begin{equation}
\ell \longrightarrow \overline{\ell} = - \frac{\xi}{2} a^{4-D} \Pi_0
	+ \partial_i A_i - (D\!-\!2) \mathcal{H} A_0 \, .
\end{equation}
is precisely motivated by this observation. This choice defines the gauge-fixed canonical action,
\begin{align}
\mathscr{S}_\star \bigl[ A_0, \Pi_0, A_i, \Pi_i \bigr]
	\equiv
	\mathscr{S} \bigl[ A_0, \Pi_0, A_i, \Pi_i, \ell \!\to\! \overline{\ell} \, \bigr]
	=
	\int\! d^{D\!}x \, \Bigl[
		\Pi_0 \partial A_0 + \Pi_i \partial A_i - \mathscr{H}_\star
		\Bigr] 
		\, ,
\label{gauge-fixed action}
\end{align}
with the gauge-fixed Hamiltonian,
\begin{equation}
\mathscr{H}_\star = \frac{ a^{4-D} }{2} \bigl( \Pi_i \Pi_i - \xi \Pi_0 \Pi_0 \bigr)
	+ \Pi_i \partial_i A_0 + \Pi_0 \partial_i A_i
	- (D\!-\!2) \mathcal{H} \Pi_0 A_0
	+ \frac{a^{D-4}}{2} F_{ij} F_{ij}
	\, .
\end{equation}
The the corresponding configuration space formulation of the gauge-fixed theory is 
consequently given by,
\begin{equation}
S_\star [A_\mu] = \int\! d^{D\!}x \, \sqrt{-g} \, \biggl[
	- \frac{1}{4} g^{\mu\rho} g^{\nu\sigma} F_{\mu\nu} F_{\rho\sigma} 
	- \frac{1}{2\xi} \bigl( \nabla^\mu A_\mu \bigr)^{\!2}
	\biggr]
	\, ,
\label{fixed action}
\end{equation}
which is precisely the gauge-invariant action~(\ref{invariant action})
with the general covariant gauge-fixing term~(\ref{gauge}) added.
The gauge-fixed action now defines the gauge-fixed dynamics, in the sense that the 
ambiguity present in the original equations of motion are now fixed.
For the gauge-fixed equations,
\begin{subequations}
\begin{align}
\partial_0 A_0 ={}&
	- \xi a^{4-D} \Pi_0 + \partial_i A_i - (D\!-\!2) \mathcal{H} A_0
	\, ,
\\
\partial_0 \Pi_0 ={}&
	\partial_i \Pi_i + (D\!-\!2) \mathcal{H} \Pi_0
	\, ,
\\
\partial_0 A_i ={}&
	a^{4-D} \Pi_i + \partial_i A_0
	\, ,
\\
\partial_0 \Pi_i ={}&
	\partial_i \Pi_0 + a^{D-4} \partial_j F_{ji}
	\, ,
\end{align}
\label{classical EOMs}%
\end{subequations}
the initial value problem is well defined once the initial conditions are specified.

However, the gauge-fixed action is not a complete gauge-fixed description of the system,
as it makes no reference to the first-class constraints~(\ref{primary constraint}) 
and~(\ref{secondary constraint}) that are present in the original gauge-invariant system.
In fact, the set of equations~(\ref{classical EOMs}) admits many more solutions than the 
original gauge-invariant system of equations, regardless of the gauge freedom.
That is why the two constraints~(\ref{primary constraint}) 
and~(\ref{secondary constraint}) have to be required to hold
as subsidiary conditions on the initial value surface,
i.e. on the initial conditions that are propagated by the evolution 
equations in~(\ref{classical EOMs}),
\begin{equation}
\Psi_1(\eta_0, \vec{x}) = \Pi_0 (\eta_0,\vec{x}) = 0 \, ,
\qquad \qquad
\Psi_2(\eta_0,\vec{x}) = \partial_i \Pi_i(\eta_0,\vec{x}) = 0 \, .
\label{initial constraints}
\end{equation}
Equations of motion preserve these conditions so that they are valid for all times once 
imposed initially. Therefore, it is the gauge-fixed action~(\ref{gauge-fixed action}) 
{\it and} the constraints~(\ref{initial constraints}) that define the gauge-fixed version of the 
theory.

\subsection{Quantized photon in FLRW}
\label{subsec: Quantized photon in FLRW}

The gauge-fixing procedure for the classical theory described in the preceding section
lends itself well to canonical quantization in the Heisenberg picture. The 
gauge-fixed dynamics is quantized straightforwardly by promoting vector potential
components and their conjugate momenta into field 
operators,~$A_0 \!\to\! \hat{A}_0$,~$\Pi_0 \!\to\! \hat{\Pi}_0$,~$A_i \!\to\! \hat{A}_i$,~$\Pi_i \!\to\! \hat{\Pi}_i$,
and their nonvanishing Poisson brackets~(\ref{Poisson brackets}) to canonical 
commutation relations,
\begin{equation}
\bigl[ \hat{A}_0(\eta,\vec{x}) , \hat{\Pi}_0(\eta, \vec{x}^{\,\prime}) \bigr] = i \delta^{D-1}(\vec{x} \!-\! \vec{x}^{\,\prime})
\, ,
\qquad
\bigl[ \hat{A}_i(\eta,\vec{x}) , \hat{\Pi}_j(\eta, \vec{x}^{\,\prime}) \bigr] = \delta_{ij} \, i \delta^{D-1}(\vec{x} \!-\! \vec{x}^{\,\prime})
\, .
\label{canonical commutators}
\end{equation}
The equations of motion the field operators satisfy are just the same as the classical 
equations~(\ref{classical EOMs}), with fields promoted to field operators.
The algebra of field operators is assumed represented on some space of states.
However, this space cannot be the usual Fock space because of the subsidiary 
conditions~(\ref{initial constraints}) that need to be quantized as well.

The generalization of the subsidiary condition to the quantized theory 
firstly requires that the expectation values of Hermitian operators associated to
classical constraints,~$\hat{\Psi}_1 \!=\! \hat{\Pi}_0$ and~$\hat{\Phi}_2 \!=\! \partial_i \hat{\Pi}_i$,
vanish at initial time on account of the correspondence principle,
\begin{equation}
\bigl\langle \Omega \bigr| \hat{\Psi}_1(\eta_0,\vec{x}) \bigl| \Omega \bigr\rangle = 0 \, ,
\qquad \quad
\bigl\langle \Omega \bigr| \hat{\Psi}_2(\eta_0,\vec{x}) \bigl| \Omega  \bigr\rangle = 0 \, .
\label{1pt constraint correlators}
\end{equation}
Secondly, all the correlators of constraints have to likewise vanish, 
e.g.~two-point correlators,
\begin{subequations}
\begin{align}
&
\bigl\langle \Omega \bigr| \hat{\Psi}_1(\eta_0,\vec{x}) \hat{\Psi}_1(\eta_0,\vec{x}^{\,\prime}) \bigl| \Omega \bigr\rangle = 0 \, ,
\\
&
\bigl\langle \Omega \bigr| \hat{\Psi}_1(\eta_0,\vec{x}) \hat{\Psi}_2(\eta_0,\vec{x}^{\,\prime}) \bigl| \Omega \bigr\rangle = 0 \, ,
\\
&
\bigl\langle \Omega \bigr| \hat{\Psi}_2(\eta_0,\vec{x}) \hat{\Psi}_2(\eta_0,\vec{x}^{\,\prime}) \bigl| \Omega \bigr\rangle = 0 \, ,
\end{align}
\label{2pt constraint correlators}%
\end{subequations}
which ensures no fluctuations of constraints can be measured. The evolution equations 
then ensure that
these correlators then vanish for all times. The conditions in~(\ref{1pt constraint correlators})
and~(\ref{2pt constraint correlators}) are naturally implemented as conditions 
on the state vectors forming the space of states, in the form of an operator annihilating 
the state,
\begin{equation}
\hat{K}(\vec{x}) \bigl| \Omega \bigr\rangle = 0 \, ,
\qquad
\forall \vec{x} \, . 
\label{K condition}
\end{equation}
Just as in the classical theory, the subsidiary conditions cut out a subspace of physically 
allowed initial conditions. In the quantized theory this condition is meant to
cut out a subset of physically allowed states from the vector space of states.
In order for the condition in~(\ref{K condition}) to be consistent with canonical
commutation relations in~(\ref{canonical commutators}) the operator~$\hat{K}$
cannot be Hermitian. Rather, it is some non-Hermitian linear combination of Hermitian constraint
operators,
\begin{equation}
\hat{K}(\vec{x}) = \int \! d^{D-1}x' \, 
	\Bigl[
	f_1(\eta_0 ; \vec{x}, \vec{x}^{\,\prime}) \hat{\Psi}_1(\eta_0,\vec{x})
	+
	f_2(\eta_0 ; \vec{x}, \vec{x}^{\,\prime}) \hat{\Psi}_2(\eta_0,\vec{x})
	\Bigr]
	\, .
\label{position K constraint}
\end{equation}
In fact, the equations of motion guarantee that such a time-independent 
linear combination can be formed from Hermitian constraints at any point in time,
\begin{equation}
\hat{K}(\vec{x}) = \int \! d^{D-1}x' \, 
	\Bigl[
	f_1(\eta ; \vec{x}, \vec{x}^{\,\prime}) \hat{\Psi}_1(\eta,\vec{x})
	+
	f_2(\eta ; \vec{x}, \vec{x}^{\,\prime}) \hat{\Psi}_2(\eta,\vec{x})
	\Bigr]
	\, ,
\end{equation}
where the time-dependent coefficient functions satisfy their own equations of motion,
\begin{equation}
\partial_0 f_1 = - \nabla^2 f_2 - (D\!-\!2) \mathcal{H} f_1 \, ,
\qquad \quad
\partial_0 f_2 = - f_1 \, ,
\end{equation}
There is a great deal of freedom in choosing initial conditions,
and hence in choosing the subsidiary constraint operator,
and particular choices can lead to more favourable expressions.

\section{Field operator dynamics}
\label{sec: Field operator dynamics}

The gauge-fixed dynamics of the linear theory is completely determined by solving the 
operator equations of motion, which are  just the classical equations~(\ref{classical EOMs})
with fields promoted to field operators, and by requiring the solutions to respect canonical
commutation relations~(\ref{canonical commutators}). This task is best accomplished
if we first employ a couple of decompositions of the field operators, after which some
of the equations decouple, and some can be brought into the 
form of the scalar mode equation~(\ref{mode eq}). This section is devoted to introducing
the necessary decompositions and to solving the resulting equations of motion.

Firstly, it is convenient to split the spatial components of field operators  
into transverse and longitudinal parts,
\begin{equation}
\hat{A}_i = \hat{A}_i^{\scr T} + \hat{A}_i^{\scr L} \, ,
\qquad \qquad
\hat{\Pi}_i = \hat{\Pi}_i^{\scr T} + \hat{\Pi}_i^{\scr L} \, ,
\end{equation}
such that~$\partial_i \hat{A}_i^{\scr T} \!=\! 0 \!=\! \partial_i \hat{\Pi}_i^{\scr T}$.
This is best accomplished with the use of transverse and longitudinal projectors,
\begin{equation}
\mathbb{P}^{\scr T}_{ij} = \delta_{ij} - \frac{\partial_i \partial_j}{\nabla^2} \, ,
\qquad \qquad
\mathbb{P}^{\scr L}_{ij} = \frac{\partial_i \partial_j}{\nabla^2} \, .
\end{equation}
These projectors are orthogonal,~$\mathbb{P}_{ij}^{\scr T} \mathbb{P}_{jk}^{\scr L} \!=\! \mathbb{P}_{ij}^{\scr L} \mathbb{P}_{jk}^{\scr T} \!=\! 0 $, and 
idempotent,~$\mathbb{P}_{ij}^{\scr T} \mathbb{P}_{jk}^{\scr T} \!=\! \mathbb{P}_{ik}^{\scr T} $,
and the field operator components are then projected out with correct 
properties,~$\hat{A}_i^{\scr T} \!=\! \mathbb{P}_{ij}^{\scr T} \hat{A}_j$,~$\hat{\Pi}_i^{\scr T} \!=\! \mathbb{P}_{ij}^{\scr T} \hat{\Pi}_j$,~$\hat{A}_i^{\scr L} \!=\! \mathbb{P}_{ij}^{\scr L} \hat{A}_j$,~$\hat{\Pi}_i^{\scr L} \!=\! \mathbb{P}_{ij}^{\scr L} \hat{\Pi}_j$.
After this split the equations of motion for the transverse operators 
decouple from the rest.

Secondly, given the isotropy of spatially flat cosmological spaces, it is convenient to 
expand the field operators in Fourier modes of the comoving momentum space,
\begin{subequations}%
\begin{align}
\hat{A}_0(\eta,\vec{x})
	={}&
	a^{- \frac{D-2}{2} } \!
	\int\! \frac{ d^{D-1}k }{ (2\pi)^{\frac{D-1}{2}} } \,
	e^{i \vec{k} \cdot \vec{x} }
	\, \hat{\mathcal{A}}_0(\eta,\vec{k}) \, ,
\\
\hat{\Pi}_0(\eta,\vec{x})
	={}&
		a^{\frac{D-2}{2} } \!
	\int\! \frac{ d^{D-1}k }{ (2\pi)^{\frac{D-1}{2}} } \,
	e^{i \vec{k} \cdot \vec{x} }
	\, \hat{\pi}_{0}(\eta,\vec{k}) \, ,
\\
\hat{A}_i^{\scr L}(\eta,\vec{x})
	={}&
	a^{- \frac{D-2}{2} } \!
	\int\! \frac{ d^{D-1}k }{ (2\pi)^{\frac{D-1}{2}} } \,
	e^{i \vec{k} \cdot \vec{x} }
	\frac{(-i ) k_i}{k}
	\, \hat{\mathcal{A}}_{\scr L}(\eta,\vec{k}) \, ,
\\
\hat{\Pi}_i^{\scr L}(\eta,\vec{x})
	={}&
	a^{\frac{D-2}{2} } \!
	\int\! \frac{ d^{D-1}k }{ (2\pi)^{\frac{D-1}{2}} } \,
	e^{i \vec{k} \cdot \vec{x} }
	\frac{(-i ) k_i}{k}
	\, \hat{\pi}_{\scr L}(\eta,\vec{k}) \, ,
\\
\hat{A}_i^{\scr T}(\eta,\vec{x})
	={}&
	a^{- \frac{D-4}{2} } \!
	\int\! \frac{ d^{D-1}k }{ (2\pi)^{\frac{D-1}{2}} } \,
	e^{i \vec{k} \cdot \vec{x} }
	\sum_{\sigma=1}^{D-2} \varepsilon_i^{\scr T}(\sigma,\vec{k})
	\, \hat{\mathcal{A}}_{{\scr T},\sigma}(\eta,\vec{k}) \, ,
\\
\hat{\Pi}_i^{\scr T}(\eta,\vec{x})
	={}&
	a^{\frac{D-4}{2} } \!
	\int\! \frac{ d^{D-1}k }{ (2\pi)^{\frac{D-1}{2}} } \,
	e^{i \vec{k} \cdot \vec{x} }
	\sum_{\sigma=1}^{D-2} \varepsilon_i^{\scr T}(\sigma,\vec{k})
	\, \hat{\pi}_{{\scr T},\sigma}(\eta,\vec{k}) \, ,
\end{align}
\label{Fourier transforms}%
\end{subequations}
where the powers of the scale factor have been factored out for later convenience.
Here we introduced transverse polarization tensors with the following properties,
\begin{align}
&
k_i \, \varepsilon_i(\sigma, \vec{k}) = 0 \, ,
\qquad
&&
\bigl[ \varepsilon_i (\sigma,\vec{k}) \bigr]^*
	= \varepsilon_i(\sigma, - \vec{k})
	\, ,
\\
&
\varepsilon_i^*(\sigma,\vec{k}) \, \varepsilon_i(\sigma', \vec{k})
	= \delta_{\sigma \sigma'} \, ,
\qquad
&&
\sum_{\sigma=1}^{D-2} \varepsilon_i^*(\sigma, \vec{k}) \, \varepsilon_j(\sigma, \vec{k})
	= \delta_{ij} - \frac{ k_i k_j }{ k^2 }
	\, ,
\end{align}
where~$k \!=\! \| \vec{k} \|$.
The Fourier transforms of field operators in~(\ref{Fourier transforms}) 
are themselves Hermitian, which in momentum space implies they behave under conjugation 
as~$\hat{\mathcal{O}}^\dag(\vec{k}) \!=\! \hat{\mathcal{O}}(-\vec{k})$.
The canonical commutators of the momentum space field operators are now,%
\begin{subequations}
\begin{align}%
&
\bigl[ \hat{\mathcal{A}}_0(\eta,\vec{k} ) , \hat{\pi}_0(\eta,\vec{k}^{\,\prime} ) \bigr]
	= \bigl[ \hat{\mathcal{A}}_{\scr L}(\eta,\vec{k} ) , \hat{\pi}_{\scr L}(\eta,\vec{k}^{\,\prime} ) \bigr]
	= i \delta^{D-1} ( \vec{k} \!+\! \vec{k}^{\,\prime} ) \, ,
\\
&
	\bigl[ \hat{\mathcal{A}}_{{\scr T},\sigma}(\eta,\vec{k} ) , \hat{\pi}_{{\scr T},\sigma'}(\eta,\vec{k}^{\,\prime} ) \bigr]
	= i \delta_{\sigma\sigma'}\delta^{D-1} ( \vec{k} \!+\! \vec{k}^{\,\prime} ) \, .
\end{align}
\label{momentum commutators}%
\end{subequations}
The equations of motion for the transverse sector are
\begin{align}
\partial_0 \hat{\mathcal{A}}_{{\scr T},\sigma}
	={}&
	\hat{\pi}_{{\scr T},\sigma} + \frac{1}{2} (D\!-\!4) \mathcal{H} \hat{\mathcal{A}}_{{\scr T},\sigma} \, ,
\label{AT eq}
\\
\partial_0 \hat{\pi}_{{\scr T},\sigma}
	={}&
	- k^2 \hat{\mathcal{A}}_{{\scr T},\sigma}
	- \frac{1}{2} (D\!-\!4) \mathcal{H} \hat{\pi}_{{\scr T},\sigma} \, .
\label{PiT eq}
\end{align}
while the equations of motion for the scalar sector, including longitudinal and temporal components, are
\begin{align}
\partial_0 \hat{\mathcal{A}}_{0}
	={}&
	- \xi a^2 \hat{\pi}_{0}
	+ k \hat{\mathcal{A}}_{\scr L}
	- \frac{1}{2} (D\!-\!2) \mathcal{H} \hat{\mathcal{A}}_{0} 
	\, ,
\label{A0 eq}
\\
\partial_0 \hat{\pi}_{0}
	={}&
	k \hat{\pi}_{\scr L}
	+ \frac{1}{2} (D\!-\!2) \mathcal{H} \hat{\pi}_{0} 
	\, ,
\label{Pi0 eq}
\\
\partial_0 \hat{\mathcal{A}}_{\scr L}
	={}&
	a^2 \hat{\pi}_{\scr L}
	- k \hat{\mathcal{A}}_{0}
	+ \frac{1}{2} (D\!-\!2) \mathcal{H} \hat{\mathcal{A}}_{\scr L}
	\, ,
\label{AL eq}
\\
\partial_0 \hat{\pi}_{\scr L}
	={}&
	- k \hat{\pi}_{0}
	- \frac{1}{2} (D\!-\!2) \mathcal{H} \hat{\pi}_{\scr L}
	\, .
\label{PiL eq}
\end{align}
%

\subsection{Transverse sector}
\label{subsec: Transverse sector}

The equations of motion~(\ref{AT eq}) and~(\ref{PiT eq}) 
of the transverse sector
combine into a single second order equation,
\begin{align}
\biggl[ \partial_0^2 + k^2 
	- \Bigl( \nu_{\scr T}^2 \!-\! \frac{1}{4} \Bigr) (1\!-\!\epsilon)^2 \mathcal{H}^2 \biggr]
		\hat{ \mathcal{A} }_{ {\scr T} , \sigma} ={}& 0 
\label{T 2nd order}
	\, ,
\\
\hat{\pi}_{ {\scr T} ,\sigma} ={}& \biggl[
	\partial_0 - \Bigl( \nu_{\scr T} \!+\! \frac{1}{2} \Bigr) (1\!-\!\epsilon) \mathcal{H}
	\biggr] \hat{\mathcal{A}}_{ {\scr T} ,\sigma}
	\, ,
\label{T other eq}
\end{align}
where we define the transverse sector index,
\begin{equation}
\nu_{\scr T} 
	= 
	\frac{D\!-\!3 \!-\! \epsilon}{ 2(1\!-\!\epsilon) } 
	\, .
\label{nuT def}
\end{equation}
Equation~(\ref{T 2nd order}) is easily recognized as the scalar mode equation~(\ref{mode eq})
whose solution is then given by~(\ref{general mode function}), upon which Eq.~(\ref{T other eq}) 
can be recognized as the recurrence relation from~(\ref{mode recurrence}). The solutions for 
transverse field operators are thus,
\begin{align}
\hat{\mathcal{A}}_{ {\scr T} ,\sigma}(\eta,k)
	={}&
	U_{\nu_{\scr T}}(\eta,k) \, \hat{b}_{\scr T}(\sigma, \vec{k})
	+
	U_{\nu_{\scr T}}^*(\eta,k) \, \hat{b}_{\scr T}^\dag(\sigma,-\vec{k})
	\, ,
\label{AT final solution}
\\
\hat{\pi}_{ {\scr T} ,\sigma}(\eta,k)
	={}&
	- i k U_{\nu_{\scr T}-1}(\eta,k) \, \hat{b}_{\scr T}(\sigma, \vec{k})
	+
	i k U_{\nu_{\scr T}-1}^*(\eta,k) \, \hat{b}_{\scr T}^\dag(\sigma,-\vec{k})
	\, ,
\label{PiT final solution}
\end{align}
where the non-Hermitian operators~$\hat{b}_{\scr T}(\vec{k},\sigma)$ can be seen as initial conditions.
Their commutation relations follow from the canonical commutation relations~(\ref{momentum commutators}).
The non-vanishing ones are,
\begin{equation}
\bigl[ \hat{b}_{\scr T} (\sigma, \vec{k}) , \hat{b}_{\scr T}^\dag(\sigma', \vec{k}^{\,\prime}) \bigr]
	=
	\delta_{\sigma \sigma'} \delta^{D-1}(\vec{k} \!-\! \vec{k}^{\,\prime} )
	\, ,
\label{bT comm}
\end{equation}
which are just the commutation relations for creation and annihilation operators.
Therefore the transverse sector of the space of states is naturally constructed as
a Fock space. It is natural to consider the vacuum state of this sector to be the 
state that minimizes the energy mode-per-mode at an asymptotic past, by analogy with
the CTBD state for the scalar field. Here this state corresponds to the one that is
annihilated by the annihilation operator given in the solutions~(\ref{AT final solution})
and~(\ref{PiT final solution}),
\begin{equation}
\hat{b}_{\scr T}(\sigma,\vec{k}) \bigl| \Omega \bigr\rangle = 0 \, ,
\qquad
\forall \vec{k}, \sigma
\, .
\label{transverse state}
\end{equation}
This is the state that we shall consider in Sec.~\ref{sec: Two-point function}
when computing the photon two-point function.

\subsection{Scalar sector}
\label{subsec: Scalar sector}

The second and the fourth equations~(\ref{Pi0 eq}) and~(\ref{PiL eq})
of the scalar sector decouple from the rest, and combine into a singe second order one,
\begin{align}
\biggl[ \partial_0^2 + k^2 
	- \Bigl( \nu^2 \!-\! \frac{1}{4} \Bigr) (1\!-\!\epsilon)^2 \mathcal{H}^2 \biggr]
		\hat{\pi}_{\scr L} ={}& 0 
	\, ,
\\
\hat{\pi}_0 ={}& - \frac{1}{k} \biggl[
	\partial_0 + \Bigl( \nu\!+\! \frac{1}{2} \Bigr) (1\!-\!\epsilon) \mathcal{H}
	\biggr] \hat{\pi}_{\scr L}
	\, ,
\end{align}
where we introduce the index of the scalar sector,
\begin{equation}
\nu = \frac{ D \!-\! 3 \!+\! \epsilon }{ 2(1\!-\!\epsilon) } 
\, .
\label{nu def}
\end{equation}
According to considerations in Sec.~\ref{subsec: Scalar mode functions}
the solutions are immediately written as,
\begin{align}
\hat{\pi}_{\scr L}(\eta, \vec{k}) 
	={}&
	k U_\nu (\eta,k) \hat{b}_{\scr P}(\vec{k})
		+ k U_\nu^*(\eta,k) \hat{b}_{\scr P}^\dag(-\vec{k}) 
	\, ,
\label{piL solution}
\\
\hat{\pi}_{0}(\eta, \vec{k}) 
	={}&
	i k U_{\nu+1} (\eta,k) \hat{b}_{\scr P}(\vec{k})
		- i k U_{\nu+1}^*(\eta,k) \hat{b}_{\scr P}^\dag(-\vec{k}) 
	\, ,
\label{pi0 solution}
\end{align}
where the non-Hermitian operators~$\hat{b}_{\scr P}(\vec{k})$ play the role of 
initial conditions/integration constants.
The remaining two equations~(\ref{A0 eq}) and~(\ref{AL eq})
combine into a sourced second-order equation,
\begin{align}
\biggl[ \partial_0^2 + k^2 
	- \Bigl( \nu^2 \!-\! \frac{1}{4} \Bigr) (1\!-\!\epsilon)^2 \mathcal{H}^2 \biggr]
	\hat{\mathcal{A}}_{0} 
	={}& 
	(1\!-\!\xi) a^2 k \hat{\pi}_{\scr L}
	- 2 \xi a^2 \mathcal{H} \hat{\pi}_0
	\, ,
\label{A0 2nd order eq}
\\
\hat{\mathcal{A}}_{\scr L} 
	={}&
	\frac{1}{k} \biggl[
		\partial_0 + \Bigl( \nu\!+\! \frac{1}{2} \Bigr) (1\!-\!\epsilon) \mathcal{H}
		\biggr] \hat{\mathcal{A}}_{0} 
		+ \frac{\xi a^2}{k} \hat{\pi}_0 \, ,
\label{AL other eq}
\end{align}
We can write the solutions to these equations as,
\begin{align}
\hat{\mathcal{A}}_0 (\eta,\vec{k})
	={}&
	U_\nu(\eta,k) \hat{b}_{\scr H}(\vec{k})
	+ U_\nu^*(\eta,k) \hat{b}_{\scr H}^\dag(-\vec{k})
\nonumber \\
&	\hspace{4cm}
	+ v_0(\eta,k) \hat{b}_{\scr P}(\vec{k})
	+ v_0^*(\eta,k) \hat{b}_{\scr P}^\dag(-\vec{k})
	\, ,
\label{A0 final solution}
\\
\hat{\mathcal{A}}_{\scr L} (\eta,\vec{k})
	={}&
	- i U_{\nu+1}(\eta,k) \hat{b}_{\scr H}(\vec{k})
	+ i U_{\nu+1}^*(\eta,k) \hat{b}_{\scr H}^\dag(-\vec{k})
\nonumber \\
&	\hspace{4cm}
	- i v_{\scr L}(\eta,k) \hat{b}_{\scr P}(\vec{k})
	+ i v_{\scr L}^*(\eta,k) \hat{b}_{\scr P}^\dag(-\vec{k})
	\, ,
\label{AL final solution}
\end{align}
where the homogeneous parts depend on operators~$\hat{b}_{\scr H}(\vec{k})$
standing for integration constants, while the particular parts depend on the 
integration constants for canonical momentum operators
introduced in solutions~(\ref{piL solution}) and~(\ref{pi0 solution}).
The particular mode functions~$v_0$ and~$v_{\scr L}$ 
in the solutions~(\ref{A0 final solution}) and~(\ref{AL final solution})
satisfy equations descending from~(\ref{A0 2nd order eq}) and~(\ref{AL other eq}),
\begin{align}
\MoveEqLeft[3.5]
\biggl[ \partial_0^2 + k^2 
	- \Bigl( \nu^2 \!-\! \frac{1}{4} \Bigr) (1\!-\!\epsilon)^2 \mathcal{H}^2 \biggr]
	v_{0} 
\nonumber \\
	={}&
	\frac{ - i \xi k a^2 \mathcal{H} }{ \bigl[ \nu(1\!-\!\epsilon) \!+\! 1 \bigr] }
		\biggl[ 2 \bigl[ \nu(1 \!-\! \epsilon) \!+\! 1 \bigr] U_{\nu+1} 
			- (1\!+\!\epsilon) \frac{i k}{\mathcal{H}} U_\nu \biggr]
	+ \Bigl( 1 \!-\! \frac{ \xi }{ \xi_s } \Bigr) k^2 a^2 U_\nu
	\, ,
\label{particular eq 1}
\\	
v_{\scr L} 
	={}&
	\frac{i}{k} \biggl[
		\partial_0 + \Bigl( \nu\!+\! \frac{1}{2} \Bigr) (1\!-\!\epsilon) \mathcal{H}
		\biggr] v_{0} 
		-\xi a^2 U_{\nu+1} 
		\, .
\label{particular eq 2}
\end{align}
where we introduced the parameter,
\begin{equation}
\xi_s = \frac{ \nu(1\!-\!\epsilon) \!+\! 1 }{ \nu(1\!-\!\epsilon) \!-\! \epsilon }
	= \frac{ D \!-\! 1 \!+\! \epsilon }{ D \!-\! 3 \!-\! \epsilon } 
	\, ,
\label{simple covariant gauge}
\end{equation}
that we refer to as the {\it simple covariant gauge}. This name is suggestive of the simplification
that happens when the gauge fixing parameter takes this value,~$\xi\!=\!\xi_s$.
The second part of the source in~(\ref{particular eq 1}) drops out, and the 
identity~(\ref{mode id1}) immediately gives the solutions,
\begin{align}
v_0 
	\xrightarrow{\xi\to\xi_s}{}& 
	\frac{ - i k }{ (D \!-\! 3 \!-\! \epsilon) } 
		\biggl[ \frac{a^2}{\mathcal{H}} U_{\nu+1} - \frac{1}{H_0} U_\nu \biggr]
	\, ,
\label{v0 simple}
\\
v_{\scr L} 
	\xrightarrow{\xi\to\xi_s}{}& 
	\frac{ - i k }{ (D \!-\! 3 \!-\! \epsilon) } 
		\biggl[ \frac{a^2}{\mathcal{H}} U_\nu - \frac{1}{H_0} U_{\nu+1} \biggr]
	\, .
\label{vL simple}
\end{align}
These solutions for particular mode functions are chosen to satisfy the Wronskian-like relation,
\begin{equation}
{\rm Re} \Bigl( v_0 U_{\nu+1}^* + v_{\scr L} U_\nu^* \Bigr) = 0 \, ,
\label{Wronskian-like}
\end{equation}
in addition to reducing to the corresponding de Sitter space solutions found in~\cite{Glavan:2022dwb}.
The Wronskian relations~(\ref{Wronskian}) and~(\ref{Wronskian-like}) are in fact sufficient
to compute the commutation relations between the momentum space operators 
introduced appearing in solutions for the field operators, without solving for the particular 
mode functions. They follow from the canonical commutation relations~(\ref{momentum commutators}),
\begin{equation}
\bigl[ \hat{b}_{\scr H}(\vec{k}) , \hat{b}_{\scr H}^\dag(\vec{k}^{\,\prime}) \bigr]
	=
	\bigl[ \hat{b}_{\scr P}(\vec{k}) , \hat{b}_{\scr P}^\dag(\vec{k}^{\,\prime}) \bigr]
	= 0 \, ,
\qquad 
\bigl[ \hat{b}_{\scr H}(\vec{k}) , \hat{b}_{\scr P}^\dag(\vec{k}^{\,\prime}) \bigr]
	= - \delta^{D-1}( \vec{k} \!-\! \vec{k}^{\,\prime}) \, ,
\label{scalar commutation relations}
\end{equation}
These commutators are not canonical, in the sense that they are not the ones 
of creation/annihilation operators. 
Nevertheless, these operators
are used to construct a basis
of the space of states~\cite{Glavan:2022pmk}.

Finding particular mode functions for~$\xi\!\neq\! \xi_s$ is considerably more complicated
and Appendix~\ref{sec: Particular mode functions} is devoted to the detailed derivation.
Requiring that the mode functions reduce to the de Sitter solutions from~\cite{Glavan:2022dwb} for an arbitrary gauge-fixing parameter,
\begin{subequations}
\begin{align}
v_0 \xrightarrow{\epsilon\to0}{}&
	\frac{ -i \xi k }{ 2(\nu_0 \!+\! 1) H_0 }
	\biggl[
	\frac{\mathcal{H} }{H_0} U_{\nu_0+1} - U_{\nu_0}
	\bigg]
	-
	\Bigl( 1 \!-\! \frac{\xi}{\xi_s^0} \Bigr)
	\frac{ik}{2H_0}
	\biggl[
	\frac{ik}{ \nu_0 H_0} \frac{\partial U_{\nu_0} }{ \partial \nu_0 }
	+ U_{\nu_0}
	\biggr]
	\, ,
\\
v_{\scr L} \xrightarrow{\epsilon\to0}{}&
	\frac{ -i \xi k }{ 2(\nu_0 \!+\! 1) H_0 }
	\biggl[
	\frac{\mathcal{H} }{H_0} U_{\nu_0} - U_{\nu_0+1}
	\bigg]
\nonumber \\
&	\hspace{1cm}
	-
	\Bigl( 1 \!-\! \frac{\xi}{\xi_s^0} \Bigr)
	\frac{ik}{2H_0}
	\biggl[
	\frac{ik}{ \nu_0 H_0} \frac{\partial U_{\nu_0+1} }{ \partial \nu_0 }
	+ \frac{\mathcal{H}}{\nu_0 H_0} U_{\nu_0}
	+ U_{\nu_0+1}
	\biggr]
	\, ,
\end{align}
\label{v dS}%
\end{subequations}
where,
\begin{equation}
\nu \xrightarrow{\epsilon \to 0} \frac{D \!-\! 3}{2} \equiv \nu_0 \, ,
\qquad \quad
\xi_s \xrightarrow{\epsilon \to 0} 
	\frac{ D \!-\! 1 }{ D\!-\! 3}
	\equiv \xi_s^0
	\, ,
\end{equation}
and that in the flat space limit they reduce directly to,
\begin{subequations}
\begin{align}
v_0 \xrightarrow{H_0\to0}{}&
	\frac{1}{4} \Bigl[ (1\!+\!\xi) + 2 (1\!-\!\xi) ik (\eta \!-\! \eta_0) \Bigr]
	\frac{ e^{-ik(\eta-\eta_0)} }{ \sqrt{2k} }
	\, ,
\\
v_{\scr L} \xrightarrow{H_0\to0}{}&
	\frac{1}{4} \Bigl[ - (1\!+\!\xi) + 2 (1\!-\!\xi) ik (\eta \!-\! \eta_0) \Bigr]
	\frac{ e^{-ik(\eta-\eta_0)} }{ \sqrt{2k} }
	\, ,
\end{align}
\label{v flat}%
\end{subequations}
in addition to satisfying~(\ref{Wronskian-like}), essentially uniquely fixes the solutions to be,
\begin{align}
v_0(\eta,k)
	={}& 
	\frac{ - i \xi k }{ 2 \bigl[ \nu(1\!-\!\epsilon) \!+\! 1 \bigr] } 
		\biggl[ \frac{a^2}{\mathcal{H}} U_{\nu+1}(\eta,k) 
			- \frac{1}{H_0} U_\nu(\eta,k) \biggr]
\nonumber \\
&	\hspace{1.5cm}
	+ \Bigl( 1 \!-\! \frac{ \xi }{ \xi_s } \Bigr) 
		\biggl[
		\mathscr{Q}(\eta,k)  U_\nu(\eta,k)
		-
		\widetilde{\mathscr{Q}}(\eta,k) U_\nu^*(\eta,k)
		\biggr]
	\, ,
\\
v_{\scr L}(\eta,k)
	={}& 
	\frac{ - i \xi k }{2 \bigl[ \nu(1\!-\!\epsilon) \!+\! 1 \bigr] } 
		\biggl[ \frac{a^2}{\mathcal{H}} U_\nu(\eta,k) 
			- \frac{1}{H_0} U_{\nu+1}(\eta,k) \biggr]
\nonumber \\
&	\hspace{1.5cm}
	+ \Bigl( 1 \!-\! \frac{ \xi }{ \xi_s } \Bigr) 
		\biggl[
		\mathscr{Q}(\eta,k)  U_{\nu+1}(\eta,k)
		+
		\widetilde{\mathscr{Q}}(\eta,k) U_{\nu+1}^*(\eta,k)
		\biggr]
	\, .
\end{align}
The coefficient functions here are given as linear combinations,
\begin{align}
\mathscr{Q}(\eta,k) ={}&
	\frac{ i }{ 4 \pi }
	\Bigl[ \frac{ k }{ (1\!-\!\epsilon) H_0 } \Bigr]^{ \frac{2}{1-\epsilon} }
	\biggl[ 
	- 2 \cos(\pi\nu) 
	\mathscr{J}_2\Bigl( \frac{ -2\epsilon }{ 1\!-\!\epsilon } , \nu; 
			\frac{k}{(1\!-\!\epsilon) \mathcal{H} } \Bigr)
\label{Q res 1 main}
\\
&	
	+
	\mathscr{J}_1\Bigl( \frac{ -2\epsilon }{ 1\!-\!\epsilon } , \nu; 
			\frac{k}{(1\!-\!\epsilon) \mathcal{H} } \Bigr)
	+ 
	\mathscr{J}_1\Bigl( \frac{ -2\epsilon }{ 1\!-\!\epsilon } , -\nu; 
			\frac{k}{(1\!-\!\epsilon) \mathcal{H} } \Bigr)
		\biggr]
	+ \mathscr{A}
	\, ,
\nonumber \\
\widetilde{\mathscr{Q}}(\eta,k) ={}&
	\frac{ e^{ \frac{- 2i k}{ (1-\epsilon) H_0 } }  }{4\pi}
	\Bigl[ \frac{k}{(1 \!-\! \epsilon)H_0} \Bigr]^{ \frac{2}{1-\epsilon} }
	\biggl[
	-
	2 \mathscr{J}_2 \Bigl( \frac{-2\epsilon}{1\!-\!\epsilon},\nu; \frac{k}{(1\!-\!\epsilon) \mathcal{H}} \Bigr) 
\label{Qtilde res 1 main}
\\
&	\hspace{0cm}
	+
	e^{-  i\pi\nu} \!
	\mathscr{J}_1 \Bigl( \frac{-2\epsilon}{1\!-\!\epsilon},\nu; \frac{k}{(1 \!-\! \epsilon) \mathcal{H}} \Bigr) 
	+
	e^{i\pi\nu } \!
	\mathscr{J}_1 \Bigl( \frac{-2\epsilon}{1 \!-\! \epsilon},-\nu; \frac{k}{(1 \!-\! \epsilon) \mathcal{H}} \Bigr) 
	\biggr]
	+ \mathscr{B}
	\, ,
\nonumber  
\end{align}
of special functions defined in terms of generalized hypergeometric functions,
\begin{align}
\mathscr{J}_1(\rho,\lambda;z) ={}&
	\frac{ - \Gamma^2(-\lambda) z^{\rho+2\lambda} }
		{ 4^\lambda (\rho \!+\! 2\lambda ) }  \,
	{}_2F_3 \biggl(  \Bigl\{ \frac{\rho}{2} \!+\!\lambda , \frac{ 1 }{2} \!+\! \lambda  \Bigr\} , 
		\Bigl\{ \frac{2 \!+\! \rho }{2} \!+\!\lambda , 1\!+\!\lambda , 1 \!+\! 2\lambda \Bigr\} , - z^2 \biggr) 
		\, ,
\\
\mathscr{J}_2(\rho,\lambda;z) ={}&
	\frac{ \Gamma(\lambda) \, \Gamma(-\lambda) z^{\rho} }{ \rho } \,
	{}_2F_3 \biggl(  \Bigl\{ \frac{\rho }{2} , \frac{ 1 }{2} \Bigr\} ,
		\Bigl\{ \frac{2 \!+\! \rho }{2}  , 1\!+\!\lambda , 1\!-\!\lambda  \Bigr\} , - z^2 \biggr) 
		\, ,
\end{align}
and where the two integration constants are given by,
\begin{align}
\mathscr{A}
	={}&
	\frac{-ik}{2(1\!+\!\epsilon)H_0}
	+
	\frac{ i\, \Gamma\bigl( \frac{-\epsilon}{1-\epsilon} \bigr) \, \Gamma\bigl( \frac{1}{2} \!+\! \frac{\epsilon}{1-\epsilon} \bigr)
		\, \Gamma\bigl( \frac{-\epsilon}{1-\epsilon} \!+\! \nu \bigr) \, \Gamma\bigl( \frac{-\epsilon}{1-\epsilon} \!-\! \nu \bigr) }
	{ 4 \, \Gamma\bigl( \frac{1}{2} \bigr) \, \Gamma\bigl( \frac{1}{2} \!+\! \nu \bigr) \, \Gamma\bigl( \frac{1}{2} \!-\! \nu \bigr) }
			\Bigl[ \frac{k}{(1\!-\!\epsilon)H_0} \Bigr]^{\frac{2}{1-\epsilon}}
	\, ,
\label{Amain}
\\
\mathscr{B}
	={}&
	\frac{ \Gamma\bigl( \frac{-\epsilon}{1-\epsilon} \bigr)  \,
		\Gamma\bigl( \frac{-\epsilon}{1-\epsilon} \!+\! \nu \bigr) \, \Gamma\bigl( \frac{-\epsilon}{1-\epsilon} \!-\! \nu \bigr) }
			{ 4 \, \Gamma\bigl( \frac{1}{2} \bigr) \, 
				\Gamma\bigl( \frac{1}{2} \!-\! \frac{\epsilon}{1-\epsilon} \bigr) }
			\Bigl[ \frac{k}{(1\!-\!\epsilon)H_0} \Bigr]^{\frac{2}{1-\epsilon}}
	e^{\frac{-2ik}{(1-\epsilon) H_0} } 
	e^{ \frac{ - i \pi \epsilon}{(1-\epsilon)} }
		\, .
\label{Bmain}
\end{align}
The huge difference in the complexity of particular mode functions in the general covariant gauge
versus the simple covariant gauge~$\xi\!=\!\xi_s$ makes the former case rather intractable to work with.
This is why from Sec.~\ref{sec: Two-point function} on we consider only the simple covariant gauge when 
computing the position-space two-point functions and the observables.

\subsection{Constraints}
\label{subsec: Constraints}

The subsidiary condition on the space of states~(\ref{K condition}) takes an analogous
expression in momentum space,
\begin{equation}
\hat{\mathcal{K}}(\vec{k}) \bigl| \Omega \bigr\rangle = 0 \, ,
\qquad
\forall \vec{k} \, ,
\end{equation}
where the non-Hermitian constraint operator is given
as a local linear combination,
\begin{equation}
\hat{\mathcal{K}}(\vec{k})
	= c_1(\eta,k) \hat{\psi}_1(\eta,\vec{k})
	+ c_{2}(\eta,k) \hat{\psi}_2(\eta,\vec{k})
	\, ,
\label{momentum K}
\end{equation}
of the momentum space Hermitian constraint operators,
\begin{align}
 \hat{\psi}_1(\eta,\vec{k}) = \hat{\pi}_0(\eta,\vec{k})
	\, ,
\qquad \quad
\hat{\psi}_2(\eta,\vec{k}) = \hat{\pi}_{\scr L}(\eta,\vec{k})
	\, .
\end{align}
The conservation of~(\ref{momentum K}) implies equations 
for the coefficients,
\begin{align}
\partial_0 c_1 - k c_2 + \frac{1}{2} (D\!-\!2) \mathcal{H} c_1 = 0 \, ,
\qquad \quad
\partial_0 c_2 + k c_1 - \frac{1}{2} (D\!-\!2) \mathcal{H} c_2 = 0 \, .
\end{align}
They combine into a second order equation,
\begin{align}
\biggl[ \partial_0^2 + k^2 
	- \Bigl( \nu^2 \!-\! \frac{1}{4} \Bigr) (1\!-\!\epsilon)^2 \biggr] c_1 ={}& 0 \, ,
\\
c_2 ={}&
	\frac{1}{k} \biggl[ \partial_0 
		+ \Bigl( \nu\!+\! \frac{1}{2} \Bigr) (1\!-\!\epsilon) \mathcal{H} \biggr] c_1 \, ,
\end{align}
that is the scalar mode equation~(\ref{mode eq}) with the general solution
given in~(\ref{general mode function}).
The general solution for the non-Hermitian constraint respecting the
symmetries of the background,
\begin{equation}
\hat{\mathcal{K}}(\vec{k} ) =
	e^{i\theta(k)} \Bigl( e^{-i\varphi(k) } \ch[\rho(k)] \hat{b}_{\scr P}(\vec{k})
		+
		e^{i\varphi(k) } \sh[\rho(k)] \hat{b}_{\scr P}^\dag(\vec{k})
		 \Bigr)
		 \, ,
\label{K solution}
\end{equation}
can be parametrized by three real functions~$\theta$,~$\varphi$, and~$\rho$ of the 
momentum modulus.
The space of states has to admit at least one vector annihilated by the operator above.
To fully specify this state we need another operator that will annihilate it,
\begin{equation}
\hat{\mathcal{B}}(\vec{k}) \bigl| \Omega \bigr\rangle = 0 \, ,
\qquad
\forall \vec{k}
\, ,
\end{equation}
and the choice preserving commutation 
relations~(\ref{scalar commutation relations}),
\begin{equation}
[\hat{\mathcal{K}}(\vec{k}) , \hat{\mathcal{K}}^\dag(\vec{k}^{\,\prime})] 
	=
	[\hat{\mathcal{B}}(\vec{k}) , \hat{\mathcal{B}}^\dag(\vec{k}^{\,\prime})] 
	=
	0 \, ,
\qquad
[\hat{\mathcal{K}}(\vec{k}) , \hat{\mathcal{B}}^\dag(\vec{k}^{\,\prime})] 
	= - \delta^{D-1}(\vec{k} \!-\! \vec{k}^{\,\prime})
\end{equation}
is given by,
\begin{equation}
\hat{\mathcal{B}}(\vec{k})
	= e^{i\theta(k)}
	\Bigl( e^{-i\varphi(k) } \ch[\rho(k)] \hat{b}_{\scr H}(\vec{k})
		+
		e^{i\varphi(k) } \sh[\rho(k)] \hat{b}_{\scr H}^\dag(\vec{k})
		 \Bigr) \, .
\label{B solution}
\end{equation}
The choice consistent with  de Sitter symmetries~\cite{Glavan:2022dwb}
in the limit~$\epsilon\!\to\!0$, and with Poincar\'{e} 
symmetries~\cite{Gupta:1949rh,Bleuler:1950cy}
in the flat space limit~$H_0\!\to\!0$ would be for all the three real functions
$\theta,\varphi$, and $\rho$ to vanish,
and that would be the natural choice to consider here. Nevertheless, anticipating issues
with IR convergence of the two-point function, we take for the Bogolyubov-like coefficients
of the scalar sector to be,
\begin{align}
\rho(k) = \theta(k) = \varphi(k) = 0 \, ,
\qquad
k \geq k_0 \, ,
\end{align}
where~$k_0\!\ll\!H_0$ is some deep IR scale, which implies that the physical state 
has to satisfy,
\begin{equation}
\hat{b}_{\scr P}(\vec{k}) \bigl| \Omega \bigr\rangle = 0 \, ,
\qquad
\hat{b}_{\scr H}(\vec{k}) \bigl| \Omega \bigr\rangle = 0 \, ,
\qquad
k \geq k_0 \, .
\label{scalar state}
\end{equation}
Below the IR scale~$k_0$ we keep the general form of the Bogolyubov coefficients
in~(\ref{K solution}) and~(\ref{B solution}), but with an additional assumption.
They are chosen such that the contribution of deep IR modes to the two-point 
function contains only terms that vanish in the limit~$k_0\!\to\!0$. That means 
their contribution to the two-point function can safely be neglected if~$k_0/H_0\!\ll\!1$
is assumed small enough. The logic behind such a choice mirrors the one from 
Sec.~\ref{subsubsec: IR divergent scalar two-point functions} for the scalar
two-point functions.

\section{Two-point function}
\label{sec: Two-point function}

The two-point functions of a free field theory are the
building blocks of loop expansions in quantum field theory.
In this section we use the solutions for field operators found in the preceding section
to evaluate the expectation values that define the two-point function. We first recount
the general properties that two-point functions must satisfy regardless of the quantum
state, and then proceed to define the natural Gaussian state with respect to which we
compute the two-point function as a sum-over modes. Due to the complicated form
of the scalar sector mode functions we only evaluate the sum-over-modes in the
simple covariant gauge~(\ref{simple covariant gauge}) in which the mode functions are tractable.
Nonetheless, the final covariantized expression for the two-point functions we find
is rather complicated, and does not exhibit a simple structure of being composed of 
scalar two-point functions and their derivatives. This owes to the necessity of
explicitly evaluating the inverse Laplace operator acting on the scalar-two 
point function which results in a combination of
Appell's fourth functions, as shown in Appendix~\ref{app: Inverting Laplacian}.

\subsection{Generalities}
\label{subsec: Generalities}

The Schwinger-Keldysh formalism for nonequilibrium quantum field 
theory~\cite{Schwinger:1960qe,Mahanthappa:1962ex,Bakshi:1962dv,Bakshi:1963bn,Keldysh:1964ud,Jordan:1986ug,Calzetta:1986ey}
(see e.g.~\cite{Berges:2004yj,NoneqLectures} for introductions) necessitates the
use of several different two-point functions in perturbative (loop) computations.
The positive-frequency Wightman function for the photon field is defined as
the expectation value of the off-coincident product of two field operators,
\begin{equation}
i \bigl[ \tensor*[_\mu^{\scr - \!}]{\Delta}{_\nu^{\scr \! +}} \bigr](x;x')
	=
	\bigl\langle \Omega \bigr| \hat{A}_\mu(x) \hat{A}_\nu(x') \bigl| \Omega \bigr\rangle \, ,
\label{Wightman def}
\end{equation}
while its complex 
conjugate,~$i \bigl[ \tensor*[_\mu^{\scr + \!}]{\Delta}{_\nu^{\scr \! -}} \bigr](x;x') \!=\!
	\bigl\{ i \bigl[ \tensor*[_\mu^{\scr - \!}]{\Delta}{_\nu^{\scr \! +}} \bigr](x;x') \bigr\}^{*}$,
is the negative-frequency Wightman function.
The expectation value of the time-ordered product is the Feynman propagator,
\begin{align}
i \bigl[ \tensor*[_\mu^{\scr + \!}]{\Delta}{_\nu^{\scr \! +}} \bigr](x;x')
	={}&
	\bigl\langle \Omega \bigr| \mathcal{T} \Bigl( \hat{A}_\mu(x) \hat{A}_\nu(x') \Bigr)
		\bigl| \Omega \bigr\rangle 
\nonumber \\
	={}&
	\theta(\eta \!-\! \eta') \,
	i \bigl[ \tensor*[_\mu^{\scr - \!}]{\Delta}{_\nu^{\scr \! +}} \bigr](x;x')
	+
	\theta( \eta' \!-\! \eta ) \,
	i \bigl[ \tensor*[_\mu^{\scr + \!}]{\Delta}{_\nu^{\scr \! -}} \bigr](x;x')
	\, ,
\label{Feynman def}
\end{align}
and its complex 
conjugate,~$i \bigl[ \tensor*[_\mu^{\scr - \!}]{\Delta}{_\nu^{\scr \! -}} \bigr](x;x') \!=\!
	\bigl\{ i \bigl[ \tensor*[_\mu^{\scr + \!}]{\Delta}{_\nu^{\scr \! +}} \bigr](x;x') \bigr\}^* $,
is the Dyson propagator.
The position space equations of motion~(\ref{classical EOMs}) satisfied
by field operators can be written in the more familiar covariant form,
\begin{equation}
{\mathcal{D}_\mu}^\nu \, \hat{A}_\nu(x) = 0 \, ,
\qquad \quad
{\mathcal{D}_\mu}^\nu
	= \nabla_\rho \nabla^\rho \delta_\mu^\nu
	- \Bigl( 1 \!-\! \frac{1}{\xi} \Bigr) \nabla_\mu \nabla^\nu
	- {R_\mu}^\nu \, .
\label{D operator}
\end{equation}
These equations of motion are inherited by the two-point functions,
\begin{equation}
{\mathcal{D}_\mu}^\rho \, i \bigl[ \tensor*[_\rho^{\tt a \! }]{\Delta}{^{\tt \! b}_\nu} \bigr] (x;x')
	= 
	{\tt S}^{\tt ab} \, g_{\mu\nu} \frac{i \delta^D(x\!-\!x')}{ \sqrt{-g} } 
	\, ,
\label{2pt EOM}
\end{equation}
where the sign symbol~${\tt S}^{\tt ab}$ is defined in~(\ref{SK sign}).
The local source in the equations for the Feynman and Dyson propagators descends
from time-ordering in the definition~(\ref{Feynman def}), and from the canonical
commutation relations~(\ref{canonical commutators}). In multiplier gauges
the equations of motion are not the only equations that two-point functions
must satisfy. There are also subsidiary conditions~(\ref{2pt constraint correlators}) 
that descend from the classical first-class constraints. 
These subsidiary conditions are all captured by the Ward-Takahashi identity,
\begin{equation}
\nabla^\mu \, i \bigl[ \tensor*[_\mu^{\tt a \!}]{\Delta}{_\nu^{\tt \! b }} \bigr](x;x')
	= - \xi \partial'_\nu \, i \bigl[ \tensor*[^{\tt a \!}]{\Delta}{^{\tt \! b  }} \bigr]_{\nu+1}(x;x')
	\, ,
\label{WT identity}
\end{equation}
where the quantity on the right-hand-side is the 
massless, minimally coupled scalar (MMCS) two-point function,
\begin{equation}
\dalembertian \, 
	i \bigl[ \tensor*[^{\tt a \!}]{\Delta}{^{\tt \! b }} \bigr]_{\nu+1} (x;x') 
	=
	{\tt S}^{\tt ab} \, \frac{i \delta^D(x\!-\!x')}{ \sqrt{-g} } 
	\, ,
\end{equation}
discussed in Sec.~\ref{sec: Preliminaries}.
We compute the photon two-point functions in the following section using 
the sum-over-modes representation with  the mode functions worked 
out in sections~\ref{subsec: Transverse sector} and~\ref{subsec: Scalar sector}.
In appendix~\ref{app: Checks for two-point function} the results for two-point functions
are checked to satisfy the appropriate equation of motion~(\ref{2pt EOM})
and the Ward-Takahashi identity~(\ref{WT identity}).

\subsection{Evaluating sum-over-modes}
\label{subsec: Evaluating sum-over-modes}

In Sec.~\ref{sec: Field operator dynamics} we have found 
solutions~(\ref{AT final solution}),~(\ref{A0 final solution}), and~(\ref{AL final solution})
for the vector field operators in momentum space,
and have specified the quantum state that we consider in~(\ref{transverse state})
and~(\ref{scalar state}). Together with the commutation relations~(\ref{bT comm})
and~(\ref{scalar commutation relations}), this is all we need to express the two-point 
function~(\ref{Wightman def}) as a sum-over-modes,
\begin{align}
i \bigl[ \tensor*[_0^{\scr - \! }]{\Delta}{_0^{\scr \! +}} \bigr] (x;x')
	={}&
	(aa')^{\frac{2-D}{2}} \! \int\! \frac{ d^{D-1} k }{ (2\pi)^{D-1} } \, 
	e^{i \vec{k} \cdot \Delta \vec{x}}
	\, \theta(k \!-\! k_0)
\nonumber \\
&	\hspace{2.cm}
	\times
	\biggl[
	- U_\nu(\eta,k) v_0^*(\eta',k) 
	- v_0(\eta,k) U_\nu^*(\eta',k)
	\biggr] 
	\, ,
\label{som 00}
\\
i \bigl[ \tensor*[_0^{\scr - \! }]{\Delta}{_i^{\scr \! +}} \bigr] (x;x')
	={}&
	(aa')^{\frac{2-D}{2}} \! \int\! \frac{ d^{D-1} k }{ (2\pi)^{D-1} } \,
	e^{i \vec{k} \cdot \Delta \vec{x}} \,
		\theta(k \!-\! k_0)
\nonumber \\
&	\hspace{2.cm}
	\times
	\frac{k_i}{k} 
	\biggl[
	U_{\nu}(\eta,k) v_{\scr L}^*(\eta',k) 
	+ v_0(\eta,k) U_{\nu+1}^*(\eta',k)
	\biggr] 
\label{som 0i}
	\, ,
\\
i \bigl[ \tensor*[_{i\,}^{\scr - \! }]{\Delta}{_0^{\scr \! +}} \bigr] (x;x')
	={}&
	(aa')^{\frac{2-D}{2}} \! \int\! \frac{ d^{D-1} k }{ (2\pi)^{D-1} } \,
	e^{i \vec{k} \cdot \Delta \vec{x}}
	\, \theta(k \!-\! k_0)
\nonumber \\
&	\hspace{2.cm}
	\times
	\frac{k_i}{k}
	\biggl[
	U_{\nu+1}(\eta,k) v_0^*(\eta',k) 
	+ v_{\scr L}(\eta,k) U_\nu^*(\eta',k)
	\biggr] 
\label{som i0}
	\, ,
\\
i \bigl[ \tensor*[_{i\,}^{\scr - \! }]{\Delta}{_j^{\scr \! +}} \bigr] (x;x')
	={}&
	(aa')^{\frac{4-D}{2}} \! \int\! 
		\frac{ d^{D-1}k }{ (2\pi)^{D-1} } \, e^{i\vec{k} \cdot \Delta\vec{x} } \,
	\Bigl( \delta_{ij} \!-\! \frac{ k_i k_j }{ k^2 } \Bigr) \,
	U_{\nu_{\scr T}}(\eta,k) U_{\nu_{\scr T}}^*(\eta',k)
\label{som ij}
\\
&	\hspace{-2.5cm}
	- (aa')^{\frac{2-D}{2} } \!\! \int\! \frac{ d^{D-1}k }{ (2\pi)^{D-1} } \, e^{i\vec{k} \cdot \Delta\vec{x} } \,
	\theta(k \!-\! k_0)
	\frac{ k_i k_j }{ k^2 }
	\biggl[
	U_{\nu+1}(\eta,k) v_{\scr L}^*(\eta',k) 
	\!+\! v_{\scr L}(\eta,k) U_{\nu+1}^*(\eta',k)
	\biggr]
	\, .
	\qquad
\nonumber 
\end{align}
Note that the reason why step functions~$\theta(k\!-\!k_0)$ appear is given in 
Sec.~\ref{subsec: Constraints}. Because not all sum-over modes of the scalar sector 
will be IR finite for a natural definition of the state,
we need to consider Bogolyubov coefficients in the IR that ameliorate this problem
and regulate the behaviour of the mode function.
We choose to implement this in a way that is least sensitive to this regulating procedure
and can be parametrized effectively by a deep IR scale~$k_0\!\ll\!H_0$ that effectively
manifests itself as an IR cutoff on the sums-over-modes descending from the scalar sector.
There is no need for such regulation of the transverse sector.
Note also that the~$i\delta$-prescriptions are implicit in~(\ref{som 00})--(\ref{som ij})
in the same way as for the scalar 
propagator~(\ref{int over modes}).

In the following subsections we evaluate the integrals in~(\ref{som 00})--(\ref{som ij})
for the case of simple covariant gauge~$\xi\!=\!\xi_s$ defined 
in~(\ref{simple covariant gauge}), 
for which the solutions for the
particular mode functions~(\ref{v0 simple}) and~(\ref{vL simple}) 
of the scalar sector are tractable. Upon plugging in these mode functions 
we can utilize recurrence relations~(\ref{mode recurrence}) 
to recognize that the components of
the photon two-point function can all be expressed in terms of certain operators
acting on sums-over-modes~(\ref{scalar 2pt cutoff def})  that represent the 
scalar two-point functions, 
\begin{align}
i \bigl[ \tensor*[_0^{\scr - \! }]{\Delta}{_0^{\scr \! +}} \bigr] (x;x')
	={}&
	\frac{ - 1 }{ (D \!-\! 3 \!-\! \epsilon) } 
	\biggl[
		\frac{a^2}{\mathcal{H}} \partial_0  
		+ \frac{a'^2}{\mathcal{H}'} \partial_0' 
		+ ( D \!-\! 2 ) \bigl( a^2 \!+\! a'^2 \bigr)
		\biggr]
		i \bigl[ \tensor*[^{\scr \! - \! }]{\Delta}{^{\scr \! + \! }} \bigr]_\nu (x;x')
	\, ,
\label{00 organized}
\\
i \bigl[ \tensor*[_0^{\scr - \! }]{\Delta}{_i^{\scr \! +}} \bigr] (x;x')
	={}&
	\frac{ \partial'_i }{ (D \!-\! 3 \!-\! \epsilon) } 
	\biggl[
	\frac{a^2}{\mathcal{H}} \,
	i \bigl[ \tensor*[^{\scr \! - \! }]{\Delta}{^{\scr \! + \! }} \bigr]_{\nu+1} (x;x')
	-
	\frac{a'^2}{\mathcal{H}' } \,
	i \bigl[ \tensor*[^{\scr \! - \! }]{\Delta}{^{\scr \! + \! }} \bigr]_\nu (x;x')
	\biggr] \, ,
\label{0i organized}
\\
i \bigl[ \tensor*[_{i\,}^{\scr - \! }]{\Delta}{_0^{\scr \! +}} \bigr] (x;x')
	={}&
	\frac{ \partial_i  }{ (D \!-\! 3 \!-\! \epsilon) } 
	\biggl[
	\frac{a'^2}{\mathcal{H}' } \,
	i \bigl[ \tensor*[^{\scr \! - \! }]{\Delta}{^{\scr \! + \! }} \bigr]_{\nu+1} (x;x')
	-
	\frac{a^2}{\mathcal{H}} \,
	i \bigl[ \tensor*[^{\scr \! - \! }]{\Delta}{^{\scr \! + \! }} \bigr]_\nu (x;x')
	\biggr]
	\, ,
\label{i0 organized}
\\
i \bigl[ \tensor*[_{i\,}^{\scr - \! }]{\Delta}{_j^{\scr \! +}} \bigr] (x;x')
	={}&
	\delta_{ij}
	aa'
	i \bigl[ \tensor*[^{\scr \! - \! }]{\Delta}{^{\scr \! + \! }} \bigr]_{\nu_{\scr T} } (x;x')
	+
	\frac{ aa' \partial_i \partial'_j }{ (1\!-\!\epsilon)^2 \mathcal{H} \mathcal{H}' }
	i \bigl[ \tensor*[^{\scr \! - }]{\Xi}{^{\scr + \! }} \bigr]_{\nu_{\scr T} } (x;x')
\label{ij organized}
\\
&	\hspace{-2.7cm}
	-
	\frac{ 1 }{ (D \!-\! 3 \!-\! \epsilon) } 
	\frac{ \partial_i \partial'_j }{ (1\!-\!\epsilon)^2 \mathcal{H} \mathcal{H}'  }
	\biggl[
		\frac{a'^2}{\mathcal{H}' } \partial_0 
		+ \frac{a^2}{\mathcal{H}} \partial_0' 
		+ ( D \!-\! 3 \!+\! \epsilon ) \Bigl( 
			\frac{ a^2 \mathcal{H}' }{ \mathcal{H} }
			\!+\!
			\frac{ a'^2 \mathcal{H} }{ \mathcal{H}' }
			\Bigr)
			\biggr] 
	i \bigl[ \tensor*[^{\scr \! - }]{\Xi}{^{\scr + \! }} \bigr]_\nu (x;x')
	\, .
\nonumber 
\end{align}
Thus we have accomplished expressing the photon two-point function in terms 
of derivatives  acting on scalar two-point functions, and on what we call
Laplace-inverted two-point functions that result from inverting the Laplace
operator on the scalar two-point function,
\begin{equation}
	i \bigl[ \tensor*[^{\scr \! -}]{\Xi}{^{\scr + \! }} \bigr]_\lambda(x;x')
	\equiv
	\frac{ (1\!-\!\epsilon)^2 \mathcal{H} \mathcal{H}' }{ \nabla^2 }
	i \bigl[ \tensor*[^{\scr \!-\!}]{\Delta}{^{\scr \!+\!}} \bigr]_\lambda(x;x')
	\, ,
\label{Xi def}
\end{equation}
where the time-dependent factor on the right-hand-side is chosen for convenience.

In the de Sitter limit the combination of temporal derivatives in the brackets of
the second line in~(\ref{ij organized}) conspire to produce a Laplacian that
eliminates its inverse~\cite{Glavan:2022dwb}.  
However, here we are not as fortunate, and have to consider 
computing~(\ref{Xi def}) explicitly. This is accomplished in a form of an asymptotic series 
of derivative operators acting on the scalar two-point function
in the next subsection, while in Appendix~\ref{app: Inverting Laplacian}
a detailed analysis of the sum-over-modes representation of~(\ref{Xi def}) is
given and a closed form solution in terms of Appell's fourth function is derived.

For the remainder of this subsection it is not necessary to evaluate~(\ref{Xi def}).
It is sufficient to note that
the Laplace-inverted two-point function inherits from the scalar two-point function 
the symmetry under the exchange
of coordinates, and the cosmological symmetries, and that it can be considered as
a function of bi-local 
variables,~$i \bigl[ \tensor*[^{\tt a}]{\Xi}{^{\tt b }} \bigr]_\lambda(x;x')\!=\! i \Xi_\lambda \bigl( y_{\tt ab} ,u_{\tt ab},v_{\tt ab} \bigr)$, 
just as for the scalar two-point function.
Thus we act explicitly with temporal derivatives in~(\ref{00 organized}--\ref{ij organized})
to write the components of the two-point functions as,
\begingroup
\allowdisplaybreaks
\begin{align}
&
i \bigl[ \tensor*[_0^{\scr - \! }]{\Delta}{_0^{\scr \! +}} \bigr] (x;x')
	=
	\frac{ \mathcal{H} \mathcal{H}'
		e^{ \frac{\epsilon u}{1-\epsilon} } }{ 2 \nu_{\scr T} H_0^2 } 
	\biggl[
		2 \, \ch \Bigl( \frac{v}{1 \!-\! \epsilon} \Bigr)
			\biggl( ( 2 \!-\! y ) \frac{\partial}{\partial y}
				- \frac{\partial}{\partial u}
				- 2\nu - 1 \biggr)
\nonumber \\
&	\hspace{4.7cm}
		-
		4 \, \ch \Bigl( \frac{ \epsilon v}{1 \!-\! \epsilon} \Bigr)
				\frac{\partial}{\partial y}
		-
		2 \, \sh \Bigl( \frac{v}{1 \!-\! \epsilon} \Bigr)
		\frac{\partial}{\partial v}
		\biggr]
		i \Delta_\nu (y,u,v)
	\, ,
\label{00 yuv}
\\
&
i \bigl[ \tensor*[_0^{\scr - \! }]{\Delta}{_i^{\scr \! +}} \bigr] (x;x')
	=
	\frac{ \bigl(\partial'_i y\bigr) \mathcal{H} e^{ \frac{\epsilon u }{1-\epsilon} }}
		{ 2 \nu_{\scr T} (1\!-\!\epsilon)  H_0^2 } 
	\biggl[
	e^{ \frac{ \epsilon v }{1-\epsilon} }
	\frac{\partial}{\partial y} i\Delta_{\nu+1} (y,u,v)
	- 
	e^{\frac{-v}{1-\epsilon}}
	\frac{\partial}{\partial y} i\Delta_{\nu} (y,u,v)
	\biggr]
	\, ,
\label{0i yuv}
\\
&
i \bigl[ \tensor*[_{i\,}^{\scr - \! }]{\Delta}{_0^{\scr \! +}} \bigr] (x;x')
	=
	\frac{ \bigl( \partial_i y \bigr) \mathcal{H}' e^{ \frac{\epsilon u }{1-\epsilon} } }
		{ 2 \nu_{\scr T} (1\!-\!\epsilon) H_0^2 } 
	\biggl[
	e^{ \frac{ - \epsilon v}{1-\epsilon} }
	\frac{\partial}{\partial y}
	i \Delta_{\nu+1} (y,u,v)
	-
	e^{ \frac{v}{1-\epsilon} }
	\frac{\partial}{\partial y}
	i \Delta_{\nu} (y,u,v)
	\biggr]
	\, ,
\label{i0 yuv}
\\
&
i \bigl[ \tensor*[_{i\,}^{\scr - \! }]{\Delta}{_j^{\scr \! +}} \bigr] (x;x')
	=
	\frac{\mathcal{H} \mathcal{H}'}{H_0^2} \delta_{ij}
	e^{ \frac{\epsilon u}{1 - \epsilon} }
	i \Delta_{\nu_{\scr T}} (y,u,v)
	+
	\frac{ e^{ \frac{\epsilon u}{1 - \epsilon} } \partial_i \partial'_j }
		{ 2 \nu_{\scr T} (1\!-\!\epsilon)^2 H_0^2 } 
	\Biggl\{
	2 \nu_{\scr T} i \Xi_{\nu_{\scr T} }(y,u,v)
\label{ij yuv}
\\
&	\hspace{0.2cm}
	+\!
	\biggl[
	2 \, \ch\Bigl( \frac{\epsilon v}{1 \!-\! \epsilon} \Bigr) \!
		\biggl( \!
		(2 \!-\! y) \frac{\partial}{\partial y}
		\!-\!
		\frac{\partial}{\partial u}
		\!-\!
		2 \nu
		\!
		\biggr)
	\!-\!
	4 \, \ch \Bigl( \frac{v}{1 \!-\! \epsilon} \Bigr) \frac{\partial}{\partial y}
	\!+\!
	2 \, \sh\Bigl( \frac{\epsilon v}{1 \!-\! \epsilon} \Bigr)
		\frac{\partial}{\partial v}
			\biggr]
	i \Xi_\nu(y,u,v) \!
	\Biggr\}
	.
\nonumber 
\end{align}
\endgroup
Even though we cannot fully eliminate the complicated Laplace-inverted function, 
it is nonetheless advantageous to apply some simplifications.
Introducing the notation~$I[f(y)] \!\equiv\! \int^y\! dy' \, f(y')$
for the primitive function with respect to variable~$y$,
we can make use of the two identities for Laplace operators acting on some 
arbitrary bi-scalar,
\begin{align}
I\Bigl[ \nabla^2 f(y,u,v) \Bigr]
	={}&
	\nabla^2 I\bigl[ f(y,u,v) \bigr]
	-
	4 (1\!-\!\epsilon)^2 \mathcal{H} \mathcal{H}' 
		f(y,u,v) 
	\, ,
\\
\nabla^2 I\bigl[ f(y,u,v) \bigr]
	={}&
	4 (1\!-\!\epsilon)^2 \mathcal{H} \mathcal{H}'
	\biggl(
	\Bigl[ y + 4 \, \sh^2\Bigl( \frac{v}{2} \Bigr) \Bigr]
	\frac{\partial }{\partial y }
	+
	\frac{(D\!-\!1)}{2}
	\biggr) f(y,u,v)
	\, ,
\end{align}
that are proven by explicitly acting the derivatives,
to derive another useful identity,
\begin{equation}
	\biggl( \Bigl[ y + 4 \, \sh^2\Bigl( \frac{v}{2} \Bigr) \Bigr] \frac{\partial }{\partial y }
	+
	\frac{D\!-\!3}{2} \biggr) i\Xi_\lambda(y,u,v)
	=
	\frac{1}{ 4 } I \bigl[ i \Delta_\lambda(y,u,v) \bigr]
	\, .
\end{equation}
Applying it to the~$(ij)$ component in~(\ref{ij yuv}) results in,
\begin{align}
i \bigl[ \tensor*[_{i\,}^{\scr - \! }]{\Delta}{_j^{\scr \! +}} \bigr] (x;x')
	={}&
	\frac{\mathcal{H} \mathcal{H}'}{H_0^2} \delta_{ij}
	e^{ \frac{\epsilon u}{1 - \epsilon} }
	i \Delta_{\nu_{\scr T}} (y,u,v)
\nonumber \\
&	\hspace{-0.9cm}
	+
	\frac{ e^{ \frac{\epsilon u}{1 - \epsilon} } \partial_i \partial'_j }
		{ 2 \nu_{\scr T} (1\!-\!\epsilon)^2 H_0^2 } 
	\biggl[
	- \frac{1}{ 2 } \, \ch\Bigl( \frac{\epsilon v}{1 \!-\! \epsilon} \Bigr)
		I  \bigl[ i \Delta_\nu(y,u,v) \bigr]
	+
	I \bigl[ i\Upsilon(y,u,v) \bigr]
	\biggr]
	\, ,
\label{ij yuv simplified}
\end{align}
where for notational simplicity in the remainder of the
paper it is useful to define
\begin{align}
i\Upsilon(y,u,v) \equiv{}&
	2 \nu_{\scr T}\, 
	\frac{\partial}{\partial y} i\Xi_{\nu_{\scr T} }(y,u,v)
	-
	2 \, \ch\Bigl( \frac{\epsilon v}{1 \!-\! \epsilon} \Bigr)
		\biggl(
		\frac{\partial}{\partial u}
		+
		2 \nu
		-
		\frac{D\!-\!3}{2}
		\biggr)
		\frac{\partial}{\partial y} i \Xi_\nu(y,u,v)
\nonumber \\
&	\hspace{-0.cm}
	-
		2 \,\sh\Bigl( \frac{\epsilon v}{1 \!-\! \epsilon} \Bigr)
		\biggl(
		2 \, \sh(v) 
		\frac{\partial }{\partial y}
		-
		\frac{\partial}{\partial v} 
		\biggr)
		\frac{\partial}{\partial y} i \Xi_\nu(y,u,v)
		\, .
\label{Upsilon}
\end{align}
This form~(\ref{ij yuv simplified}) of the~$(ij)$ component
is useful because it makes manifest that in de Sitter limit all the complicated functions
all cancel since~$i\Upsilon \! \xrightarrow{\epsilon\to0} \! 0$; this will be discussed
further in Sec.~\ref{subsec: De Sitter limit}. 

\medskip

It is also interesting to note down the expression for the two-point function of just 
the transverse components of the spatial vector potential,
\begin{align}
i \bigl[ \tensor*[_{i\,}^{\scr - \! }]{\Delta}{_j^{\scr \! +}} \bigr]^{\rm \scr ph} (x;x')
	={}&
	\bigl\langle \Omega \bigr| \hat{A}_i^{\scr T}(x) \hat{A}_j^{\scr T}(x')
		\bigl| \Omega \bigr\rangle
	=
	e^{ \frac{ u}{1 - \epsilon} } 
	\Bigl( \delta_{ij} \!+\! \frac{ \partial_i \partial'_j }{ \nabla^2 } \Bigr)
	i \Delta_{\nu_{\scr T} }(y,u,v)
\nonumber \\
	={}&
	e^{ \frac{ u}{1 - \epsilon} } 
	\biggl[
	\delta_{ij}
	i \Delta_{\nu_{\scr T}} (y,u,v)
	+
	\partial_i \partial'_j
	\frac{ i \Xi_{\nu_{\scr T} }(y,u,v) }{ (1\!-\!\epsilon)^2 \mathcal{H} \mathcal{H}' }
	\biggr]
	\, .
\label{phys ij}
\end{align}
This can rightfully be called the
physical two-point function, as it corresponds to the two-point function
in the Coulomb gauge, in 
which~$\partial_i\hat{A}_i \!=\!0$ and~$\hat{A}_0\!=\!0$ stand as an operator 
equalities.

\subsection{Laplace-inverted two-point function}
\label{subsec: Laplace-inverted two-point function}

The Laplace-inverted two-point function in~(\ref{Xi def})
is a formal solution of the Poisson equation with the scalar Wightman function 
as the source. There are in fact such equations for all four combinations
of Schwinger-Keldysh polarities,
\begin{equation}
\nabla^2 \frac{ i \bigl[ \tensor*[^{\tt a }]{\Xi}{^{\tt b}} \bigr]_\lambda(x;x') }
	{(1\!-\!\epsilon)^2 \mathcal{H} \mathcal{H}'}
	=
	i \bigl[ \tensor*[^{\tt a \! }]{\Delta}{^{\tt \! b}} \bigr]_\lambda(x;x')
	\, ,
\label{Poisson equation}
\end{equation}
determined by different scalar two-point functions appearing as sources on the
right-hand side. However, this is not the only 
equation that Laplace-inverted two-point functions have to satisfy. 
Using the equation of motion~(\ref{scalar EOM}) 
for the scalar two-point function it follows that the Laplace-inverted
two-point  functions satisfy a fourth order equation,
\begin{align}
\biggl[ \dalembertian - (1\!-\!\epsilon)^2 H^2
	\biggl( \Bigl[ \frac{D\!-\!1\!-\!\epsilon}{2(1\!-\!\epsilon)} \Bigr]^2 \!\!-\! \lambda^2 
		\biggr) \biggr] 
	\nabla^2 \frac{ i \bigl[ \tensor*[^{\tt a }]{\Xi}{^{\tt b}} \bigr]_\lambda(x;x') }
	{(1\!-\!\epsilon)^2 \mathcal{H} \mathcal{H}'}
	= {\tt S}^{\tt ab} \, \frac{ i \delta^D(x\!-\!x') }{ \sqrt{-g}} \, .
\end{align}
Given that the Laplacian commutes with the scalar d'Alembertian, and that the 
Poisson equation for a point charge has a unique solution, it follows that
the Laplace-inverted two-point functions satisfy another second order equation,
\begin{align}
\MoveEqLeft[9]
\biggl[ \dalembertian - (1\!-\!\epsilon)^2 H^2
	\biggl( \Bigl[ \frac{D\!-\!1\!-\!\epsilon}{2(1\!-\!\epsilon)} \Bigr]^2 \!\!-\! \lambda^2 
		\biggr) \biggr] 
	\frac{ i \bigl[ \tensor*[^{\tt a}]{\Xi}{^{\tt b}} \bigr]_\lambda (x;x') }
		{ (1\!-\!\epsilon)^2 \mathcal{H} \mathcal{H}' }
\nonumber \\
={}&
	{\tt S}^{\tt ab} \nabla^{-2} \frac{i \delta^D(x\!-\!x')}{ \sqrt{-g} }
	= 
	\frac{ -
	{\tt S}^{\tt ab}\, \Gamma\bigl( \frac{D-3}{2} \bigr) \, i \delta(\eta\!-\!\eta') }
		{ 4 \pi^{\frac{D-1}{2}} \| \Delta\vec{x} \|^{D-3} \sqrt{-g} }
		\, .
\label{second Xi eq}
\end{align}
Naturally, Laplace-inverted two-point functions also satisfies
the primed equation obtained by replacing~$\dalembertian \!\to\! \dalembertian'$
and~$H \!\to\! H'$ in the equation above.
Also, using recurrence relations~(\ref{mode recurrence}) for the mode functions,
Laplace-inverted two-point functions can be
seen to satisfy their own raising and lowering equations, respectively,
\begin{align}
\MoveEqLeft[10]
	\biggl[ \partial_0 + \Bigl( 
			\frac{D \!-\! 1 \!-\! \epsilon}{2(1 \!-\! \epsilon)} 
			\!+\! \lambda \Bigr) 
		(1\!-\!\epsilon) \mathcal{H} \biggr]
	\biggl[ \partial_0' + \Bigl( 
			\frac{D \!-\! 1 \!-\! \epsilon}{2(1 \!-\! \epsilon)}	
			\!+\! \lambda \Bigr) 
		(1\!-\!\epsilon) \mathcal{H}' \biggr]
	\frac{ i \bigl[ \tensor*[^{\tt a}]{\Xi}{^{\tt b} } \bigr]_\lambda (x;x') }
		{ (1\!-\!\epsilon)^2 \mathcal{H} \mathcal{H}' }
\nonumber \\
&
=
	- 
	i \bigl[ \tensor*[^{ \tt a \!}]{\Delta}{^{\tt \!b } } \bigr]_{\lambda+1} (x;x')
	-
	\frac{ {\tt S}^{\tt ab} \, \Gamma \bigl( \frac{D - 3}{2} \bigr) \, i \delta(\eta \!-\! \eta') }
		{ 4 \pi^{ \frac{D-1}{2} } \| \Delta\vec{x}\|^{ D-3 } a^{D-2} }
		\, ,
\label{Xi spacetime raising}
\\
\MoveEqLeft[10]
	\biggl[ \partial_0 + \Bigl( 
			\frac{D \!-\! 1 \!-\! \epsilon}{2(1 \!-\! \epsilon)} 
			\!-\! \lambda \Bigr) 
		(1\!-\!\epsilon) \mathcal{H} \biggr]
	\biggl[ \partial_0' + \Bigl( 
			\frac{D \!-\! 1 \!-\! \epsilon}{2(1 \!-\! \epsilon)}	
			\!-\! \lambda \Bigr) 
		(1\!-\!\epsilon) \mathcal{H}' \biggr]
	\frac{ i \bigl[ \tensor*[^{\tt a}]{\Xi}{^{\tt b} } \bigr]_\lambda (x;x') }
		{ (1\!-\!\epsilon)^2 \mathcal{H} \mathcal{H}' }
\nonumber \\
&
=
	- 
	i \bigl[ \tensor*[^{ \tt a \!}]{\Delta}{^{\tt \!b } } \bigr]_{\lambda-1} (x;x')
	-
	\frac{ {\tt S}^{\tt ab} \, \Gamma \bigl( \frac{D - 3}{2} \bigr) \, i \delta(\eta \!-\! \eta') }
		{ 4 \pi^{ \frac{D-1}{2} } \| \Delta\vec{x}\|^{ D-3 } a^{D-2} }
		\, .
\label{Xi spacetime lowering}
\end{align}

Different two-point functions with different Schwinger-Keldysh polarities 
can be represented by one function of bi-local variables~(\ref{bilocal variables}) 
with different~$i\delta$-prescriptions,
\begin{equation}
i \bigl[ \tensor*[^{\tt a}]{\Xi}{^{\tt b}} \bigr]_\lambda (x;x')
	=
	i \Xi_\lambda\bigl( y_{\tt ab} , u_{\tt ab} ,v_{\tt ab} \bigr)
	\, .
\end{equation}
while the function on the right-hand-side then satisfies equations
without local source terms.
The first one follows from the defining Poisson equation~(\ref{Poisson equation}),
\begin{equation}
\biggl(
	4 \Bigl[ y + 4 \, \sh^2\Bigl( \frac{v}{2} \Bigr) \Bigr] \frac{\partial}{\partial y}
	+ 2 (D\!-\!1)
	\biggr)
	\frac{\partial}{\partial y} 
	i \Xi_\lambda (y,u,v)
	=
	i \Delta_\lambda (y,u,v)
	\, .
\label{Xi eq}
\end{equation}
 The second equation~(\ref{second Xi eq}) and its primed counterpart
are best given as two linear combinations, producing an even equation,
\begin{align}
&
\Biggl[
	\bigl( 4y \!-\! y^2 \bigr) \frac{\partial^2}{\partial y^2}
	+ D(2 \!-\! y) \frac{\partial}{\partial y}
	- 2 \Bigl[ y \!+\! 4 \, \sh^2 \Bigl( \frac{v}{2} \Bigr) \Bigr]
		\biggl( \frac{\partial}{\partial u} 
			\!+\! \frac{ D\epsilon \!-\! 2 }{ 2 (1\!-\!\epsilon) } \biggr) \frac{\partial}{\partial y}
\label{Xi even equation}
\\
&	\hspace{1cm}
	- \biggl( \frac{\partial}{\partial u}
		\!+\! \frac{ D \!-\! 3 \!+\! \epsilon }{ 1 \!-\! \epsilon } \biggr) 
		\frac{\partial}{\partial u}
	+ \biggl( 4 \, \sh(v) \frac{\partial}{\partial y}
		\!-\! \frac{\partial}{\partial v} \biggr) \frac{\partial}{\partial v} 
	+
	\lambda^2 \!-\! \Bigl( \frac{ D \!-\! 3 \!+\! \epsilon }{ 2(1\!-\!\epsilon) } \Bigr)^{\!2}
	\Biggr]
	i \Xi_\lambda
	=
	0
	\, ,
\nonumber
\end{align}
and an odd equation,
\begin{equation}
\Biggl[
	\biggl( \Bigl[ y + 4 \, \sh^2\Bigl( \frac{v}{2} \Bigr) \Bigr]
		\frac{\partial}{\partial y}
	+
	\frac{\partial}{\partial u}
	+ 
	\frac{ D \!-\! 3 \!+\! \epsilon }{ 2(1\!-\!\epsilon) }
		\biggr) \frac{\partial}{\partial v} 
	- 2 \, \sh(v)  \biggl( \frac{\partial}{\partial u}
		+ \frac{ D\epsilon \!-\! 2 }{2 (1\!-\!\epsilon)} \biggr) 
		 \frac{\partial}{\partial y}
	\Biggr]
	i \Xi_\lambda
	=
	0
	\, .
\label{Xi odd equation}
\end{equation}
Lastly, the raising and lowering equations~(\ref{Xi spacetime raising}) 
and~(\ref{Xi spacetime lowering}) take the following form,
\begin{equation}
\Biggl[
	\biggl(
	\Bigl[ y + 4 \, \sh^2\Bigl( \frac{v}{2} \Bigr) \Bigr] \frac{\partial}{\partial y} 
	+ 
	\frac{\partial}{\partial u}
	+ 
	\frac{D \!-\! 3 \!+\! \epsilon}{2(1 \!-\! \epsilon)}
	+
	\lambda 
	\biggr)^{\!\!2}
	\!
	- 
	\biggl( 2 \, \sh(v) \frac{\partial}{\partial y} - \frac{ \partial }{\partial v} \biggr)^{\!\!2} \,
	\Biggr]
	i \Xi_\lambda
	=
	- i \Delta_{\lambda+1}
	\, ,
\label{Xi raising eq}
\end{equation}
\begin{equation}
\Biggl[
	\biggl(
	\Bigl[ y + 4 \, \sh^2\Bigl( \frac{v}{2} \Bigr) \Bigr] \frac{\partial}{\partial y} 
	+ 
	\frac{\partial}{\partial u}
	+ 
	\frac{D \!-\! 3 \!+\! \epsilon}{2(1 \!-\! \epsilon)}
	\! -
	\lambda 
	\biggr)^{\!\!2}
	- 
	\biggl( 2 \, \sh(v) \frac{\partial}{\partial y} - \frac{ \partial }{\partial v} \biggr)^{\!\!2} \,
	\Biggr]
	i \Xi_\lambda
	=
	- i \Delta_{\lambda-1}
	\, .
\label{Xi lowering eq}
\end{equation}

Solving for~$i\Xi_\lambda$ is best done by considering the first equation~(\ref{Xi eq})
out of the five above. 
From it we can readily solve for a derivative of~$i\Xi_\lambda$
as a power series, by considering the expression in parentheses on the left-hand side
of~(\ref{Xi eq}) as an expansion parameter,
\begin{equation}
\frac{\partial}{\partial y} i \Xi_\lambda (y,u,v)
	=
	\sum_{n=0}^{\infty}
	\Bigl(
	\frac{-2}{D\!-\!1}
	\Bigr)^{\!n}
	\biggl(
	\Bigl[ y + 4 \, \sh^2 \Bigl( \frac{v}{2} \Bigr) \Bigr] \frac{\partial}{\partial y}
	\biggr)^{\!\!n}
	\frac{i\Delta_\lambda(y,u,v)}{2 (D\!-\!1)}
	\, .
\label{M series solution}
\end{equation}
It is convenient to commute the derivatives so they act only on~$i\Delta_\lambda$,
so that the operator is written as,
\begin{equation}
\biggl(
	\Bigl[ y + 4 \, \sh^2 \Bigl( \frac{v}{2} \Bigr) \Bigr] \frac{\partial}{\partial y}
	\biggr)^{\!\!n}
	=
	\sum_{\ell=1}^{n}
	\biggl\{ \begin{matrix} n \\ \ell \end{matrix} \biggr\}
	\Bigl[ y + 4 \, \sh^2 \Bigl( \frac{v}{2} \Bigr) \Bigr]^\ell 
	\frac{\partial^\ell }{\partial y^\ell}
	\, ,
\label{commuting ders}
\end{equation}
where the coefficients~$\bigl\{ {n \atop \ell} \bigr\}$
in this expansion satisfy the following recurrence relations,
\begin{equation}
\biggl\{ \begin{matrix} n \!+\! 1 \\ \ell \end{matrix} \biggr\}
	=
	\ell \, \biggl\{ \begin{matrix} n \\ \ell \end{matrix} \biggr\}
	+
	\biggl\{ \begin{matrix} n \\ \ell \!-\! 1 \end{matrix} \biggr\}
	\, ,
\qquad \quad
\biggl\{ \begin{matrix} n \\ n \end{matrix} \biggr\} = 
\biggl\{ \begin{matrix} n \\ 1 \end{matrix} \biggr\} = 1 \, ,
\qquad \quad
n \ge \ell \ge 1 \, .
\end{equation}
Therefore they are recognized to be Stirling numbers of the second 
kind~(c.f.~\S26.8 of~\cite{Olver:2010,Olver:web}),
that admit an explicit sum representation,
\begin{equation}
\biggl\{ \begin{matrix} n \\ \ell \end{matrix} \biggr\}
	=
	\sum_{j=0}^{\ell}
	\frac{ (-1)^{\ell - j} j^n }{ j! \, (\ell \!-\! j)! }
	\, .
\end{equation}
Plugging this expansion back into the series solution~(\ref{M series solution}),
and reorganizing the series by grouping together all derivatives of the same order,
\begin{equation}
\frac{\partial}{\partial y} i \Xi_\lambda
	=
	\frac{i \Delta_\lambda}{2 (D\!-\!1)}
	+
	\sum_{\ell=1}^{\infty}
	\biggl[
	\sum_{n=\ell}^{\infty}
	\Bigl(
	\frac{-2}{D\!-\!1}
	\Bigr)^{\!n}
	\biggl\{ \begin{matrix} n \\ \ell \end{matrix} \biggr\}
	\biggr]
	\!\times\!
	\Bigl[ y + 4 \, \sh^2 \Bigl( \frac{v}{2} \Bigr) \Bigr]^\ell 
	\frac{\partial^\ell }{\partial y^\ell}
	\frac{i\Delta_\lambda }{2 (D\!-\!1)}
	\, .
\end{equation}
The series over~$n$ can be recognized as the generating function
for the Stirling numbers of the second kind (Eq.~26.8.11 from~\cite{Olver:2010,Olver:web}),
that has a closed form expression,
\begin{equation}
\sum_{n=\ell}^{\infty}
	\biggl\{ \begin{matrix} n \\ \ell \end{matrix} \biggr\}
	z^{-n}
	=
	\frac{ (-1)^\ell \, \Gamma ( 1 \!-\! z ) }{ \Gamma ( 1 \!-\! z \!+\! \ell \, ) }
	\, .
\label{Stirling generatrix}
\end{equation}
This now yields,
\begin{equation}
\frac{\partial}{\partial y} i\Xi_\lambda(y,u,v)
	=
	\sum_{n=0}^{\infty}
	\frac{ (-1)^n \, \Gamma \bigl( \frac{D - 1}{2} \bigr) }
		{ 4 \, \Gamma \bigl( \frac{D+1}{2} \!+\! n \, \bigr) }
	\Bigl[ y + 4 \, \sh^2 \Bigl( \frac{v}{2} \Bigr) \Bigr]^n
	\frac{\partial^n }{\partial y^n }
	i\Delta_\lambda(y,u,v)
	\, .
\label{check Xi solution}
\end{equation}
Determining this derivative of~$i\Xi_\lambda$ is sufficient for the photon two-point 
function due to the two spatial derivatives in~(\ref{ij yuv}). Nonetheless, we can use 
the identity that is proved by simple partial integration,
\begin{equation}
\sum_{n=0}^{\infty} c_n (y \!+\! \alpha)^n \frac{\partial^n}{\partial y^n} I\bigl[ f(y) \bigr]
	=
I \biggl[ \, \sum_{n=0}^{\infty} \Bigl( c_n \!+\! (n\!+\!1) c_{n+1} \Bigr) (y \!+\! \alpha)^n
	\frac{\partial^n}{\partial y^n} f(y) \, \biggr]
	\, ,
\label{series identity}
\end{equation}
to obtain the final expression for the Laplace-inverted two-point function,
\begin{equation}
i\Xi_\lambda(y,u,v)
=
\sum_{n=0}^{\infty}
	\frac{ (-1)^n \, \Gamma \bigl( \frac{D - 3}{2} \bigr) }
		{ 4 \, \Gamma\bigl( \frac{D - 1}{2} \!+\! n \, \bigr) }
	\Bigl[ y + 4 \, \sh^2 \Bigl( \frac{v}{2} \Bigr) \Bigr]^n
	\frac{\partial^{n} }{\partial y^{n} }
	I \bigl[ i\Delta_\lambda(y,u,v) \bigr]
	\, .
\label{Xi solution}
\end{equation}
Here the primitive function of the scalar two-point function with respect to~$y$
is defined as a term-by-term integral of the power series~(\ref{power series})
and~(\ref{W result}),
\begin{equation}
I\bigl[ i \Delta_\lambda(y,u,v) \bigr]
	= e^{- \frac{(D-2)\epsilon}{2(1-\epsilon)} u }
	\Bigl(
	I\bigl[ \mathcal{F}_\lambda(y) \bigr]
	+
	I\bigl[ \mathcal{W}(y,u,v) \bigr]
	\Bigr)
	\, ,
\end{equation}
such that no $y$-independent constants of integration are generated,
neither in the bulk part,
\begin{align}
\MoveEqLeft[1]
I\bigl[ \mathcal{F}_\lambda(y) \bigr]
	=
	- 4 \, \frac{ \bigl[ (1\!-\!\epsilon) H_0 \bigr]^{D-2} }{ (4\pi)^{\frac{D}{2} } }
	\frac{ \Gamma\bigl( \frac{D-3}{2} \!+\! \lambda \bigr) \, \Gamma\bigl( \frac{D-3}{2} \!-\! \lambda \bigr) }
		{ \Gamma\bigl( \frac{D-2}{2} \bigr) }
\nonumber \\
&	\hspace{3cm}
	\times
	{}_2F_1\biggl( \Bigl\{ \frac{D\!-\!3}{2} \!+\! \lambda , \frac{D\!-\!3}{2} \!-\! \lambda \Bigr\} ,
	\Bigl\{ \frac{D\!-\!2}{2} \Bigr\} , 1 \!-\! \frac{y}{4} \biggr)
	\, .
\end{align}
nor in the IR part,
\begin{align}
\MoveEqLeft[2]
I \bigl[ \mathcal{W}_\lambda(y,u,v) \bigr]
	=
	-
	\frac{ \bigl[ (1 \!-\! \epsilon) H_0 \bigr]^{D-2 }  }
		{ (4\pi)^{\frac{D}{2}} } 
	\frac{ \Gamma(\lambda) \, \Gamma(2\lambda)  }
		{ \Gamma\bigl( \frac{1}{2} \!+\! \lambda \bigr) \, \Gamma\bigl( \frac{D-1}{2} \bigr)  } 
	\sum_{N= 0}^{ \lfloor \lambda - \frac{D-1}{2} \rfloor }
	\sum_{n=0}^{N}
	\sum_{\ell=0}^{N-n} 
	\frac{ c_{Nn\ell} }{ \bigl( \frac{D-1}{2} \!+\! N \!-\! \lambda \bigr) }
\nonumber  \\
&
	\times
	\Bigl[ \frac{ k_0^2 e^{-u} }{ (1\!-\!\epsilon)^2 H_0^2 } 
		\Bigr]^{ \frac{D-1}{2} -\lambda + N}
	\frac{1}{(n\!+\!1)}
	\Bigl[ y + 4 \, {\rm sh}^2\Bigl( \frac{v}{2} \Bigr) \Bigr]^{n+1}
	\,
	\ch \bigl[ ( N \!-\! n \!-\! 2\ell ) v  \bigr]
	\, .
\end{align}
Note that in deriving expression~(\ref{Xi solution}) 
we had not relied on the specific form of the 
source in Eq.~(\ref{Xi solution}), and thus it is valid for an arbitrary source for which
the sum in (5.31) converges for some region of parameters.
In principle there is a question of the homogeneous part of the solution, which is 
missed by the iterative method that generates the series solution. However, this 
contribution is not permitted as it would correspond to the Coulomb potential-like
contribution that would generate a source which is not there. 
Checking that the solution in~(\ref{Xi solution}) also satisfies the remaining 
equations~(\ref{Xi even equation})--(\ref{Xi lowering eq})
is a tedious, but straightforward task of acting
derivatives and applying equations for the scalar two-point function from 
Sec.~\ref{subsubsec: IR divergent scalar two-point functions}.
Note, however, that the series solution in~(\ref{Xi solution}) is asymptotic
and that its practical utility is limited to small spatial separations.
This corresponds to an implicit assumption that~$\bigl[  y \!+\! 4 \, \sh^2\bigl( \frac{v}{2} \bigr)\bigr] \!=\! (1\!-\!\epsilon)^2 \mathcal{H} \mathcal{H}' \| \Delta \vec{x} \|$ is used as an expansion
parameter when 
solving~(\ref{Xi eq}) iteratively. In general one should use the closed form solutions
for the Laplace-inverted two-point function expressed in terms of Appell's fourth function,
that we derive in 
Appendix~\ref{app: Inverting Laplacian}.

\medskip

It is very useful to note that some of the equations that the Laplace-inverted 
two-point function satisfies are inherited by~$i\Upsilon$ defined in~(\ref{Upsilon}).
Firstly, it follows from Eq.~(\ref{Xi eq}) that~$i\Upsilon$ satisfies,
\begin{align}
\MoveEqLeft[5]
\biggl(
	4 \Bigl[ y + 4 \, \sh^2\Bigl( \frac{v}{2} \Bigr) \Bigr] \frac{\partial}{\partial y}
	+ 2 (D\!-\!1)
	\biggr)
	i\Upsilon(y,u,v)
\nonumber \\
	={}&
	2 \nu_{\scr T} \, i \Delta_{\nu_{\scr T}} (y,u,v)
	-
	2 \, \ch\Bigl( \frac{\epsilon v}{1 \!-\! \epsilon} \Bigr)
		\biggl(
		\frac{\partial}{\partial u}
		+
		2 \nu
		-
		\frac{D\!-\!3}{2}
		\biggr)
	i \Delta_{\nu} (y,u,v)
\nonumber \\
&	\hspace{-0.cm}
	-
	2 \,\sh\Bigl( \frac{\epsilon v}{1 \!-\! \epsilon} \Bigr)
		\biggl(
		2 \, \sh(v) 
		\frac{\partial }{\partial y}
		-
		\frac{\partial}{\partial v} 
		\biggr)
		i \Delta_{\nu} (y,u,v)
		\, .
\label{Upsilon eom}
\end{align}
The remaining four equations~(\ref{Xi even equation})--(\ref{Xi lowering eq})
combine with the one above
to form three additional equations that~$i\Upsilon$ satisfies.
The even equation~(\ref{Xi even equation}) can be regarded as giving rise to
another even equation,
\begingroup
\allowdisplaybreaks
\begin{align}
\MoveEqLeft[1]
\Biggl[
	\bigl( 4y \!-\! y^2 \bigr) \frac{ \partial^2 }{\partial y^2}
	+ 
	(D\!+\!2) (2\!-\!y) \frac{\partial}{\partial y}
	-
	2 \Bigl[ y + 4 \, \sh^2\Bigl( \frac{v}{2} \Bigr) \Bigr]
		\biggl( \frac{\partial}{\partial u}
			+ \frac{D\epsilon}{2(1\!-\!\epsilon)} \biggr)
			\frac{\partial}{\partial y}
\nonumber \\
&	\hspace{1.5cm}
	-
	\biggl(
	\frac{\partial}{\partial u}
	+
	\frac{D \!-\! 1 \!-\! \epsilon}{1\!-\!\epsilon}
	\biggr)
	\frac{\partial}{\partial u}
	+
	\biggl(
	4 \, \sh(v) \frac{\partial}{\partial y} 
	-
	\frac{\partial}{\partial v}
	\biggr) \frac{\partial}{\partial v}
	+
	\frac{3 \!-\! D(2\!-\!\epsilon)}{(1\!-\!\epsilon)^2}
	\Biggr]
	i \Upsilon
\nonumber \\
&
	=
	\frac{ \sh\bigl( \frac{\epsilon v}{1-\epsilon} \bigr) }{ 1\!-\!\epsilon }
	\biggl[
	2 \epsilon \, \sh(v)
		\frac{\partial}{\partial y} i \Delta_{\nu}
	-
	(1\!+\!\epsilon)
		\frac{\partial}{\partial v} i \Delta_{\nu}
	\biggr]
	+
	\frac{ \ch\bigl( \frac{\epsilon v}{1-\epsilon} \bigr) }{ 1\!-\!\epsilon }
	\biggl[
	- 
	2 \epsilon \frac{\partial}{\partial y} i \Delta_{\nu+1}
\nonumber \\
&	\hspace{0.5cm}
	-
	2 \epsilon \, \ch(v) \frac{\partial}{\partial y} i \Delta_{\nu}
	+
	\biggl( 
	\frac{\partial}{\partial u}
	+
	\frac{(D\!-\! 2) \epsilon}{2(1 \!-\! \epsilon)}
	\biggr) i \Delta_{\nu}
	+
	(1 \!-\! \epsilon) \nu i \Delta_{\nu}
	\biggr]
	- 
	\frac{\nu_{\scr T} \, i \Delta_{\nu_{\scr T}} }{ 1\!-\!\epsilon }
	\, ,
\end{align}
\endgroup
the odd equation~(\ref{Xi odd equation}) as giving rise to another odd equation,
\begin{align}
&
\Biggl[
	\biggl(
	\Bigl[ y + 4 \, \sh^2\Bigl( \frac{v}{2} \Bigr) \Bigr]
		\frac{\partial}{\partial y}
	+
	\frac{\partial}{\partial u}
	+
	\frac{ D \!-\! 1 \!-\! \epsilon }{2(1\!-\!\epsilon)}
	\biggr)
	\frac{\partial}{\partial v}
	- 
	2 \, \sh(v)
		\biggl( \frac{\partial}{\partial u}
			+ \frac{ D\epsilon \!-\! 2}{2(1\!-\!\epsilon)} 
			\biggr)
			\frac{\partial}{\partial y}
	\Biggr]
	i \Upsilon
\nonumber \\
&	\hspace{0.5cm}
	=
	-
	\biggl[
	\frac{\epsilon \, \sh(v) }{ 1\!-\!\epsilon }
	\ch\Bigl( \frac{\epsilon v}{1\!-\!\epsilon} \Bigr)
	+
	\ch(v)
	\,
	\sh\Bigl( \frac{\epsilon v}{1\!-\!\epsilon} \Bigr)
	\biggr]
		\frac{\partial}{\partial y} i \Delta_{\nu}
\nonumber \\
&	\hspace{2.5cm}
	+
	\frac{ \epsilon \, \sh\bigl( \frac{\epsilon v}{1-\epsilon} \bigr) }{2(1\!-\!\epsilon)}
	\biggl[
	\frac{ 1 }{\sh(v)} 
	\frac{\partial}{\partial v} i \Delta_{\nu+1}
	-
	\frac{ \ch(v) }{\sh(v)} 
		\frac{\partial}{\partial v} i \Delta_{\nu}
	-
	\nu \, i \Delta_{\nu}
	\biggr]
	\, ,
\end{align}
and the rising and lowering equations~(\ref{Xi raising eq}) 
and~(\ref{Xi raising eq}) as producing
\begin{align}
&
\biggl( \frac{\partial}{\partial u} 
	+ \frac{ (D\!-\!2) \epsilon }{ 2 (1\!-\!\epsilon) } 
	- \nu_{\scr T} \biggr)
\biggl( \frac{\partial}{\partial u} 
	+ \frac{ (D\!-\!2) \epsilon }{ 2 (1\!-\!\epsilon) } \biggr)
	i \Upsilon
\nonumber \\
&	\hspace{1cm}
	=
	\frac{1}{2}
	\biggl( \frac{\partial}{\partial u} 
		+ \frac{(D\!-\!2)\epsilon}{2(1\!-\!\epsilon)} 
		- \nu_{\scr T}
		\biggr) 
	\Biggl\{
	\ch\Bigl( \frac{\epsilon v}{1\!-\!\epsilon} \Bigr)
	\biggl[
		\frac{1}{\sh(v)}
		\frac{\partial}{\partial v} i \Delta_{\nu+1}
	-
	\frac{ \ch(v) }{\sh(v)} \frac{\partial}{\partial v} i \Delta_\nu
\nonumber \\
&	\hspace{3cm}
	+
	\biggl(
	\frac{\partial}{\partial u}
	+
	\frac{ (D\!-\!2) \epsilon}{ 2(1 \!-\! \epsilon) } 
	\biggr)
	i \Delta_\nu
	\biggr]
	-
	\sh\Bigl( \frac{\epsilon v}{1\!-\!\epsilon} \Bigr)
	\frac{\partial}{\partial v} i \Delta_\nu
	\Biggr\}
	\, .
\label{Upsilon last eq}
\end{align}
The association of the three equations above to their 
antecedents~(\ref{Xi even equation})--(\ref{Xi lowering eq}) is only
qualitative since each of the new equations is a combination of essentially
all of the antecedent ones.
Additionally, in deriving the last equation~(\ref{Upsilon last eq}) we used that,
\begin{equation}
\frac{\partial}{\partial u} i \Delta_{\nu_{\scr T}} 
	= - \frac{(D\!-\!2)\epsilon}{2(1\!-\!\epsilon)} i \Delta_{\nu_{\scr T}} 
	\, ,
\qquad \qquad
\frac{\partial}{\partial v} i \Delta_{\nu_{\scr T}} = 0 \, .
\end{equation}
owing to the transverse sector index being~$\nu_{\scr T} \!<\! (D\!-\!1)/2$.

\subsection{Covariantization}
\label{subsec: Covariantization}

The components of the photon two-point function found in 
Sec.~\ref{subsec: Evaluating sum-over-modes}
can be written in a more systematic covariantized form,
\begin{align}
\MoveEqLeft[1.5]
i \bigl[ \tensor*[_\mu^{\scr - \!}]{\Delta}{_\nu^{\scr \! +}} \bigr] (x;x')
	=
	\bigl( \partial_\mu \partial'_\nu y \bigr) \, \mathcal{C}_1(y,u,v)
	+ \bigl( \partial_\mu y \bigr) \bigl( \partial'_\nu y \bigr) \, \mathcal{C}_2(y,u,v)
\nonumber 
\\
&
	+ \Bigl[ \bigl( \partial_\mu y \bigr) \bigl( \partial'_\nu u \bigr) 
		\!+\! \bigl( \partial_\mu u \bigr) \bigl( \partial'_\nu y \bigr) \Bigr] \,
		 \mathcal{C}_3(y,u,v)
	+ \Bigl[ \bigl( \partial_\mu y \bigr) \bigl( \partial'_\nu u \bigr) \!-\! \bigl( \partial_\mu u \bigr) \bigl( \partial'_\nu y \bigr) \Bigr] \, \overline{\mathcal{C}}_3(y,u,v)
\nonumber \\
&
	+ \bigl( \partial_\mu u \bigr) \bigl( \partial'_\nu u \bigr) \, \mathcal{C}_4(y,u,v) 
	\, ,
\label{cov rep}
\end{align}
where the elements of the appropriate bi-tensor basis for this form are constructed out of
independent derivatives of the bi-scalar variables~(\ref{bilocal variables}),
and are multiplied by five scalar structure functions.
This is the tensor basis from~\cite{Glavan:2020zne} supplemented by the 
odd tensor structure with an accompanying structure function~$\overline{\mathcal{C}}_3$.
We need to determine the structure functions by matching the form~(\ref{cov rep}) 
to the results in~(\ref{00 yuv})--(\ref{i0 yuv}) and~(\ref{ij yuv simplified}).
This is facilitated by writing out explicitly the components of the covariantized form,
\begin{align}
i \bigl[ \tensor*[_{i\,}^{\scr - \!}]{\Delta}{_j^{\scr \! +}} \bigr] (x;x')
	={}&
	2 \delta_{ij} (1\!-\!\epsilon)^2 \mathcal{H} \mathcal{H}' 
		\Bigl\{ - \mathcal{C}_1 + I[\mathcal{C}_2] \Bigr\}
	+ \partial_i \partial'_j I^2 [ \mathcal{C}_2 ] \, ,
\\
i \bigl[ \tensor*[_{0}^{\scr - \!}]{\Delta}{_{i}^{\scr \! +}} \bigr] (x;x')
	={}&
	(1\!-\!\epsilon) \mathcal{H} \bigl( \partial'_i y \bigr) \Bigl\{
		\mathcal{C}_1 - \bigl[ 2\!-\!y \!-\! 2 e^{-v} \bigr] \mathcal{C}_2 + \mathcal{C}_3 - \overline{\mathcal{C}}_3 
		\Bigr\} \, ,
\\
i \bigl[ \tensor*[_{i\,}^{\scr - \!}]{\Delta}{_{0}^{\scr \! +}} \bigr] (x;x')
	={}&
	(1\!-\!\epsilon) \mathcal{H}' \bigl( \partial_i y \bigr) \Bigl\{
		\mathcal{C}_1 - \bigl[ 2\!-\!y \!-\! 2 e^{v} \bigr] \mathcal{C}_2 + \mathcal{C}_3 + \overline{\mathcal{C}}_3 
		\Bigr\} \, ,
\\
i \bigl[ \tensor*[_{0}^{\scr - \!}]{\Delta}{_{0}^{\scr \! +}} \bigr] (x;x')
	={}&
	(1\!-\!\epsilon)^2 \mathcal{H} \mathcal{H}' \Bigl\{
		- \bigl[ 2\!-\!y \!-\! 4 \, \ch(v) \bigr] \mathcal{C}_1 
		- \bigl[ 4y \!-\! y^2 \!-\! 8 \!+\! 4 (2\!-\!y) \, \ch(v) \bigr] \mathcal{C}_2
\nonumber \\
&
		- 2 \bigl[ 2\!-\!y\!-\! 2 \, \ch(v) \bigr] \mathcal{C}_3
		- 4 \, \sh(v) \, \overline{\mathcal{C}}_3
		+ \mathcal{C}_4
		\Bigr\} 
		\, .
\end{align}
A straightforward comparison then yields the structure functions,
\begingroup
\allowdisplaybreaks
\begin{align}
\mathcal{C}_1 ={}&
	\frac{ e^{ \frac{\epsilon u}{1 - \epsilon} } }
		{ 2 \nu_{\scr T} (1\!-\!\epsilon)^2 H_0^2 } 
	\Biggl\{
	- \nu_{\scr T} i \Delta_{\nu_{\scr T}}
	- \frac{1}{ 2 } \, \ch\Bigl( \frac{\epsilon v}{1 \!-\! \epsilon} \Bigr) i \Delta_\nu
	+ i\Upsilon
	\Biggr\}
	\, ,
\label{C1 result}
\\
\mathcal{C}_2 ={}&
	\frac{ e^{ \frac{\epsilon u}{1 - \epsilon} } }
		{ 2 \nu_{\scr T} (1\!-\!\epsilon)^2 H_0^2 } 
	\Biggl\{
	- \frac{1}{ 2 } \, \ch\Bigl( \frac{\epsilon v}{1 \!-\! \epsilon} \Bigr)
		\frac{\partial}{\partial y} i \Delta_\nu
	+
	\frac{\partial}{\partial y} i\Upsilon
	\Biggr\}
	\, ,
\label{C2 result}
\\
\mathcal{C}_3 ={}&
	\frac{ e^{ \frac{\epsilon u }{1-\epsilon} }}
		{ 2 \nu_{\scr T} (1\!-\!\epsilon)^2 H_0^2 } 
	\Biggl\{
	\frac{\nu_{\scr T}}{2} i \Delta_{\nu_{\scr T}}
	+
	\frac{ 1 }{ 2 \, \sh(v) } \,
	\ch \Bigl( \frac{ \epsilon v }{1 \!-\! \epsilon} \Bigr)
	\biggl[
	\frac{\partial}{\partial v} i \Delta_{\nu+1}
	-
	\ch(v) \frac{\partial}{\partial v} i \Delta_\nu
	\biggr]
\nonumber \\
&	\hspace{0.3cm}
	-
	\frac{ ( D\!-\!3 ) }{4} \, \ch\Bigl( \frac{\epsilon v}{1 \!-\! \epsilon} \Bigr)
		i \Delta_{\nu}
	-
	\frac{1}{2} \,\sh\Bigl( \frac{\epsilon v}{1 \!-\! \epsilon} \Bigr)
		\frac{\partial}{\partial v} 
		i \Delta_{\nu}
	+
	\frac{(D\!-\!3)}{2} i\Upsilon
	\Biggr\}
	\, ,
\label{C3result}
\\
\overline{\mathcal{C}}_3 ={}&
	\frac{ e^{ \frac{\epsilon u }{1-\epsilon} }}
		{ 2 \nu_{\scr T} (1\!-\!\epsilon)^2  H_0^2 } 
	\Biggl\{
	-
	\,
	\sh \Bigl( \frac{ \epsilon v }{1 \!-\! \epsilon} \Bigr)
	\frac{\partial}{\partial y}
	\biggl[
	i\Delta_{\nu+1}
	+
	\ch(v) 
	i \Delta_{\nu}
	\biggr]
	-
	2 \,\sh(v) \frac{\partial }{\partial y} i \Upsilon
	\Biggr\} 
	\, ,
\label{C3bar result}
\\
\mathcal{C}_4 ={}&
	\frac{ e^{ \frac{\epsilon u}{1-\epsilon} } }{ 2 \nu_{\scr T} (1\!-\!\epsilon)^2 H_0^2 } 
	\Biggl\{
	\ch\Bigl( \frac{\epsilon v}{1 \!-\! \epsilon} \Bigr)
	\biggl[
	\frac{\partial}{\partial u}
	+
	\frac{ \bigl[ 2\!-\!y \!-\! 2 \, \ch(v) \bigr]  }{ 2 \, \sh(v) } \frac{\partial}{\partial v} 
	\biggr]
	i \Delta_{\nu+1}
\label{C4 result}
\\
&	\hspace{0.3cm}
	+
	\frac{1}{2} \,
	\ch\Bigl( \frac{\epsilon v}{1 \!-\! \epsilon} \Bigr)
	\biggl[
	(2\!-\!y) \biggl(
		\frac{\partial}{\partial u}
		\!+\! 2\nu \!-\! D\!+\!3
		\biggr)
	-
	\frac{ \bigl[ ( 2\!-\!y ) \, \ch(v) \!-\! 2 \bigr] }{ \sh(v) }
	\frac{\partial}{\partial v}
	\biggr]
	i \Delta_\nu
\nonumber \\
&	\hspace{0.3cm}
	-
	\sh \Bigl( \frac{ \epsilon v }{1 \!-\! \epsilon} \Bigr)
	\frac{\partial}{\partial v} 
	\Bigl[
	(2\!-\!y) i \Delta_{\nu} 
	+
	2 i \Delta_{\nu+1}
	\Bigr]
	+
	\Bigl[ 
	\bigl( 4y\!-\!y^2 \bigr) \frac{\partial}{\partial y}
	+
	(D\!-\!2) (2\!-\!y) 
	\Bigr]
	i\Upsilon
	\Biggr\}
	\, ,\;
\nonumber
\end{align}
\endgroup
where in simplifying the final expressions we made use of generalized recurrence 
relations~(\ref{generalized recurrence 1}) and~(\ref{generalized recurrence 2})
for scalar two-point functions, the reduction formula~(\ref{Upsilon eom}), and
addition formulas for hyperbolic functions. These structure functions are the main
result of our paper, as they determine the photon two-point function
 in the simple covariant gauge~(\ref{simple covariant gauge})
in the covariant representation~(\ref{cov rep}). The remaining two-point functions
defined in Sec.~\ref{subsec: Generalities} are obtained from this result by 
changing the~$i\delta$-prescription for the  bi-local variables in the structure functions
and the tensor basis to the one appropriate according to~(\ref{Delta x2 prescription})
and~(\ref{v prescription}).

Explicitly checking that our result for the photon two-point function~(\ref{cov rep})
with solutions for the structure functions above satisfies both the 
the equation of motion~(\ref{2pt EOM}) and the Ward-Takahashi 
identity~(\ref{WT identity}) is important, particularly 
in view of recent findings~\cite{Glavan:2022pmk}
that  in de Sitter space a number of the results from the literature
fails to satisfy the latter condition.
Our result indeed satisfies all the necessary equations; the details necessary
to perform the checks are given in Appendix~\ref{app: Checks for two-point function}.

\section{Various limits}
\label{sec: Various limits}

In this section we derive several limits of the covariant form of the two-point 
function~(\ref{cov rep}) with structure functions~(\ref{C1 result})--(\ref{C4 result}),
and compare them to the literature where possible.

\subsection{De Sitter limit}
\label{subsec: De Sitter limit}

In the de Sitter limit
the bi-local variables~(\ref{bilocal variables}) simply take their values with~$\epsilon\!=\!0$,
and accordingly so do the basis tensor introduced in~(\ref{cov rep}).
All the limits of the scalar two-point functions and derived quantities
are manifestly finite in the de Sitter limit, due to the IR sum. Therefore, 
all the terms in structure functions~(\ref{C1 result})--(\ref{C4 result}) multiplied
by~$\epsilon$ automatically vanish in the de Sitter limit. The few terms that 
remain have simple limits.
The mode function indices~(\ref{nuT def}) and~(\ref{nu def}) 
from the transverse and scalar sectors become degenerate,
\begin{equation}
\nu, \ \nu_{\scr T} \xrightarrow{\epsilon \to 0} \nu_0 = \frac{D\!-\!3}{2} \, .
\label{dS indices}
\end{equation}
The scalar two-point functions carrying those indices 
reduce to one and the same de Sitter invariant scalar two-point function,
\begin{equation}
i \Delta_\nu(y,u,v) , \
	i \Delta_{\nu_{\scr T}}(y,u,v)
	\xrightarrow{\epsilon \to 0} \mathcal{F}_{\nu_0}(y) 
	\, ,
\end{equation}
Their derivatives with respect to~$y$ persist, but derivatives with
respect to~$u$ and~$v$ vanish. For the Laplace-inverted two-point 
function~(\ref{check Xi solution}) 
it is only relevant that the derivative with respect to~$u$ vanishes,
\begin{equation}
\frac{\partial}{\partial u} 
	\frac{\partial}{\partial y}
	i \Xi_\nu(y,u,v) \xrightarrow{\epsilon \to 0} 0  \, ,
\label{check Xi flat}
\end{equation}
as this implies that all of them cancel completely in~(\ref{Upsilon}),
\begin{equation}
i \Upsilon(y,u,v) \xrightarrow{\epsilon \to 0} 0 \, .
\label{Upsilon dS}
\end{equation}
The limits above encompass all the terms appearing in~(\ref{C1 result})--(\ref{C4 result}),
except for few instances where the scalar propagator with the index~$\nu\!+\!1$ appears.
Because in de Sitter limit we have for the index~$\nu\!+\!1 \! \xrightarrow{\epsilon\to0}\! \nu_0 \!+\! 1 \!=\! (D\!-\!1)/2$, the two-point function will reduce to the MMCS two-point
function, which is known not to be de Sitter invariant. All instances when it appears 
include derivatives, which are evaluated using the 
form given in Sec.~\ref{subsubsec: IR divergent scalar two-point functions},
\begin{equation}
\frac{\partial}{\partial u} i \Delta_{\nu+1}(y,u,v) 
	\xrightarrow{ \epsilon \to 0 }
	\frac{ H_0^{D-2 }  }{ (4\pi)^{\frac{D}{2}} } 
	\frac{  \Gamma(D \!-\! 1)  }{ \Gamma\bigl( \frac{D}{2} \bigr) } 
	\, ,
\qquad \quad
\frac{\partial}{\partial v} i \Delta_{\nu+1}(y,u,v) 
	\xrightarrow{ \epsilon \to 0 }
	0
	\, .
\label{du Delta}
\end{equation}
Finally, taking the de Sitter value of the simple covariant gauge,
\begin{equation}
\xi_s \xrightarrow{\epsilon \to 0} \xi_s^0 
	= \frac{ \nu_0 \!+\! 1 }{ \nu_0 }
	=  \frac{ D \!-\! 1 }{ D \!-\! 3 } \, ,
\end{equation}
leads to the de Sitter limit of structure functions,
\begin{align}
\mathcal{C}_1 
	\xrightarrow{\epsilon\to0}{}&
	\frac{ 1 }{ 2 \nu_0 H_0^2 } 
	\biggl[
	- \Bigl( \nu_0 \!+\! \frac{1}{2} \Bigr) \mathcal{F}_{\nu_0} (y)
	\biggr]
	\, ,
\qquad \quad
\mathcal{C}_2 
	\xrightarrow{\epsilon\to0}
	\frac{ 1 }{ 2 \nu_0 H_0^2 } 
	\biggl[
	- \frac{1}{ 2 }
		\frac{\partial}{\partial y} \mathcal{F}_{\nu_0} (y)
	\biggr]
	\, ,
\nonumber \\
\mathcal{C}_3 
	\xrightarrow{\epsilon\to0}{}&
	0
	\, ,
\qquad \quad
\overline{\mathcal{C}}_3 
	\xrightarrow{\epsilon\to0}
	0
	\, ,
\qquad \quad
\mathcal{C}_4
	\xrightarrow{\epsilon\to0}
	\xi_s^0
	\times
	\frac{ H_0^{D-4 }  }{ (4\pi)^{\frac{D}{2}} } 
	\frac{  \Gamma(D \!-\! 1)  }{ (D\!-\!1) \, \Gamma\bigl( \frac{D}{2} \bigr) } 
	\, .
\end{align}
These correctly reproduce the de Sitter limit,
including the de Sitter breaking non-vanishing structure function~$\mathcal{C}_4$
recently found in~\cite{Glavan:2022dwb,Glavan:2022nrd}. The latter is here
a consequence of the first expression in~(\ref{du Delta}) not vanishing, that is due to the 
non-existence of IR finite de Sitter invariant two-point functions for MMCS in the CTBD 
state~\cite{Allen:1985ux,Allen:1987tz}.

\medskip

While the covariant gauge two-point function has a relatively simple de Sitter limit, 
compared to its form in power-law inflation, no such simplifications
happen for the physical two-point function~(\ref{phys ij}),
\begin{align}
i \bigl[ \tensor*[_{i\,}^{\scr - \! }]{\Delta}{_j^{\scr \! +}} \bigr]^{\rm \scr ph} (x;x')
	\xrightarrow{\epsilon\to0}& \,
	e^{ u } 
	\biggl[
	\delta_{ij}
	i \Delta_{\nu_{0}} (y,u,v)
	+
	\partial_i \partial'_j
	\frac{ i \Xi_{\nu_{0} }(y,u,v) }{ \mathcal{H} \mathcal{H}' }
	\biggr]
	\, ,
\end{align}
since~$i\Xi_{\nu_{\scr T}}$ retains its form~(\ref{Xi solution}) without any simplifications.
It is really the transverse and the longitudinal contributions to the~$(ij)$ component
of the two-point function that conspire so that the complicated parts cancel between 
them. In practice this is due to two non-commensurate mode function indices
reducing to one and the same in the de Sitter limit~(\ref{dS indices}).

\subsection{Flat space limit}
\label{subsec: Flat space limit}

In the Minkowski space limit,~$H_0 \!\to\! 0$, the three bi-local 
variables~(\ref{bilocal variables}) reduce to,
\begin{eqnarray}
y_{\tt ab} \overset{H_0 \to 0}{\longsim} (1\!-\!\epsilon)^2 H_0^2 \Delta x_{\tt ab}^2 
	\, ,
\quad \
u_{\tt ab} \overset{H_0 \to 0}{\longsim}
	(1\!-\!\epsilon) H_0 \bigl( \eta \!+\! \eta' \!-\! 2 \eta_0 \bigr)
	 ,
\quad \
v_{\tt ab} \overset{H_0 \to 0}{\longsim} (1\!-\!\epsilon) H_0 \Delta\eta_{\tt ab}
	\,,
\nonumber\\
\end{eqnarray}
and accordingly the tensor structures reduce to,
\begin{subequations}
\begin{align}
\bigl( \partial_\mu \partial'_\nu y \bigr)
	\, \overset{H_0 \to 0}{\longsim}{}&
	- 2 \bigl[ (1\!-\!\epsilon) H_0 \bigr]^2 \eta_{\mu\nu} 
	\, ,
\\
\bigl( \partial_\mu y \bigr) \bigl( \partial'_\nu y \bigr)
	\, \overset{H_0 \to 0}{\longsim}{}&
	- 4 \bigl[ (1\!-\!\epsilon) H_0 \bigr]^4 \Delta x_\mu \Delta x_\nu 
	\, ,
\\
\Bigl[ \bigl( \partial_\mu y \bigr) \bigl( \partial'_\nu u \bigr)
		\pm \bigl( \partial_\mu u \bigr) \bigl( \partial'_\nu y \bigr) \Bigr]
	\, \overset{H_0 \to 0}{\longsim}{}& \,
	2 \bigl[ (1\!-\!\epsilon) H_0\bigr]^3 
		\bigl( \Delta x_\mu \delta_\nu^0 \mp \delta_\mu^0 \Delta x_\nu \bigr)
	\, ,
\\
\bigl( \partial_\mu u \bigr) \bigl( \partial'_\nu u \bigr)
	\, \overset{H_0 \to 0}{\longsim}{}& \,
	\bigl[ (1\!-\!\epsilon) H_0\bigr]^2 \delta_\mu^0 \delta_\nu^0
	\, ,
\quad
\end{align}
\end{subequations}
where~$\Delta x_\mu \!=\! x_\mu \!-\! x'_\mu$.
For derivatives appearing in the scalar structure functions we have,
\begin{equation}
\frac{\partial}{\partial y} 
	\, \overset{H_0 \to 0}{\longsim} \,
	\frac{1}{ (1\!-\!\epsilon)^2 H_0^2 } \frac{ \partial }{ \partial (\Delta x^2)}
	\, ,
\qquad
\qquad
\frac{\partial}{\partial v} 
	\, \overset{H_0 \to 0}{\longsim} \, 
	\frac{1}{(1\!-\!\epsilon) H_0} \frac{\partial}{\partial \Delta\eta}
	\, .
\end{equation}
The scalar two-point functions all reduce to the massless scalar two-point function,
\begin{equation}
i \Delta_{\lambda}(y,u,v)
	\xrightarrow{H_0 \to 0 }
	\frac{\Gamma\bigl( \frac{D-2}{2} \bigr) }
		{ 4\pi^{ \frac{D}{2} } \bigl( \Delta x^2 \bigr)^{\!\frac{D-2}{2}} } 
		\equiv 
		i \check{\Delta} \bigl( \Delta x^2 \bigr)
		\, ,
\quad
\label{Delta flat}
\end{equation}
and its derivatives with respect to remaining bi-local variables reduce to,
\begin{equation}
\frac{\partial}{\partial u} i \Delta_\lambda(y,u,v)
	\xrightarrow{ H_0 \to 0}
	- \frac{ (D\!-\!2) \epsilon}{ 2 (1\!-\!\epsilon) } \,
	i \check{\Delta} \bigl( \Delta x^2 \bigr)
	\, ,
\qquad
\frac{\partial}{\partial v} i \Delta_\lambda(y,u,v)
	\xrightarrow{ H_0 \to 0}
	0
	\, .
\end{equation}
Note that here we additionally have to assume that~$k_0\!\to\!0$,
such that~$k_0/H_0\!\ll\!1$ remains satisfied.
Using these we can also infer the flat space limit of~(\ref{check Xi solution}),
\begin{align}
\MoveEqLeft[2]
\frac{\partial}{\partial y}
i \Xi_\lambda(y,u,v)
	\xrightarrow{H_0 \to 0}
	\sum_{n=0}^{\infty}
	\frac{ (-1)^n \, \Gamma \bigl( \frac{D - 1}{2} \bigr) }
		{ 4 \, \Gamma \bigl( \frac{D+1}{2} \!+\! n \bigr) }
	\bigl( \Delta x^2  \!+\! \Delta\eta^2 \bigr)^n
	\frac{\partial^n }{\partial (\Delta x^2)^n } \,
	i \check{\Delta} \bigl( \Delta x^2 \bigr)
\nonumber \\
&
	=
	\sum_{n=0}^{\infty}
	\frac{ \Gamma \bigl( \frac{D - 1}{2} \bigr) \, \Gamma\bigl( \frac{D-2}{2} \!+\! n \bigr) }
		{ 4 \, \Gamma \bigl( \frac{D+1}{2} \!+\! n \bigr) }
	\frac{ \bigl( \Delta x^2  \!+\! \Delta\eta^2 \bigr)^n }
		{ 4\pi^{ \frac{D}{2} } \bigl( \Delta x^2 \bigr)^{\!\frac{D-2}{2} + n} } 
\label{check Xi flat limit} \\
&
	=
	\frac{ i \check{\Delta} \bigl( \Delta x^2 \bigr) }{ 2 (D\!-\!1) } \!\times\!
	{}_2F_1\biggl( \Bigl\{ 1 , \frac{D \!-\! 2}{2} \Bigr\} , \Bigl\{ \frac{D \!+\! 1}{2} \Bigr\} , 
		1 \!+\! \frac{ \Delta\eta^2 }{ \Delta x^2 }  \biggr)
	\equiv 
	\frac{\partial}{\partial (\Delta x^2)}
	i \check{\Xi} \bigl( \Delta x^2, \Delta\eta \bigr)
	\, ,
\nonumber
\end{align}
and of its derivative with respect to~$u$,
\begin{equation}
\frac{\partial}{\partial u} 
\frac{\partial}{\partial y}
	i \Xi_\lambda(y,u,v)
	\xrightarrow{H_0 \to 0}
	- \frac{ (D\!-\!2) \epsilon}{ 2 (1\!-\!\epsilon) }
	\frac{\partial}{\partial (\Delta x^2)}
	i \check{\Xi} \bigl( \Delta x^2, \Delta\eta \bigr)
	\, .
\end{equation}
It is then straightforward to derive the equation
that~(\ref{check Xi flat limit}) satisfies,
\begin{align}
&
\biggl[
	2 (\Delta\eta)^2
	\frac{ \partial }{ \partial (\Delta x^2)}
	-
	\Delta\eta
	\frac{\partial}{\partial \Delta\eta}
	+
	1
	\biggr]
		\frac{\partial}{\partial (\Delta x^2)}
	i \check{\Xi} \bigl( \Delta x^2, \Delta\eta \bigr)
	=
	\frac{1}{2}
	i \check{\Delta} \bigl( \Delta x^2 \bigr)
	\, ,
\label{du check Xi flat}
\end{align}
which, together with identities~(\ref{check Xi flat limit})
and~(\ref{du check Xi flat}), implies the flat space limit of~(\ref{Upsilon}),
\begin{equation}
i \Upsilon(y,u,v) \xrightarrow{H_0 \to 0}
	\frac{ - \epsilon }{( 1 \!-\! \epsilon) } \,  i \check{\Delta} \bigl( \Delta x^2 \bigr)
		\, .
\end{equation}
Given all the listed flat space limits, it is only the first two structure functions 
out of~(\ref{C1 result})--(\ref{C4 result})
that contribute to the flat space limit,
\begin{equation}
i \bigl[ \tensor*[_\mu^{\scr - \!}]{\Delta}{_\nu^{\scr \! +}} \bigr] (x;x')
	\xrightarrow{H_0\to0}
	\biggl[
	\eta_{\mu\nu} 
	- \frac{ (1 \!-\! \xi_s ) }{2} 
	\biggl(
	\eta_{\mu\nu}
	-
	(D\!-\! 2)
	\frac{ \Delta x_\mu \Delta x_\nu  }{ \Delta x^2 }
	\biggr)
	\biggr]
	i \check{\Delta} \bigl( \Delta x^2 \bigr) \, ,
\label{2pt flat limit}
\end{equation}
that can be written in a form more often encountered,
\begin{equation}
i \bigl[ \tensor*[_\mu^{\scr - \!}]{\Delta}{_\nu^{\scr \! +}} \bigr] (x;x')
	\xrightarrow{H_0\to0}
	\biggl[
	\eta_{\mu\nu} 
	- (1 \!-\! \xi_s )
	\frac{ \partial_\mu \partial_\nu  }{ \partial^2 }
	\biggr]
	i \check{\Delta} \bigl( \Delta x^2 \bigr)
	\, ,
\end{equation}
where the gauge-fixing parameter corresponding to the simple covariant gauge
defined in~(\ref{simple covariant gauge}). The two-point function in~(\ref{2pt flat limit})
of course contains the correct~$i\delta$-prescription~\cite{Falceto:2022wgj},
inherited from the bi-local variables adapted for FLRW.

\medskip

The flat-space limit of the physical two-point function~(\ref{phys ij}) 
is inferred from the flat space limits of the scalar two-point function~(\ref{Delta flat})
and of the Laplace-inverted two-point function~(\ref{M flat}),
\begin{align}
i \bigl[ \tensor*[_{i\,}^{\scr - \! }]{\Delta}{_j^{\scr \! +}} \bigr]^{\rm \scr ph} (x;x')
	\xrightarrow{H_0 \to 0 }{}&
	\frac{ \Gamma\bigl( \frac{D-2}{2} \bigr) }
		{ 4 \pi^{ \frac{D}{2} } }
	\Biggl[
	\frac{ \delta_{ij} }{ \bigl( \Delta x^2 \bigr)^{\!\frac{D-2}{2}} } 
	-
	\frac{ \bigl( - \Delta\eta^2 \bigr)^{ \! \frac{4-D}{2} } }
		{ 2 (D\!-\!3) (D\!-\!4) }
	\times 
\nonumber \\
&	\hspace{0.5cm}
	\times
	\partial_i \partial'_j \,
	{}_2F_1\biggl( \Bigl\{ \frac{D\!-\!3}{2} ,
		\frac{ D \!-\! 4 }{2} \Bigr\} ,
		\Bigl\{ \frac{ D \!-\! 1 }{2} \Bigr\} , 
		1 \!+\! \frac{ \Delta x^2}{ \Delta\eta^{2} } \biggr)
	\Biggr]
	\, .\;
\label{ph flat}
\end{align}
It corresponds to the Coulomb gauge two-point function in~$D$-dimensional 
Minkowski space. Of course, in four spacetime dimensions, due to conformal invariance,
the Coulomb gauge two-point function reduces to the flat space 
expression~\cite{Cotaescu:2008hv}. In position space this expression reads,
\begin{align}
i \bigl[ \tensor*[_{i\,}^{\scr - \! }]{\Delta}{_j^{\scr \! +}} \bigr]^{\rm \scr ph} (x;x')
	\xrightarrow{D \to 4 }  {}&
	\frac{ 1 }{ 4 \pi^2 }
	\Biggl\{
	\frac{ \delta_{ij} }{ \Delta x^2 } 
	+
	\frac{ \partial_i \partial'_j }{ 2 }
	\biggl[
	\ln \bigl( \mu^2 \Delta x^2 \bigr)
	+
	\frac{ \Delta\eta }{ \|\Delta \vec{x} \| } 
	\ln \biggl( \frac{ \Delta\eta \!+\! \| \Delta \vec{x} \| }
		{ \Delta\eta \!-\! \| \Delta \vec{x} \| } \biggr)
	\biggr]
	\Biggr\}
	\, .
\label{FourDimFlat}
\end{align}
Note that the seeming~$1/(D\!-\!4)$ divergence is removed by the spatial derivatives
in the second line of~(\ref{ph flat}).
Also note that the dependence on an arbitrary scale~$\mu$ drops out of the expression 
due to spatial derivatives, and its role is just to make the argument of the 
logarithm dimensionless.

\section{Simple observables}
\label{sec: Simple observables}

Our propagator in the simple covariant gauge given in~(\ref{cov rep}) 
with structure functions~(\ref{C1 result})--(\ref{C4 result}) 
can now be used in loop computations.
As a simple consistency check here we consider two simplest observables:
the  tree-level field strength correlators, and the one-loop energy-momentum tensor.

\subsection{Field strength correlator}
\label{subsec: Field strength correlators}

The tree-level correlator of the field stress tensor,
\begin{equation}
\bigl\langle \Omega \bigr| \hat{F}_{\mu\nu}(x) \hat{F}_{\rho\sigma}(x') \bigl| \Omega \bigr\rangle
	= 4 \bigl( \delta^\alpha_{[\mu} \partial_{\nu]} \bigr) 
		\bigl( \delta^\beta_{[\rho} \partial'_{\sigma]} \bigr)
		i \bigl[ \tensor*[_\alpha^{\scr - \!}]{\Delta}{_\beta^{\scr \! +}} \bigr](x;x')
\label{F correlator def}
\end{equation}
is gauge independent and constitutes an observable.
It can thus serve as a simple consistency check for the photon two-point function we 
found in Sec~\ref{sec: Two-point function}.
Acting the derivatives in~(\ref{F correlator def}) onto the covariantized
representation of the two-point function~(\ref{cov rep}) organizes itself
in the following tensor basis~\cite{Glavan:2020zne},
\begin{align}
\MoveEqLeft[5]
\bigl\langle \Omega \bigr| \hat{F}_{\mu\nu}(x) \hat{F}_{\rho\sigma}(x') \bigl| \Omega \bigr\rangle
	= \bigl( \partial_\mu \partial'_{[\rho} y \bigr) \bigl( \partial'_{\sigma]} \partial_\nu y \bigr)
		\mathcal{G}_1 
	+
	\bigl( \partial_{[\mu} y \bigr) \bigl( \partial_{\nu]} \partial'_{[\sigma} y \bigr)
		\bigl( \partial'_{\rho]} y \bigr) \mathcal{G}_2
\nonumber \\
&
	+\Bigl[ \bigl( \partial_{[\mu} y \bigr) \bigl( \partial_{\nu]} \partial'_{[\sigma} y \bigr) 
			\bigl( \partial'_{\rho]} u \bigr)
		\!+\! \bigl( \partial_{[\mu} u \bigr) \bigl( \partial_{\nu]} \partial'_{[\sigma} y \bigr) 
			\bigl( \partial'_{\rho]} y \bigr) \Bigr] \mathcal{G}_3
\nonumber \\
&
	+\Bigl[ \bigl( \partial_{[\mu} y \bigr) \bigl( \partial_{\nu]} \partial'_{[\sigma} y \bigr) 
			\bigl( \partial'_{\rho]} u \bigr)
		\!-\! \bigl( \partial_{[\mu} u \bigr) \bigl( \partial_{\nu]} \partial'_{[\sigma} y \bigr) 
			\bigl( \partial'_{\rho]} y \bigr) \Bigr] \overline{\mathcal{G}}_3
\nonumber \\
&
	+ \bigl( \partial_{[\mu} u \bigr) \bigl( \partial_{\nu]} \partial'_{[\sigma} y \bigr)
			\bigl( \partial'_{\rho]} u \bigr) \mathcal{G}_4
	+ \bigl( \partial_{[\mu} y \bigr) \bigl( \partial_{\nu]} u \bigr)
		\bigl( \partial'_{[\rho} y \bigr) \bigl( \partial'_{\sigma]} u \bigr) \mathcal{G}_5
	\, ,
\label{F tensor basis}
\end{align}
where the structure functions are expressed in terms of structure 
functions~(\ref{C1 result})--(\ref{C4 result}) of the photon two-point function,
\begin{align}
\mathcal{G}_1 ={}&
	4 \biggl(
	\frac{\partial \mathcal{C}_1 }{\partial y}
	- \mathcal{C}_2
	\biggr)
	\, ,
\qquad
\mathcal{G}_2 =
	\frac{ \partial \mathcal{G}_1 }{ \partial y }
	\, ,
\qquad
\mathcal{G}_3 =
	\frac{ \partial \mathcal{G}_1 }{ \partial u }
	\, ,
\qquad
\overline{\mathcal{G}}_3 =
	- \frac{ \partial \mathcal{G}_1}{\partial v}
	\, ,
\label{G structure functions}
\\
\mathcal{G}_4 ={}&
	4 \biggl(
	\frac{\partial^2 \mathcal{C}_1}{\partial u^2} 
	- \frac{\partial^2 \mathcal{C}_1}{\partial v^2} 
	- 2 \frac{\partial \mathcal{C}_3}{\partial u}
	- 2 \frac{\partial \overline{ \mathcal{C}}_3}{\partial v}
	+ \frac{\partial \mathcal{C}_4}{\partial y}
	\biggr)
	\, ,
\qquad
\mathcal{G}_5 =
	\frac{\partial \mathcal{G}_4}{\partial y}
	- \frac{\partial^2 \mathcal{G}_1}{\partial u^2}
	+ \frac{\partial^2 \mathcal{G}_1}{\partial v^2}
	\, .
\nonumber
\end{align}
Evaluating these structure functions is accomplished by judiciously applying
equations~(\ref{Upsilon eom})--(\ref{Upsilon last eq})
for the~$i\Upsilon$ function defined in~(\ref{Upsilon}),
as well as equations of motion~(\ref{scalar eom even}) and~(\ref{scalar eom odd}),
and recurrence relations~(\ref{generalized recurrence}) for the scalar 
two-point functions,
\begingroup
\allowdisplaybreaks
\begin{subequations}
\begin{align}
\mathcal{G}_1 ={}&
	\frac{ 2 \, e^{-\frac{(D-4)\epsilon}{2(1-\epsilon)} u } }{ (1\!-\!\epsilon)^2 H_0^2 }
		\biggl[ - \frac{\partial \mathcal{F}_{\nu_{\scr T}}}{ \partial y} \biggr] \, ,
\\
\mathcal{G}_2 ={}&
	\frac{ 2 \, e^{-\frac{(D-4)\epsilon}{2(1-\epsilon)} u } }{ (1\!-\!\epsilon)^2 H_0^2 }
		\biggl[ - \frac{\partial^2 \mathcal{F}_{\nu_{\scr T}}}{ \partial y^2 } \biggr] \, ,
\\
\mathcal{G}_3 ={}&
	\frac{ 2 \, e^{-\frac{(D-4)\epsilon}{2(1-\epsilon)} u } }{ (1\!-\!\epsilon)^2 H_0^2 }
		\biggl[ \frac{(D\!-\!4)\epsilon}{2(1\!-\!\epsilon)} \, \frac{\partial \mathcal{F}_{\nu_{\scr T}}}{ \partial y } \biggr] \, ,
\\
\overline{\mathcal{G}}_3 ={}&
	0
	\, ,
\\
\mathcal{G}_4 ={}&
	\frac{ 2 \, e^{-\frac{(D-4)\epsilon}{2(1-\epsilon)} u } }{ (1\!-\!\epsilon)^2 H_0^2 }
	\biggl[
	\frac{(D\!-\!4)\epsilon}{2(1\!-\!\epsilon)}
	\biggl( 1 \!-\! \frac{(D\!-\!4)\epsilon}{2(1\!-\!\epsilon)} \biggr)
	\mathcal{F}_{\nu_{\scr T}}
	\biggr] 
	\, ,
\\
\mathcal{G}_5 ={}&
	\frac{ 2 \, e^{-\frac{(D-4)\epsilon}{2(1-\epsilon)} u } }{ (1\!-\!\epsilon)^2 H_0^2 }
	\biggl[
	\frac{(D\!-\!4)\epsilon}{2(1\!-\!\epsilon)}
	\frac{\partial \mathcal{F}_{\nu_{\scr T}} }{ \partial y}
	\biggr] 
	\, .
\end{align}
\end{subequations}
\endgroup
The end result reveals that the only contributions come 
from the transverse sector. That means our two-point function 
produces a gauge independent result for this observable, as it should.
Furthermore, the~$D\!\to\!4$ limit reduces to the flat space vacuum result,
\begin{equation}
\bigl\langle \Omega \bigr| \hat{F}_{\mu\nu}(x) \hat{F}_{\rho\sigma}(x') \bigl| \Omega \bigr\rangle
	\xrightarrow{D\to4}
	\frac{2}{\pi^2 \bigl(\Delta x^2 \bigr)^{\!2} }
	\biggl[
	\eta_{\mu[\rho} \eta_{\sigma]\nu} 
	- 4 \eta_{\alpha [\mu} \eta_{\nu] [\sigma } \eta_{\rho] \beta}
		\frac{ \Delta x^\alpha \Delta x^\beta }{ \Delta x^2}
	\biggr]
	\, .
\label{FF limit}
\end{equation}
This is a manifestation of the conformal coupling of the photon to gravity in~$D\!=\!4$,
which allows for a natural choice of the photon vacuum state that does not sense the expansion.
The limit~(\ref{FF limit}) confirms that we have chosen precisely such a state for the photon.

\subsection{Energy-momentum tensor}
\label{subsec: Energy-momentum tensor}

The energy-momentum tensor can be given by two different
definitions: either as a variation of the gauge invariant Maxwell 
action~(\ref{invariant action}), or as a variation of the gauge-fixed 
action~(\ref{fixed action}). This is the case both in the classical and the 
quantum theory. However, it is sufficient
to consider just the former definition, 
\begin{equation}
T_{\mu\nu} (x)
	= 
	\frac{-2}{ \sqrt{-g} } \frac{\delta S}{ \delta g^{\mu\nu}(x) }
	=
	\Bigl( \delta_\mu^\rho \delta_\nu^\sigma - \frac{1}{4} g_{\mu\nu} g^{\rho\sigma} \Bigr) 
		g^{\alpha\beta} F_{\rho\alpha}(x) F_{\sigma\beta}(x)
		\, ,
\label{classical Tmunu}
\end{equation}
as the latter differs by the contribution of the gauge-fixing part of the action that is 
guaranteed to vanish on-shell. In the classical theory this is a 
simple consequence of the two constraints~(\ref{initial constraints}) vanishing on-shell.
In the quantized theory the contribution of the gauge-fixing part can be seen to vanish
on-shell in two ways: (i) either due to the proper operator ordering and to the 
subsidiary condition~(\ref{K condition})~\cite{Glavan:2022pmk},
or (ii) due to the cancellations between the gauge-fixing
part and the Faddeev-Popov ghost part that has to be added if all the operators are 
Weyl-ordered~\cite{Belokogne:2015etf,Belokogne:2016dvd,Glavan:2022nrd}.

Defining an operator associated to~(\ref{classical Tmunu}) is straightforward,
since when expressed in terms of the canonical fields all the terms are composed either 
solely of transverse fields, or solely of constraints. Therefore, we may define the operator 
to be Weyl-ordered, and the expectation value essentially reduces to the coincident field strength correlator,
\begin{equation}
\bigl\langle \Omega \bigr| \hat{T}_{\mu\nu}(x) \bigl| \Omega \bigr\rangle
	=
	\Bigl( \delta_\mu^\rho \delta_\nu^\sigma - \frac{1}{4} g_{\mu\nu} g^{\rho\sigma} \Bigr) 
		g^{\alpha\beta} \bigl\langle \Omega \bigr| \hat{F}_{\rho\alpha}(x) \hat{F}_{\sigma\beta}(x) \bigl| \Omega \bigr\rangle
		\, .
\label{F correlator}
\end{equation}
Computing the coincident correlator amounts to computing a
dimensionally regulated coincidence limit of~(\ref{F correlator def}).
This is best done in the tensor basis~(\ref{F tensor basis}) where the only
non-vanishing tensor structures in this limit are the first and the fourth one,
\begin{align}
\bigl( \partial_\mu \partial'_{[\rho} y \bigr) \bigl( \partial'_{\sigma]} \partial_\nu y \bigr)
	\xrightarrow{x' \! \to x}{}&
	4 \bigl[ (1\!-\!\epsilon) H \bigr]^4 g_{\mu [\rho} g_{\sigma] \nu}
	\, ,
\\
\bigl( \partial_{[\mu} u \bigr) \bigl( \partial_{\nu]} \partial'_{[\sigma} y \bigr)
			\bigl( \partial'_{\rho]} u \bigr)
	\xrightarrow{x' \! \to x}{}&
	- 2 \bigl[ (1\!-\!\epsilon) H \bigr]^4 
	\bigl( a^2 \delta^0_{[\mu} g_{\nu] [\sigma} \delta^0_{\rho] } \bigr)
	\, ,
\end{align}
and where the dimensionally regulated coincident limits of the corresponding 
structure functions in~(\ref{F tensor basis}) are inferred from the power-series 
representation~(\ref{power series}),
\begin{align}
\mathcal{G}_1 \xrightarrow{x' \! \to x}{}&
	\frac{\bigl[ (1\!-\!\epsilon) H \bigr]^{D-4}}{ (4\pi)^{ \frac{D}{2} } }
	\frac{ \Gamma\bigl( \frac{D+1}{2} \!+\! \nu_{\scr T} \bigr) \, \Gamma\bigl( \frac{D+1}{2} \!-\! \nu_{\scr T} \bigr) }
			{ \Gamma\bigl( \frac{1}{2} \!+\! \nu_{\scr T} \bigr) \, \Gamma\bigl( \frac{1}{2} \!-\! \nu_{\scr T} \bigr) }
	\frac{ \Gamma\bigl( \frac{2-D}{2} \bigr)}{ (-D) }
	\xrightarrow{ D \to 4}
	\frac{1}{32 \pi^2 (1\!-\!\epsilon)}
	\, ,
\\
\mathcal{G}_4  \xrightarrow{x' \to x}{}&
	\frac{\bigl[ (1\!-\!\epsilon) H \bigr]^{D-4}}{ (4\pi)^{\frac{D}{2}} }
	\frac{ \Gamma\bigl( \frac{D-1}{2} \!+\! \nu_{\scr T} \bigr) \, \Gamma\bigl( \frac{D-1}{2} \!-\! \nu_{\scr T} \bigr) }
			{ \Gamma\bigl( \frac{1}{2} \!+\! \nu_{\scr T} \bigr) \, \Gamma\bigl( \frac{1}{2} \!-\! \nu_{\scr T} \bigr) }
\nonumber \\
&	\hspace{4cm}
	\times
	\biggl[ 1 \!-\! \frac{(D\!-\!4)\epsilon}{2(1\!-\!\epsilon)} \biggr]
	\frac{4 \epsilon \, \Gamma\bigl( \frac{6-D}{2} \bigr)}{ (D\!-\!2) (1\!-\!\epsilon) }
	\xrightarrow{ D \to 4}
	0 
	\, .
\end{align}
These limits are finite in~$D\!=\!4$, and thus no counterterms are necessary 
for renormalization. Plugging the coincident correlator into~(\ref{F correlator})
and performing the remaining contractions 
finally gives a vanishing result for the gauge-invariant energy-momentum tensor,
\begin{equation}
\bigl\langle \Omega \bigr| \hat{T}_{\mu\nu}(x) \bigl| \Omega \bigr\rangle
	=
	0
	\, .
\label{Tmunu result}
\end{equation}

There is still the conformal anomaly contributing to the
energy-momentum tensor at one loop~\cite{Capper:1974ic,Brown:1977pq,Adler:1976jx},
that is not captured by the result~(\ref{Tmunu result}).
In fact, for conformally coupled fields in conformally flat backgrounds the 
conformal anomaly contribution to the energy-momentum tensor is the only nonvanishing
one~\cite{Brown:1977sj}, and its unambiguous part
is gauge independent~\cite{Endo:1984sz}.
In that sense the result in~(\ref{Tmunu result}) is
a consistency check of the two-point function we computed.
The conformal anomaly contribution, however, does not appear due to divergences 
in the one-loop diagram corresponding to the source for the graviton tadpole,
which is the energy-momentum tensor. 
As a matter of fact, there are no logarithmic divergences in that diagram,
as we have shown in this section.
The conformal anomaly contribution appears due to divergences found in other 
one-loop diagrams of the theory, and  to recover it one would need to renormalize the 
effective action, rather than just the particular diagram corresponding to the 
observable we consider here.

\section{Discussion}
\label{sec: Discussion}

Understanding how large are the effects of quantum loop corrections 
to inflationary observables is of indisputable importance. In this work we were concerned
with the building blocks necessary for quantifying these corrections in theories 
containing gauge vector fields in power-law inflation. Two-point functions
of free quantum fields --- propagators ---  are the basic ingredients for 
computing Feynman diagrams
representing loop corrections. This motivated us to consider the photon propagator 
for power-law inflation in general covariant gauges.
The main result of this work is the~$D$-dimensional position space two-point 
function for the photon in power-law inflation. We have presented it in the covariantized
form~(\ref{cov rep}) with structure functions given in~(\ref{C1 result})--(\ref{C4 result}).
We have computed this two-point function in the simple covariant 
gauge~(\ref{simple covariant gauge}) because only for this particular choice are the
photon two-point functions~(\ref{v0 simple}) and~(\ref{vL simple}) simple enough
for the evaluation of the Fourier integrals in~(\ref{som 00})--(\ref{som ij}) to be feasible.
Our two-point function satisfies both the equations of motion~(\ref{2pt EOM}),
and the Ward-Takahashi identity~(\ref{WT identity}) that follows from the canonical
quantization recalled in Sec.~\ref{sec: Photon in FLRW} and 
Sec.~\ref{sec: Field operator dynamics}.
The detailed checks of these are given in Appendix~\ref{app: Checks for two-point function}.
Furthermore, our result correctly reproduces the flat space limit, and the de Sitter
space limit, including the de Sitter breaking term~\cite{Glavan:2022dwb,Glavan:2022nrd}.
Both of these limits are worked out in Sec.~\ref{sec: Various limits}.

The complexity of the final result for the photon propagator is somewhat unexpected. 
The experience with 
non-minimally coupled massless scalar fields suggests that scalar mode functions,
and consequently scalar two-point functions are just as complicated in power-law
inflation as they are in de Sitter. The expectation was the same for photons, for which
this turns out not to be the case. The complications descend from the fact that
equations of motion couple the components of the vector potential in a way
that in general no longer yields simple CTBD scalar mode functions and their derivatives 
as solutions.
This is seen in Sec.~\ref{subsec: Scalar sector} where we give the mode function solutions
for the scalar sector of the vector potential,
that are worked out in the accompanying appendix~\ref{sec: Particular mode functions}.
Only for a special choice of the gauge-fixing parameter~$\xi\!=\!\xi_s$ 
in~(\ref{simple covariant gauge}), that we dubbed the simple covariant gauge, do the 
mode functions simplify to the well known
CTBD ones given in~(\ref{mode function}) and their derivatives. 
In this special case the momentum space two-point function of the photon
retains the level of complexity it has in de Sitter.

In position space, on the other hand, even the simple covariant gauge two-point function
takes on a considerably more complex form in power-law inflation that it does in the
de Sitter space limit~\cite{Glavan:2022dwb}. 
Here one is required to explicitly evaluate the inverse Laplace operator
acting on the scalar two-point function. In 
Sec.~\ref{subsec: Laplace-inverted two-point function} we have evaluated this
Laplace-inverted two-point function in terms of a series~(\ref{Xi solution}) of higher 
and higher derivatives acting in the scalar two-point
function, that is appropriate for sub-Hubble separations. 
In the accompanying appendix~\ref{app: Inverting Laplacian} the closed
form expressions for this object are found in terms of 
 Appell's fourth function. This function is a particular instance
of a non-factorizable double hypergeometric series, and its analytical structure is
much more complicated than the structure of the hypergeometric function that
describes scalar propagators. Incidentally, evaluating
the Laplace-inverted two-point function explicitly is also necessary to compute the
Coulomb gauge photon two-point function in position space, and we also report 
this result in~(\ref{phys ij}). 

The penultimate section of the main text is devoted to checking that our two-point
function correctly reproduces two simple observables, namely the field strength
correlator~(\ref{F correlator def}), and the energy-momentum tensor~(\ref{F correlator}).
It is confirmed that only the transverse sector contributes to these observables,
and that the gauge sector  completely drops out. It is straightforward to demonstrate such 
manifest gauge independence in these two simple examples. This is because the lowest
orders of the two observables are composed of a single photon two-point function only. 
In general this is not the case for more complicated loop computations. Our 
result passes all the tests before attempting these. The practical problem, however, 
might be the considerable complexity that our power-law inflation two-point function 
exhibits, compared to its de Sitter counterpart. Another issue is the absence of a free
gauge-fixing parameter in the simple covariant gauge, which precludes manifest checks
of gauge-independence of the final results when computing observables.
However, gauge independence {\it can} be checked when combined with 
computation using the propagator in another gauge, such as the Coulomb gauge 
two-point function we also worked out in~(\ref{phys ij}).

In conclusion, one can use our photon two-point function in the simple covariant gauge
for loop computations in power-law inflation. However, it might be advisable to 
first examine whether different linear non-covariant gauges~\cite{Glavan:2022pmk}
lead to more tractable photon two-point functions, and consequently to simpler loop
computations. It is also worth pointing out that similar complications as seen here
are expected if attempting to construct the graviton propagator in covariant gauges 
in power-law inflation, for which no results are known~(cf.~\cite{Janssen:2007ht}).

\section*{Acknowledgements}

We are grateful to Richard P. Woodard for explaining to us the origins of conformal
anomaly in dimensional regularization; to Igor Khavkine for helpful discussions about the 
analytic structure of Appell's functions; and to Jos\'{e} L.~L\'{o}pez for 
looking into asymptotic expansions of Appell's first function and sending us 
preliminary results.
DG was supported by the European Union and the Czech Ministry of Education, 
Youth and Sports 
(Project: MSCA Fellowship CZ FZU I --- CZ.02.01.01/00/22\textunderscore010/0002906).
This work is part of the Delta ITP consortium, a program of the Netherlands Organisation
for Scientific Research (NWO) that is funded by the Dutch Ministry of Education, Culture
and Science (OCW) --- NWO project number 24.001.027.

\appendix

\section{Particular mode functions}
\label{sec: Particular mode functions}

Here we solve equations~(\ref{particular eq 1}) and~(\ref{particular eq 2}) for the particular 
mode functions for an arbitrary gauge-fixing parameter~$\xi$.
After shifting the mode functions,
\begin{align}
v_0 
	={}& 
	\frac{ - i \xi k }{ 2 \bigl[ \nu(1\!-\!\epsilon) \!+\! 1 \bigr] } 
		\biggl[ \frac{a^2}{\mathcal{H}} U_{\nu+1} - \frac{1}{H_0} U_\nu \biggr]
	+ \Bigl( 1 \!-\! \frac{ \xi }{ \xi_s } \Bigr) w_0 
	\, ,
\label{v0 shifted}
\\
v_{\scr L} 
	={}& 
	\frac{ - i \xi k }{2 \bigl[ \nu(1\!-\!\epsilon) \!+\! 1 \bigr] } 
		\biggl[ \frac{a^2}{\mathcal{H}} U_\nu - \frac{1}{H_0} U_{\nu+1} \biggr]
	+ \Bigl( 1 \!-\! \frac{ \xi }{ \xi_s } \Bigr) w_{\scr L} 
	\, ,
\label{vL shifted}
\end{align}
and applying identities~(\ref{mode recurrence}) and~(\ref{mode id1}),
these equations read,
\begin{align}
\biggl[ \partial_0^2 + k^2 
	- \Bigl( \nu^2 \!-\! \frac{1}{4} \Bigr) (1\!-\!\epsilon)^2 \mathcal{H}^2 \biggr]
	w_{0} 
	={}& 
	a^2 k^2 U_\nu
	\, ,
\\
w_{\scr L} 
	={}&
	\frac{i}{k} \biggl[
		\partial_0 + \Bigl( \nu\!+\! \frac{1}{2} \Bigr) (1\!-\!\epsilon) \mathcal{H}
		\biggr] w_{0} 
		\, .
\label{wL equation}
\end{align}
We solve the first equation using the retarded Green's function,
\begin{equation}
G_{\scr R}(\eta;\eta') = \theta(\eta\!-\!\eta') G(\eta;\eta') \, ,
\qquad \! \!
G(\eta;\eta') = i \Bigl[ U_\nu(\eta,k) U_\nu^*(\eta' \! ,k) - U_\nu^*(\eta,k) U_\nu(\eta'\!,k) \Bigr] \, ,
\end{equation}
so that the solution for~$w_0$ takes the form,
\begin{align}
w_0(\eta,k) ={}& 
	k^2 \! \int_{\eta_0}^{\eta} \! d\eta' \, G(\eta;\eta') \, a^2(\eta') U_\nu(\eta'\!,k)
	+ \mathscr{A}(\epsilon, H_0, k)  U_\nu(\eta,k) 
\nonumber  \\
&	\hspace{1cm}
	- \mathscr{B}(\epsilon, H_0, k)  U_\nu^*(\eta,k)
	=
	\mathscr{Q}(\eta,k)  U_\nu(\eta,k)
	-
	\widetilde{\mathscr{Q}}(\eta,k) U_\nu^*(\eta,k)
	\, ,
\label{w0 formal}
\end{align}
where the coefficient functions are,
\begin{align}
\mathscr{Q}(\eta,k) 
	={}&
	ik^2 \! \int_{\eta_0}^{\eta} \! d\eta' \, a^2(\eta') U_\nu^*(\eta' \! ,k) U_\nu(\eta',k) 
	+ 
	\check{\mathscr{A}}(\epsilon, H_0, k) 
	\, ,
\label{Q def}
\\
\widetilde{\mathscr{Q}}(\eta, k) ={}&
	ik^2 \! \int_{\eta_0}^{\eta} \! d\eta' \, a^2(\eta')  U_\nu(\eta' \! ,k) U_\nu(\eta',k) 
	+
	\check{\mathscr{B}}(\epsilon, H_0, k) 
	\, ,
\label{Qtilde def}
\end{align}
with~$\check{\mathscr{A}}$ and~$\check{\mathscr{B}}$ arbitrary 
constants of integration.
The solution for~$w_{\scr L}$ 
follows from acting the derivatives in~(\ref{wL equation}) onto the 
solution~(\ref{w0 formal}) for~$w_0$,
\begin{equation}
w_{\scr L}(\eta,k) = 
	\mathscr{Q}(\eta,k)   U_{\nu+1}(\eta,k) 
	+
	\widetilde{\mathscr{Q}}(\eta,k) U_{\nu+1}^*(\eta,k) 
	\, ,
\label{wL formal}
\end{equation}
using the fact that~$G(\eta;\eta)\!=\!0$. Therefore, the task of computing the particular
mode functions is reduced to evaluating the two integrals from~(\ref{Q def})
and~(\ref{Qtilde def}), and choosing the accompanying constants of integration. 
This is accomplished by making use of the integral 
1.8.3.1. from~\cite{Prudnikov2},
\begin{align}
\MoveEqLeft[1.5]
\int_0^z \! dz' \, z'^{\rho-1} J_\lambda(z') J_\mu(z')
	=
	\frac{ z^{\rho+\lambda+\mu} }{ 2^{\lambda+\mu} (\rho \!+\! \lambda \!+\! \mu ) \, 
		\Gamma(1\!+\!\lambda) \, \Gamma(1\!+\!\mu) }
\\
&
	\times \!
	{}_3F_4 \biggl( \! \Bigl\{ 
		\frac{ 1 \!+\! \lambda \!+\! \mu }{2} , 
		\frac{2\!+\!\lambda\!+\!\mu}{2} , 
		\frac{\rho\!+\!\lambda\!+\!\mu}{2} \Bigr\} , \!
		\Bigl\{ 1\!+\!\lambda , 1\!+\!\mu , 1 \!+\! \lambda\!+\!\mu , 
			\frac{2 \!+\! \rho\!+\!\lambda\!+\!\mu }{2} \Bigr\} , \! - z^2 \biggr) 
			\, ,
\nonumber 
\end{align}
that is valid for~${\rm Re}(\rho \!+\! \mu \!+\! \lambda) \!>\!0$.
In fact we  need only two special cases of that result,
\begin{align}
\int_0^z \! dz' \, z'^{\rho-1} J_\lambda(z') J_\lambda(z')
	={}&
	\frac{ z^{\rho+2\lambda} }
		{ 4^\lambda (\rho \!+\! 2\lambda ) \, \Gamma^2(1\!+\!\lambda) }
\nonumber \\
&
	\times 
	{}_2F_3 \biggl(  \Bigl\{ \frac{\rho}{2} \!+\!\lambda , \frac{ 1 }{2} \!+\! \lambda  \Bigr\} , 
		\Bigl\{ \frac{2 \!+\! \rho }{2} \!+\!\lambda , 1\!+\!\lambda , 1 \!+\! 2\lambda \Bigr\} , - z^2 \biggr) 
\nonumber \\
\equiv{}&
	\frac{ \mathscr{J}_1(\rho,\lambda; z) }
		{\Gamma(1\!+\!\lambda) \, \Gamma(1 \!-\! \lambda) \, 
			\Gamma(\lambda) \, \Gamma(-\lambda)}
	\, ,
\label{J1 def}
 \\
\int_0^z \! dz' \, z'^{\rho-1} J_\lambda(z') J_{-\lambda}(z')
	={}&
	\frac{ z^{\rho} }{ \rho \, 
		\Gamma(1\!+\!\lambda) \, \Gamma(1\!-\!\lambda) }
	{}_2F_3 \biggl(  \Bigl\{ \frac{\rho }{2} , \frac{ 1 }{2} \Bigr\} ,
		\Bigl\{ \frac{2 \!+\! \rho }{2}  , 1\!+\!\lambda , 1\!-\!\lambda  \Bigr\} , - z^2 \biggr) 
\nonumber \\
\equiv{}&
	\frac{ \mathscr{J}_2(\rho,\lambda; z) }{ \Gamma(1\!+\!\lambda) \, \Gamma(1\!-\!\lambda) \, \Gamma(\lambda) \, \Gamma(-\lambda)}
	\, .
\label{J2 def}
\end{align}
The integrals above are defined for a limited range of parameters only,
the one in~(\ref{J1 def}) for~${\rm Re}(\rho\!+\!2\lambda) \!>\!0$, and the one
in~(\ref{J2 def}) for~${\rm Re}(\rho) \!>\! 0$, because of zero being the lower limit 
of integration.
Nonetheless, we take the right-hand-sides to define the two 
functions~$\mathscr{J}_1$ and~$\mathscr{J}_2$
on an unlimited range of parameters. 
This detail is essentially immaterial for our purposes since the coefficient functions 
in~(\ref{Q def}) and~(\ref{Qtilde def}) come with two integration constants.
Using the results in~(\ref{J1 def}) and~(\ref{J2 def}), 
and the expression~(\ref{mode function}) for the mode function
we can evaluate the coefficient functions,
\begin{align}
\mathscr{Q}(\eta,k) ={}&
	\frac{ i }{ 4 \pi }
	\Bigl[ \frac{ k }{ (1\!-\!\epsilon) H_0 } \Bigr]^{ \frac{2}{1-\epsilon} }
	\biggl[ 
	- 2 \cos(\pi\nu) 
	\mathscr{J}_2\Bigl( \frac{ -2\epsilon }{ 1\!-\!\epsilon } , \nu; 
			\frac{k}{(1\!-\!\epsilon) \mathcal{H} } \Bigr)
\label{Q res 1}
\\
&	
	+
	\mathscr{J}_1\Bigl( \frac{ -2\epsilon }{ 1\!-\!\epsilon } , \nu; 
			\frac{k}{(1\!-\!\epsilon) \mathcal{H} } \Bigr)
	+ 
	\mathscr{J}_1\Bigl( \frac{ -2\epsilon }{ 1\!-\!\epsilon } , -\nu; 
			\frac{k}{(1\!-\!\epsilon) \mathcal{H} } \Bigr)
		\biggr]
	+ 
	\mathscr{A}(\epsilon,H_0,k)
	\, ,
\nonumber \\
\widetilde{\mathscr{Q}}(\eta,k) ={}&
	\frac{ e^{ \frac{- 2i k}{ (1-\epsilon) H_0 } }  }{4\pi}
	\Bigl[ \frac{k}{(1 \!-\! \epsilon)H_0} \Bigr]^{ \frac{2}{1-\epsilon} }
	\biggl[
	-
	2 \mathscr{J}_2 \Bigl( \frac{-2\epsilon}{1\!-\!\epsilon},\nu; \frac{k}{(1\!-\!\epsilon) \mathcal{H}} \Bigr) 
\label{Qtilde res 1}
\\
&	\hspace{-0.8cm}
	+
	e^{-  i\pi\nu} 
	\mathscr{J}_1 \Bigl( \frac{-2\epsilon}{1\!-\!\epsilon},\nu; \frac{k}{(1 \!-\! \epsilon) \mathcal{H}} \Bigr) 
	+
	e^{i\pi\nu } 
	\mathscr{J}_1 \Bigl( \frac{-2\epsilon}{1 \!-\! \epsilon},-\nu; \frac{k}{(1 \!-\! \epsilon) \mathcal{H}} \Bigr) 
	\biggr]
	+
	\mathscr{B}(\epsilon,H_0,k)
	\, ,
\nonumber  
\end{align}
up to the arbitrary  integration constants, in which we have absorbed time independent
contributions from the lower limit of integration in~(\ref{Q def})
and~(\ref{Qtilde def}), and relabeled them to~$\mathscr{A}$ and~$\mathscr{B}$.
The two solutions for the coefficient functions above define our particular mode functions
once the integration constants are chosen.
In the following we fix them by considering three conditions:~(i) the Wronskian-like relation~(\ref{Wronskian-like}),
(ii) the de Sitter limit~(\ref{v dS}) computed in~\cite{Glavan:2022dwb},
and (iii) the Minkowski limit~(\ref{v flat}).

\bigskip

\noindent {\bf Wronskian relation.}
The Wronskian-like relation~(\ref{Wronskian-like}) for the shifted 
particular mode functions~(\ref{v0 shifted}) and~(\ref{vL shifted}) reads,
\begin{equation}
{\rm Re} \Bigl( w_0 U_{\nu+1}^* + w_{\scr L} U_\nu^* \Bigr) = 0 \, .
\end{equation}
Given the Wronskian in~(\ref{Wronskian}) this condition 
requires~${\rm Re}\bigl[ \mathscr{Q}(\eta,k) \bigr]\!=\!0$, which consequently
implies~${\rm Re}\bigl[ \mathscr{A}(\epsilon,H_0,k) \bigr] \!=\!0$.

\bigskip

\noindent {\bf De Sitter limit.}
The de Sitter limit~(\ref{v dS}) we require implies that the shifted particular mode functions in~(\ref{w0 formal}) and~(\ref{wL formal}) in the limit~$\epsilon\!\to\!0$ reduce to,
\begin{subequations}
\begin{align}
w_0(\eta,k) \xrightarrow{\epsilon \to 0}{}&
	- \frac{ik}{2H_0}
	\biggl[
	\frac{ik}{ \nu_0 H_0} \frac{\partial U_{\nu_0} }{ \partial \nu_0 }
	+ U_{\nu_0}
	\biggr]
	\, ,
\\
w_{\scr L}(\eta,k) \xrightarrow{\epsilon \to 0}{}&
	-
	\frac{ik}{2H_0}
	\biggl[
	\frac{ik}{ \nu_0 H_0} \frac{\partial U_{\nu_0+1} }{ \partial \nu_0 }
	+ \frac{a}{\nu_0} U_{\nu_0}
	+ U_{\nu_0+1}
	\biggr]
	\, .
\end{align}
\label{w dS}%
\end{subequations}
Demonstrating that this limit is correctly reproduced, and determining what are the 
conditions this imposes on the constants of integration in~(\ref{w0 formal})
and~(\ref{wL formal}) is not straightforward. This is due to the parametric derivative
of the mode function appearing in~(\ref{w dS}), that has to be tied to the
special functions appearing in the coefficient functions~(\ref{Q res 1}) 
and~(\ref{Qtilde res 1}). We start by working out the de Sitter limit of these special
functions introduced in~(\ref{J1 def}) and~(\ref{J2 def}),
\begin{align}
\mathscr{J}_1\Bigl( \frac{ -2\epsilon }{ 1\!-\!\epsilon } , \pm \nu; 
	\frac{k}{(1\!-\!\epsilon) \mathcal{H} } \Bigr)
	\ \overset{\epsilon \to 0}{\longsim} \ {}&
	\mp \frac{ \Gamma^2( \mp \nu_0) }{ 2 \nu_0  }
		\Bigl( \frac{k}{2\mathcal{H} } \Bigr)^{\! \pm 2\nu_0} \times
\nonumber \\
&	\hspace{-4.7cm}
	\times 
	{}_2F_3 \biggl(  \Bigl\{ \pm\nu_0 , \frac{ 1 }{2} \!\pm\! \nu_0  \Bigr\} , 
		\Bigl\{ 1 \!\pm\!\nu_0 , 1\!\pm\!\nu_0 , 1 \!\pm\! 2\nu_0 \Bigr\} , 
			- \frac{k^2}{\mathcal{H}^2 } \biggr) 
	\equiv
	\mathscr{J}_1^0 \Bigl( \pm \nu_0 ; \frac{k}{\mathcal{H} } \Bigr)
	\, ,
\label{J1 dS}
\\
\mathscr{J}_2\Bigl( \frac{ -2\epsilon }{ 1\!-\!\epsilon } , \nu; 
	\frac{k}{(1\!-\!\epsilon) \mathcal{H} } \Bigr)
	\ \overset{\epsilon \to 0}{\longsim} \ {}&
	-
	\frac{(1\!-\!\epsilon) \Gamma(\nu ) \, \Gamma(-\nu ) } { 2\epsilon}
	+
	\Gamma(\nu_0 ) \, \Gamma(-\nu_0 ) \times
\nonumber \\
&	\hspace{-4cm}
	\times \!
	\biggl[
	\ln\Bigl( \frac{k}{\mathcal{H}} \Bigr)
	-
	\frac{ 1 }{ (1\!-\!\nu_0^2) }
	\Bigl( \frac{k}{ 2\mathcal{H}} \Bigr)^{\!2}
	{}_3F_4\biggl( \Bigl\{  1 , 1 , \frac{3}{2} \Bigr\} , \Bigl\{ 2 , 2 , 2\!+\!\nu_0 , 2\!-\!\nu_0 \Bigr\} , 
	-\frac{k^2}{\mathcal{H}^2} \biggr)
	\biggr]
\nonumber \\
&	\hspace{-4.3cm}
	\equiv
	-
	\frac{ (1\!-\!\epsilon) \, \Gamma(\nu) \, \Gamma(-\nu) }{2\epsilon}
	+
	\mathscr{J}_2^0 \Bigl( \nu_0 ; \frac{k}{\mathcal{H} } \Bigr)
	\, .
\label{J2 dS}
\end{align}
Thus, the coefficient functions~(\ref{Q res 1}) and~(\ref{Qtilde res 1}) 
in the de Sitter limit are,
\begingroup
\allowdisplaybreaks
\begin{align}
\mathscr{Q}(\eta,k) 
	\ \overset{\epsilon \to 0}{\longsim} \ {}&
	\frac{ i }{ 4 \pi }
	\Bigl( \frac{ k }{ H_0 } \Bigr)^{\!2}
	\biggl[ 
	\mathscr{J}_1^0 \Bigl( \nu_0; \frac{k}{ \mathcal{H} } \Bigr)
	+ 
	\mathscr{J}_1^0 \Bigl( -\nu_0 ; \frac{k}{ \mathcal{H} } \Bigr)
	- 
	2 \cos(\pi\nu_0)
	\mathscr{J}_2^0 \Bigl( \nu_0 ; \frac{k}{\mathcal{H} } \Bigr)
	\biggr]
\nonumber \\
&
	-
	\frac{ i (1\!-\!\epsilon) \cot(\pi\nu) }{ 4 \epsilon \nu }
	\Bigl[ \frac{ k }{ (1\!-\!\epsilon) H_0 } \Bigr]^{ \frac{2}{1-\epsilon} }
	+ 
	\mathscr{A} \bigl( \epsilon\!\to\!0,H_0,k \bigr)
	\, ,
\label{QdS} \\
\widetilde{\mathscr{Q}}(\eta,k)
	\ \overset{\epsilon \to 0}{\longsim} \ {}&
	\frac{ e^{ - \frac{ 2i k}{ H_0 } }  }{4\pi}
	\Bigl( \frac{k}{H_0} \Bigr)^{ \!2 }
	\biggl[
	e^{-  i\pi\nu_0 } 
	\mathscr{J}_1^0 \Bigl( \nu_0 ; \frac{k}{ \mathcal{H}} \Bigr) 
	+
	e^{i\pi\nu_0 } 
	\mathscr{J}_1^0 \Bigl( -\nu_0 ; \frac{k}{ \mathcal{H}} \Bigr) 
	-
	2 \mathscr{J}_2^0 \Bigl( \nu_0 ; \frac{k}{\mathcal{H} } \Bigr)
	\biggr]
\nonumber  \\
&	\hspace{0cm}
	-
	\frac{ e^{ \frac{ - 2i k}{ (1\!-\!\epsilon)H_0 } } (1\!-\!\epsilon) }{ 4 \epsilon \nu \sin(\pi\nu)} 
	\Bigl[ \frac{ k }{ (1\!-\!\epsilon) H_0 } \Bigr]^{ \frac{2}{1-\epsilon} }
	+
	\mathscr{B}\bigl( \epsilon\!\to\!0,H_0,k \bigr)
	\, .
\label{QtildedS}
\end{align}
\endgroup
Note that in the four expressions above~$\nu$ still depends on~$\epsilon$,
as opposed to~$\nu_0$ that does not; we could have expanded that dependence as well,
but it is more convenient to keep it implicit.
Recognizing the parametric derivatives of the mode functions in the two 
expressions above is accomplished with the
help of the result in Eq.~(2.2) from~\cite{Brychkov:2016} 
for the parametric derivative of the Bessel function of the first kind,
\begin{align}
&
\frac{\partial}{\partial \lambda} J_\lambda(z)
	=
	 J_{-\lambda}(z) 
	 \frac{ \Gamma(-\lambda) }{ 2 \, \Gamma(1\!+\!\lambda)}
	 \Bigl( \frac{z}{2} \Bigr)^{\! 2 \lambda}
		{}_2F_3\biggl( \Bigl\{ \lambda , \frac{1}{2} \!+\! \lambda \Bigr\} , \Bigl\{ 1 \!+\! \lambda , 1\!+\! \lambda , 1\!+\!2\lambda \Bigr\} , - z^2 \biggr)
\nonumber  \\
&	\hspace{0.1cm}
	-
	J_\lambda(z) \biggl[
		\frac{1}{2\lambda} 
		+ \psi(\lambda)
		- \ln\Bigl( \frac{z}{2} \Bigr)
		+ \frac{z^2}{4(1 \!-\! \lambda^2) } 
		{}_3F_4\biggl( \Bigl\{ 1,1,\frac{3}{2} \Bigr\} , \Bigl\{ 2,2, 2\!+\!\lambda , 2 \!-\! \lambda \Bigr\} , -z^2 \! \biggr)
		\biggr]
		.
\label{J parametric}
\end{align}
It is helpful to first express this result in terms of functions defined
in~(\ref{J1 dS}) and~(\ref{J2 dS}),
\begin{align}
\frac{\partial}{\partial\lambda} J_\lambda(z)
	={}&
	\frac{ - \mathscr{J}_1^0(\lambda;z) }{ \Gamma(\lambda) \, \Gamma(-\lambda) }
		J_{-\lambda}(z)
	-
	\biggl[
	\frac{1}{2\lambda} + \psi(\lambda) + \ln(2)
	- \frac{ \mathscr{J}_2^0(\lambda;z) }{ \Gamma(\lambda) \, \Gamma(-\lambda) }
	\biggr]
	J_\lambda(z)
	\, .
\end{align}
This then allows us to express the parametric derivative of the CTBD mode function 
in~(\ref{mode function}) in terms of the de Sitter limit of special functions
given in~(\ref{J1 dS}) and~(\ref{J2 dS}),
\begin{align}
\frac{\partial}{\partial \lambda} U_{\lambda}(\eta,k)
	={}&
	U_\lambda(\eta,k)
	\Biggl\{
	- \frac{i \cot(\pi\lambda) }{2}
		\Bigl[
		\psi(\lambda)
		+
		\psi(-\lambda)
		+
		2 \ln(2)
		\Bigr]
\nonumber \\
&	\hspace{-0.5cm}
	+
	\frac{i \lambda }{2\pi}  \biggl[
	\mathscr{J}_1^0 \Bigl( \lambda ; \frac{k}{\mathcal{H}} \Bigr)
	+
	\mathscr{J}_1^0 \Bigl( -\lambda ; \frac{k}{\mathcal{H}} \Bigr)
	-
	2 \cos(\pi\lambda) \mathscr{J}_2^0 \Bigl( \lambda ; \frac{k}{\mathcal{H}} \Bigr) \biggr]
	\Biggr\}
\nonumber \\
&	\hspace{-1.5cm}
	-
	U_\lambda^* (\eta,k)
	\, e^{ - \frac{2ik}{H_0} }
	\Biggl\{
	-
	\frac{ 1 }{ 2 \sin(\pi\lambda) } 
		\Bigl[ i \pi + \psi(\lambda) + \psi(-\lambda) + 2 \ln(2) \Bigr]
\nonumber \\
&	\hspace{-0.5cm}
	+
	\frac{\lambda }{ 2\pi }
	\biggl[ 
	e^{-i\pi\lambda} \mathscr{J}_1^0\Bigl( \lambda; \frac{k}{\mathcal{H}} \Bigr)
	+
	e^{i\pi\lambda} \mathscr{J}_1^0\Bigl( -\lambda; \frac{k}{\mathcal{H}} \Bigr) 
	-
	2 \mathscr{J}_2^0\Bigl( \lambda; \frac{k}{\mathcal{H}} \Bigr)
	\biggr]
	\Biggr\}
	\, ,
\end{align}
where we have used the reflection formula for the digamma 
function,~$\psi(1\!-\!\lambda) \!-\! \psi(\lambda) \!=\! \pi \cot(\pi\lambda)$.
Furthermore, using~(\ref{J parametric}) and the recurrence relation~(\ref{mode recurrence})
we derive the analogous expression for a contiguous mode function,
\begingroup
\allowdisplaybreaks
\begin{align}
\frac{\partial}{\partial \lambda} U_{\lambda+1}(\eta,k)
	={}&
	\frac{i\mathcal{H}}{k} U_\lambda(\eta,k)
	+
	U_{\lambda+1}(\eta,k)
	\Biggl\{
	-
	\frac{i \cot(\pi\lambda) }{2}
		\Bigl[
		\psi(\lambda)
		+
		\psi(-\lambda)
		+
		2 \ln(2)
		\Bigr]
\nonumber \\
&
	+
	\frac{i \lambda }{2\pi}  \biggl[
	\mathscr{J}_1^0 \Bigl( \lambda ; \frac{k}{\mathcal{H}} \Bigr)
	+
	\!
	\mathscr{J}_1^0 \Bigl( -\lambda ; \frac{k}{\mathcal{H}} \Bigr)
	-
	2 \cos(\pi\lambda) \mathscr{J}_2^0 \Bigl( \lambda ; \frac{k}{\mathcal{H}} \Bigr) \biggr]
	\Biggr\}
\nonumber \\
&	\hspace{-1.5cm}
	+
	U_{\lambda+1}^* (\eta,k)
	\,
	e^{ - \frac{2ik}{H_0} }
	\Biggl\{
	-
	\frac{ 1 }{ 2 \sin(\pi\lambda) } 
		\Bigl[ i \pi + \psi(\lambda) + \psi(-\lambda) + 2 \ln(2) \Bigr]
\nonumber \\
&	\hspace{-0.5cm}
	\frac{\lambda }{ 2\pi }
	\biggl[ 
	e^{-i\pi\lambda} \mathscr{J}_1^0\Bigl( \lambda; \frac{k}{\mathcal{H}} \Bigr)
	+
	e^{i\pi\lambda} \mathscr{J}_1^0\Bigl( -\lambda; \frac{k}{\mathcal{H}} \Bigr) 
	-
	2 \mathscr{J}_2^0\Bigl( \lambda; \frac{k}{\mathcal{H}} \Bigr)
	\biggr]
	\Biggr\}
	\, ,
\end{align}
\endgroup
where we made use of derivative properties,
\begin{subequations}
\begin{align}
\frac{\partial}{\partial z} \mathscr{J}_1^0(\lambda;z)
	={}&
	\Gamma(1\!+\!\lambda) \, \Gamma(1\!-\!\lambda) \, 
		\Gamma(\lambda) \, \Gamma(-\lambda)
		\frac{ J_\lambda(z) J_\lambda(z) }{z}
		\, ,
\\
\frac{\partial}{\partial z} \mathscr{J}_2^0(\lambda;z)
	={}&
	\Gamma(1\!+\!\lambda) \, \Gamma(1\!-\!\lambda) \, 
		\Gamma(\lambda) \, \Gamma(-\lambda)
		\frac{ J_\lambda(z) J_{-\lambda}(z) }{z}
		\, ,
\end{align}
\end{subequations}
that follow from the definitions~(\ref{J1 def}) and~(\ref{J2 def}),
and de Sitter limits~(\ref{J1 dS}) and~(\ref{J2 dS}).
Finally, this allows us to write the de Sitter limit of the shifted particular mode
functions~(\ref{w0 formal}) and~(\ref{wL formal}) as,
\begin{align}
w_{0} 
	\ \overset{\epsilon \to 0}{\longsim} \ {}&
	\frac{k^2}{2\nu_0 H_0^2 } \frac{\partial U_{\nu_0}}{ \partial \nu_0 }
	+
	\biggl\{
	\frac{i k^2 \cot(\pi\nu_0)}{4\nu_0 H_0^2} 
		\Bigl[ \psi(\nu_0) + \psi(-\nu_0) + 2 \ln(2) \Bigr] 
\nonumber \\
&	\hspace{1.5cm}
	-
	\frac{ i (1\!-\!\epsilon) \cot(\pi\nu) }{ 4 \epsilon \nu }
	\Bigl[ \frac{ k }{ (1\!-\!\epsilon) H_0 } \Bigr]^{ \frac{2}{1-\epsilon} }
	+
	\mathscr{A} \bigl( \epsilon\!\to\!0,H_0,k \bigr)
	\biggr\} U_{\nu_0}
\nonumber \\
&
	-
	\biggl\{
	\frac{ e^{- \frac{2ik}{H_0}} k^2 }{ 4 \nu_0 H_0^2 \sin(\pi\nu_0) }
		\Bigl[ i \pi + \psi(\nu_0) + \psi(-\nu_0) + 2 \ln(2) \Bigr] 
\nonumber \\
&	\hspace{1.5cm}
	-
	\frac{ e^{ \frac{ - 2i k}{ (1\!-\!\epsilon) H_0 } } (1\!-\!\epsilon) }{ 4 \epsilon \nu \sin(\pi\nu)} 
	\Bigl[ \frac{ k }{ (1\!-\!\epsilon) H_0 } \Bigr]^{ \frac{2}{1-\epsilon} }
	+
	\mathscr{B}\bigl( \epsilon\!\to\!0,H_0,k \bigr)
	\biggr\} U_{\nu_0}^*
	\, ,
\\
w_{\scr L} 
	\ \overset{\epsilon \to 0}{\longsim} \ {}&
	\frac{ - i k }{2\nu_0 H_0}
	\biggl[
	\frac{ ik }{ H_0 } \frac{ \partial U_{\nu_0+1} }{ \partial \nu_0 }
	+
	a U_{\nu_0}
	\biggr]
	+
	\biggl\{
	\frac{ i k^2 \cot(\pi\nu_0) }{ 4\nu_0 H_0^2 } 
		\Bigl[ \psi(\nu_0) + \psi(-\nu_0) + 2 \ln(2) \Bigr] 
\nonumber \\
&	\hspace{1.5cm}
	-
	\frac{ i (1\!-\!\epsilon) \cot(\pi\nu) }{ 4 \epsilon \nu }
	\Bigl[ \frac{ k }{ (1\!-\!\epsilon) H_0 } \Bigr]^{ \frac{2}{1-\epsilon} }
	+
	\mathscr{A} \bigl( \epsilon\!\to\!0,H_0,k \bigr)
	\biggr\}
	U_{\nu_0+1}
\nonumber  \\
&	\hspace{0cm}
	+
	\biggl\{
	\frac{ e^{ - \frac{2ik}{H_0} } k^2 }{ 4 \nu_0 H_0^2 \sin(\pi\nu_0) }
		\Bigl[ i\pi + \psi(\nu_0) + \psi(-\nu_0) + 2 \ln(2) \Bigr] 
\nonumber \\
&	\hspace{1.5cm}
	-
	\frac{ e^{ \frac{ - 2i k}{ (1\!-\!\epsilon)H_0 } } (1\!-\!\epsilon) }{ 4 \epsilon \nu \sin(\pi\nu)} 
	\Bigl[ \frac{ k }{ (1\!-\!\epsilon) H_0 } \Bigr]^{ \frac{2}{1-\epsilon} }
	+
	\mathscr{B}\bigl( \epsilon\!\to\!0,H_0,k \bigr)
	\biggr\}
	U_{\nu_0+1}^*
	\,. 
\end{align}
Requiring these expressions to match~(\ref{w dS}) finally gives the constants of integration
in the de Sitter limit,
\begingroup
\allowdisplaybreaks
\begin{align}
\mathscr{A} \ \overset{\epsilon \to 0}{\longsim}{}& \
	-
	\frac{ik}{2H_0}
	+
	\frac{ i (1\!-\!\epsilon) \cot(\pi\nu) }{ 4 \epsilon \nu }
	\Bigl[ \frac{ k }{ (1\!-\!\epsilon) H_0 } \Bigr]^{ \frac{2}{1-\epsilon} }
\nonumber \\
&	\hspace{1cm}
	-
	\frac{ i k^2 \cot(\pi\nu_0 ) }{ 4 \nu_0 H_0^2 }
	\Bigl[ \psi(\nu_0) + \psi(-\nu_0) + 2 \ln(2) \Bigr] 
	+
	\mathcal{O}(\epsilon)
	\, ,
\label{A dS requirement}
\\
\mathscr{B} \ \overset{\epsilon \to 0}{\longsim}{}& \
	\frac{ e^{ \frac{-2ik}{(1-\epsilon)H_0} } (1\!-\!\epsilon) }{ 4 \epsilon \nu \sin(\pi\nu) } 
	\Bigl[ \frac{k}{(1\!-\!\epsilon)H_0} \Bigr]^{ \frac{2}{1-\epsilon} }
\nonumber \\
&	\hspace{1cm}
	-
	\frac{ e^{- \frac{2ik}{H_0} } k^2 }{ 4 \nu_0 H_0^2 \sin(\pi\nu_0) }
	\Bigl[ i\pi + \psi(\nu_0) + \psi(-\nu_0) + 2 \ln(2) \Bigr]
	+
	\mathcal{O}(\epsilon)
	\, ,
\label{B dS requirement}
\end{align}
\endgroup
where~$\mathscr{A}$ is purely imaginary as required by the Wronskian condition.

\bigskip

\noindent {\bf Flat space limit.}
The flat space limit~(\ref{v flat}) implies that the shifted particular mode functions
in the limit~$H_0\!\to\!0$ have to reduce to
\begin{equation}
w_0 \xrightarrow{H_0\to0}
	\frac{1}{4} \Bigl[ 1 + 2 i k (\eta \!-\! \eta_0) \Bigr] 
	\frac{ e^{-ik(\eta - \eta_0)} }{ \sqrt{2k} }
	\, ,
\quad \ \
w_{\scr L} \xrightarrow{H_0\to0}
	\frac{1}{4} \Bigl[ -1 + 2 i k (\eta \!-\! \eta_0) \Bigr] 
	\frac{ e^{-ik(\eta - \eta_0)} }{ \sqrt{2k} }
	\, .
\label{w flat}
\end{equation}
To show that this limit is reproduced by~(\ref{w0 formal})
and~(\ref{wL formal}) we need the asymptotic series for the generalized 
hypergeometric functions in the limit of large argument that are given in §16.11 of~\cite{Olver:2010,Olver:web}.
In particular we need two special cases of that asymptotic series,
\begin{align}
\mathscr{J}_1(\rho,\pm\lambda,z) 
	\ \overset{z \to \infty}{\longsim}{}& \
	\Gamma(\lambda) \, \Gamma(-\lambda) \,
			\Gamma(1\!+\!\lambda) \, \Gamma(1\!-\!\lambda)
	\Biggl\{
	\frac{ \Gamma\bigl( \frac{1-\rho}{2} \bigr) \, \Gamma\bigl( \frac{\rho}{2} \!\pm\! \lambda \bigr) }
		{ 2 \, \Gamma\bigl( \frac{1}{2} \bigr) \, \Gamma\bigl( \frac{2-\rho}{2} \bigr) \, 
			\Gamma\bigl( \frac{2-\rho}{2} \!\pm\! \lambda \bigr) }
\nonumber \\
&
	-
	\frac{ z^{\rho-1} }{\pi}
	\biggl[
	\frac{ 1 }{ (1 \!-\! \rho)}
	+
	\frac{ \cos(2z\! \mp \!\pi\lambda) }{2z}
	+
	\mathcal{O}(z^{-2})
	\biggr]
	\Biggr\}
	\, ,
\\
\mathscr{J}_2(\rho,\lambda,z) 
	\ \overset{z \to \infty}{\longsim}{}& \
	\Gamma(\lambda) \, \Gamma(-\lambda) \,
			\Gamma(1\!+\!\lambda) \, \Gamma(1\!-\!\lambda)
	\Biggl\{
	\frac{ \Gamma\bigl( \frac{\rho}{2} \bigr) \, \Gamma\bigl( \frac{1-\rho}{2} \bigr)  }
		{ 2 \, \Gamma\bigl( \frac{1}{2} \bigr) \, \Gamma\bigl( \frac{2-\rho}{2} \!+\! \lambda \bigr) \, 
			\Gamma \bigl( \frac{2-\rho}{2} \!-\! \lambda \bigr) }
\nonumber \\
&
	-
	\frac{z^{\rho-1}}{\pi}
	\biggl[
	\frac{ \cos(\pi\lambda) }{ (1 \!-\! \rho) }
	+
	\frac{ \cos(2z) }{2 z }
	+
	\mathcal{O}(z^{-2})
	\biggr]
	\Biggr\}
	\, ,
\end{align}
which impy the following flat space limits for the coefficient
functions~(\ref{Q res 1}) and~(\ref{Qtilde res 1}),
\begin{align}
\mathscr{Q}(\eta,k)
	\ \overset{ H_0\to 0}{\longsim}{}& \
	\frac{1}{2} \Bigl[ 
		\frac{ik}{(1\!+\!\epsilon)H_0} 
		+ i k (\eta \!-\! \eta_0) \Bigr]
	\Bigl[ 1 + \mathcal{O}(H_0^{2}) \Bigr] 
	+
	\mathscr{A} \bigl( \epsilon, H_0\!\to\!0, k \bigr)
	\, ,
\\
&
	-
	\frac{ i \, \Gamma\bigl( \frac{-\epsilon}{1-\epsilon} \bigr) \, \Gamma\bigl( \frac{1}{2} \!+\! \frac{\epsilon}{1-\epsilon} \bigr)
		\, \Gamma\bigl( \frac{-\epsilon}{1-\epsilon} \!+\! \nu \bigr) \, \Gamma\bigl( \frac{-\epsilon}{1-\epsilon} \!-\! \nu \bigr) }
			{ 4 \, \Gamma\bigl( \frac{1}{2} \bigr) \, \Gamma\bigl( \frac{1}{2} \!+\! \nu \bigr) \, \Gamma\bigl( \frac{1}{2} \!-\! \nu \bigr) }
	\Bigl[ \frac{k}{(1\!-\!\epsilon)H_0} \Bigr]^{\frac{2}{1-\epsilon}}
	\, ,
\nonumber \\
\widetilde{\mathscr{Q}}(\eta,k)
	\ \overset{ H_0 \to 0}{\longsim}{}& \
	-
	\frac{ e^{ - 2ik(\eta-\eta_0)  }  }{4} 
	\Bigl[ 1 + \mathcal{O}(H_0) \Bigr]
	+ \mathscr{B} \bigl( \epsilon, H_0\!\to\!0, k \bigr)
\\
&	
	-
	e^{\frac{-2ik}{(1-\epsilon) H_0} } 
	e^{ \frac{ - i \pi \epsilon}{1-\epsilon} }
	\frac{ \Gamma\bigl( \frac{-\epsilon}{1-\epsilon} \bigr)  \,
		\Gamma\bigl( \frac{-\epsilon}{1-\epsilon} \!+\! \nu \bigr) \, \Gamma\bigl( \frac{-\epsilon}{1-\epsilon} \!-\! \nu \bigr) }
			{ 4 \, \Gamma\bigl( \frac{1}{2} \bigr)  \, \Gamma\bigl( \frac{1}{2} \!-\! \frac{ \epsilon}{ 1-\epsilon } \bigr) }
			\Bigl[ \frac{k}{(1\!-\!\epsilon)H_0} \Bigr]^{\frac{2}{1-\epsilon}}
	\, .
\nonumber 
\end{align}
Thus the flat space limit of the shifted particular mode functions is given by,
\begingroup
\allowdisplaybreaks
\begin{align}
w_0 \xrightarrow{H_0\to0}{}&
	\frac{ 1 }{4} 
	\Bigl[
	1
	+
	2 i k (\eta \!-\! \eta_0)
	\Bigr]
	\frac{ e^{ - i k (\eta-\eta_0) } }{ \sqrt{2k} }
	+
	\biggl\{
	\frac{ik}{2(1\!+\!\epsilon)H_0}
	+
	\mathscr{A} \bigl( \epsilon,H_0\!\to\!0,k \bigr)
\nonumber \\
&
	- 
	\frac{ i \, \Gamma\bigl( \frac{-\epsilon}{ 1 - \epsilon } \bigr) \, 
		\Gamma\bigl( \frac{1}{2} \!+\! \frac{\epsilon}{1-\epsilon} \bigr) \, 
		\Gamma\bigl( \frac{-\epsilon}{1-\epsilon} \!+\! \nu \bigr) \, 
		\Gamma\bigl( \frac{-\epsilon}{1-\epsilon} \!-\! \nu \bigr) }
			{ 4 \, \Gamma\bigl( \frac{1}{2} \bigr) 
			\, \Gamma\bigl( \frac{1}{2} \!+\! \nu \bigr)
			\, \Gamma\bigl( \frac{1}{2} \!-\! \nu \bigr) }
			\Bigl[ \frac{k}{(1\!-\!\epsilon)H_0} \Bigr]^{\frac{2}{1-\epsilon}}
	\biggr\}
	\frac{ e^{ - i k (\eta-\eta_0) } }{ \sqrt{2k} }
\nonumber \\
&	\hspace{-1cm}
	+
	\biggl\{
	e^{\frac{-2ik}{(1-\epsilon) H_0} } 
	e^{ \frac{ - i \pi \epsilon}{1-\epsilon} }
	\frac{ \Gamma\bigl( \frac{-\epsilon}{1-\epsilon} \bigr) \, 
		\Gamma\bigl( \frac{-\epsilon}{1-\epsilon} \!+\! \nu \bigr) \, 
		\Gamma\bigl( \frac{-\epsilon}{1-\epsilon} \!-\! \nu \bigr) }
			{ 4 \, \Gamma\bigl( \frac{1}{2} \bigr) \,
				\Gamma\bigl( \frac{1}{2} \!-\! \frac{\epsilon}{1-\epsilon} \bigr) }
			\Bigl[ \frac{k}{(1\!-\!\epsilon)H_0} \Bigr]^{\frac{2}{1-\epsilon}}
\nonumber \\
&
	- \mathscr{B}\bigl( \epsilon, H_0 \!\to\!0, k \bigr)
	\biggr\}
	\frac{ e^{ i k (\eta-\eta_0) } }{ \sqrt{2k} }
	\, ,
\\
w_{\scr L} \xrightarrow{H_0 \to 0}{}&
	\frac{ 1 }{4} \Bigl[ - 1 + 2 i k (\eta \!-\! \eta_0) \Bigr]
	\frac{ e^{ - i k (\eta-\eta_0) } }{ \sqrt{2k} }
	+
	\biggl\{
	\frac{ik}{2(1\!+\!\epsilon)H_0}
	+
	\mathscr{A} \bigl( \epsilon, H_0\!\to\!0, k \bigr)
	\, ,
\nonumber \\
&
	-
	\frac{ i \, \Gamma\bigl( \frac{-\epsilon}{1-\epsilon} \bigr) \, \Gamma\bigl( \frac{1}{2} \!+\! \frac{\epsilon}{1-\epsilon} \bigr)
		\, \Gamma\bigl( \frac{-\epsilon}{1-\epsilon} \!+\! \nu \bigr) \, \Gamma\bigl( \frac{-\epsilon}{1-\epsilon} \!-\! \nu \bigr) }
			{ 4 \, \Gamma\bigl( \frac{1}{2} \bigr) \, \Gamma\bigl( \frac{1}{2} \!+\! \nu \bigr) \, \Gamma\bigl( \frac{1}{2} \!-\! \nu \bigr) }
	\Bigl[ \frac{k}{(1\!-\!\epsilon)H_0} \Bigr]^{\frac{2}{1-\epsilon}}
	\Biggr\}
	\frac{ e^{ - i k (\eta-\eta_0) } }{ \sqrt{2k} }
\nonumber \\
&	\hspace{-1cm}
	-
	\biggl\{
	e^{\frac{-2ik}{(1-\epsilon) H_0} } 
	e^{ \frac{ - i \pi \epsilon}{1-\epsilon} }
	\frac{ \Gamma\bigl( \frac{-\epsilon}{1-\epsilon} \bigr) \,
		\Gamma\bigl( \frac{-\epsilon}{1-\epsilon} \!+\! \nu \bigr) \, \Gamma\bigl( \frac{-\epsilon}{1-\epsilon} \!-\! \nu \bigr) }
			{ 4 \, \Gamma\bigl( \frac{1}{2} \bigr) \, \Gamma\bigl( \frac{1}{2} \!-\! \frac{\epsilon}{1-\epsilon} \bigr) }
			\Bigl[ \frac{k}{(1\!-\!\epsilon)H_0} \Bigr]^{\frac{2}{1-\epsilon}}
\nonumber \\
&
	-
	\mathscr{B} \bigl( \epsilon, H_0\!\to\!0, k \bigr)
	\biggr\}
	\frac{ e^{ i k (\eta-\eta_0) } }{ \sqrt{2k} }
	\, .
\end{align}
\endgroup
In order for these expressions to match~(\ref{w flat})
the integration constants in the flat space limit must satisfy,
\begin{align}
\mathscr{A} \ \overset{H_0\to0}{\longsim}{}& \
	\frac{-ik}{2(1\!+\!\epsilon)H_0}
	+
	\frac{ i \, \Gamma\bigl( \frac{-\epsilon}{ 1 - \epsilon } \bigr) \, 
		\Gamma\bigl( \frac{1}{2} \!+\! \frac{\epsilon}{1-\epsilon} \bigr) \, 
		\Gamma\bigl( \frac{-\epsilon}{1-\epsilon} \!+\! \nu \bigr) \, 
		\Gamma\bigl( \frac{-\epsilon}{1-\epsilon} \!-\! \nu \bigr) }
			{ 4 \, \Gamma\bigl( \frac{1}{2} \bigr) 
			\, \Gamma\bigl( \frac{1}{2} \!+\! \nu \bigr)
			\, \Gamma\bigl( \frac{1}{2} \!-\! \nu \bigr) }
			\Bigl[ \frac{k}{(1\!-\!\epsilon)H_0} \Bigr]^{\frac{2}{1-\epsilon}}
			\, ,
\label{A flat}
\\
\mathscr{B} \ \overset{H_0\to0}{\longsim}{}& \
		e^{\frac{-2ik}{(1-\epsilon) H_0} } 
	e^{ \frac{ - i \pi \epsilon}{1-\epsilon} }
	\frac{ \Gamma\bigl( \frac{-\epsilon}{1-\epsilon} \bigr) \, 
		\Gamma\bigl( \frac{-\epsilon}{1-\epsilon} \!+\! \nu \bigr) \, 
		\Gamma\bigl( \frac{-\epsilon}{1-\epsilon} \!-\! \nu \bigr) }
			{ 4 \, \Gamma\bigl( \frac{1}{2} \bigr) \,
				\Gamma\bigl( \frac{1}{2} \!-\! \frac{\epsilon}{1-\epsilon} \bigr) }
			\Bigl[ \frac{k}{(1\!-\!\epsilon)H_0} \Bigr]^{\frac{2}{1-\epsilon}}
			\, .
\label{B flat}
\end{align}
These two conditions, together with the two conditions~(\ref{A dS requirement})
and~(\ref{B dS requirement}) found by considering the de Sitter limit, constrain
the choice for the constants of integration that we can make.
In fact, it is most convenient to promote the flat space limit conditions~(\ref{A flat})
and~(\ref{B flat}) to the full choice for the constants. It is straightforward to
check that this choice satisfies the de Sitter limit requirements automatically.

\section{Inverting Laplacian}
\label{app: Inverting Laplacian}

The Laplace-inverted two-point function introduced
in~(\ref{Xi def}) can also be evaluated in a closed form, rather than in the 
series form~(\ref{Xi solution}) given in 
Sec.~\ref{subsec: Laplace-inverted two-point function}.
This is accomplished by first reverting to the 
sum-over-modes representation,
\begin{equation}
i \bigl[ \tensor*[^{\scr \! -}]{\Xi}{^{\scr + \!} } \bigr]_\lambda (x;x') 
	=
	- (aa')^{ \frac{2-D}{2}} \!\!
	\int\! \frac{ d^{D-1}k }{ (2\pi)^{D-1} } \,
	\frac{ (1\!-\!\epsilon)^2 \mathcal{H} \mathcal{H}' }{k^2}
	e^{i \vec{k} \cdot \Delta \vec{x} } \,
	\theta(k\!-\!k_0)
	\,
	U_\lambda(\eta,k) \, U_{\lambda}^*(\eta',k) \, ,
\label{Laplace int}
\end{equation}
where the step function effectively enforces the mode function suppression in 
the IR by introducing a cutoff~$k_0 \!\ll\! H_0$.
The UV convergence for all ranges of coordinates~$x$ and~$x'$ is ensured by the 
same~$i\delta$-prescription as for the scalar two-point function 
in~(\ref{int over modes}). In this appendix we invert the Laplacian acting on
the Wightman function, with the particular~$i\delta$-prescription implicit in all 
expressions,
\begin{equation}
\mathcal{H} \to \mathcal{H}_{\scr -+} = 
	\mathcal{H} \Bigl[ 1 \!-\! \frac{(1\!-\!\epsilon) \mathcal{H} i \delta}{2} \Bigr]
	\, ,
\qquad
\mathcal{H}' \to \mathcal{H}'_{\scr -+} = 
	\mathcal{H}' \Bigl[ 1 \!+\! \frac{(1\!-\!\epsilon) \mathcal{H}' i \delta}{2} \Bigr]
	\, . 
\label{H idelta}
\end{equation}
Generalization to other two-point functions is straightforward.
We should also note  it is more convenient to
use~$\mathcal{H}$,~$\mathcal{H}'$, and~$\| \Delta\vec{x} \|$ as variables for the two-point
functions that in this appendix, 
instead of the bi-local variables~(\ref{bilocal variables}) employed in the main text.

The integral in~(\ref{Laplace int}) can be evaluated exaclty.
Integrating over the angular coordinates
involves an integral over a~$(D\!-\!2)$-sphere and an integral
representation of the Bessel function of the first 
kind, given in~(\ref{some integrals}), and it produces,
\begin{equation}
i \bigl[ \tensor*[^{\scr \! -}]{\Xi}{^{\scr + \!} } \bigr]_\lambda (x;x') 
	=
	- \frac{ (aa')^{-\frac{D-2}{2}} (1\!-\!\epsilon)^2 \mathcal{H} \mathcal{H}'}
		{ (2\pi)^{\frac{D-1}{2}} \| \Delta\vec{x} \|^{ \frac{D-3}{2} }  } \!
\int_{k_0}^{\infty} \!\! dk \, k^{ \frac{D-5}{2} }
	J_{ \frac{D-3}{2} } \bigl( k \| \Delta\vec{x} \| \bigr)
	U_\lambda(\eta,k) U_\lambda^*(\eta',k)
	\, .
\label{Laplacian int 1D}
\end{equation}
We evaluate this resulting integral analogously to evaluating the integral for the
scalar two-point functions in Sec.~\ref{subsubsec: IR divergent scalar two-point functions}.
This entails first writing out the integration over modes as the bulk (infinite)
range and subtracting from it the infrared range,~$\int_{k_0}^{\infty} \!=\! \int_{0}^{\infty} - \int_{0}^{k_0} $.
Each part is then computed
individually by dimensionally regulating their respective infrared behaviour.
This procedure would be incorrect to apply on actually infrared divergent integrals.
However, the integral in~(\ref{Laplacian int 1D}) is not infrared divergent
to start with, and the errors we are making by dimensionally
regulating the infrared of the two individual parts cancel each other out.
Therefore, it is legitimate to split the integral in~(\ref{Laplacian int 1D})
into two contributions (cf. Eq.~(\ref{IR breaking scalar 2pt})),
\begin{equation}
i \bigl[ \tensor*[^{\scr \! -}]{\Xi}{^{\scr + \!} } \bigr]_\lambda (x;x') 
	= (aa')^{- \frac{(D-2)\epsilon }{2} }
	\Bigl[
	\mathcal{M}_\lambda(x;x')
	+
	\mathcal{X}_{\lambda}(x;x')
	\Bigr]
	\, .
\label{Xi split}
\end{equation}
that we compute separately in the following subsections.

\subsection{Bulk part}
\label{subapp: Bulk part}

The bulk part of~(\ref{Xi split}) is given by the integral representation,
\begin{align}
\mathcal{M}_\lambda (x;x')
	={}&
	-
	\frac{ (aa')^{-\frac{(D-2)(1-\epsilon)}{2}} \, (1\!-\!\epsilon)\sqrt{ \mathcal{H}\mathcal{H}' } }
		{ 8 (2\pi)^{ \frac{D-3}{2} } \, \| \Delta\vec{x}\|^{\frac{D-3}{2} } }
\label{intermediate inverse}
 \\
&	\times
	\int_{0}^{\infty} \! dk \, k^{ \frac{D-5}{2} } \,
	J_{ \frac{D-3}{2} } \bigl( k \| \Delta\vec{x} \| \bigr)
	H_\lambda^{\scr (1)} \biggl( \frac{k}{(1\!-\!\epsilon) \mathcal{H} } \biggr)
	H_\lambda^{\scr (2)} \biggl( \frac{k}{(1\!-\!\epsilon) \mathcal{H}' } \biggr)
	\, .
\nonumber
\end{align}
The convergence for all ranges of coordinates is guaranteed by the 
implicit~$i \delta$-prescription~(\ref{H idelta}).
We first evaluate the integral
for the range of coordinates where the~$i \delta$-prescriptions are
not necessary, and then analytically extended to the remaining ranges of coordinates.
The result we use is an integral over three Bessel functions and a power 
that is given in 6.578.1.~of~\cite{Gradshteyn:2007},
or in 2.12.42.5.~of~\cite{Prudnikov2}, and is originally due to
Rice~\cite{Rice:1935} and Bailey~\cite{Bailey:1936},
\begin{align}
\MoveEqLeft[9]
\int_{0}^{\infty} \! dz \, z^{\rho-1} J_{\sigma}(cz) J_{\lambda}(az) J_\mu(bz)
	= 
	\frac{ 2^{\rho-1} \, \Gamma\bigl( \frac{ \lambda + \mu + \rho + \sigma }{2} \bigr) \, a^\lambda b^\mu c^{-\lambda-\mu-\rho}  }
		{ \Gamma(1\!+\!\lambda) \, \Gamma(1\!+\!\mu) \, \Gamma\bigl( 1 \!-\! \frac{\lambda+\mu + \rho-\sigma}{2} \bigr) }
\nonumber \\
&
	\times
	F_4 \biggl( \frac{ \lambda \!+\! \mu \!+\! \rho\!+\!\sigma }{2} ,  
		\frac{ \lambda \!+\! \mu \!+\! \rho \!-\! \sigma }{2} ;
		1 \!+\! \lambda , 1 \!+\! \mu ; \frac{a^2}{c^2} , \frac{b^2}{c^2} \biggr)
	\, .
\label{F4 integral}
\end{align}
It holds when the following conditions are met,
\begin{equation}
{\rm Re} \bigl( \lambda \!+\! \mu \!+\! \rho \!+\! \sigma \bigr) > 0 \, ,
\qquad
{\rm Re}(\rho) < \frac{5}{2} \, ,
\qquad
a,b,c > 0 \, ,
\qquad
c > a\!+\! b \, .
\label{F4 ranges}
\end{equation}
The  function~$F_4$ in~(\ref{F4 integral}) is the 
Appell's fourth function, defined 
inside the region~$ \bigl| \sqrt{X} \bigr| \!+\! \bigl| \sqrt{Y} \bigr| \!<\! 1 $
by the convergent double power series,
\begin{equation}
F_4 \bigl( \alpha, \beta ; \gamma, \gamma' ; X, Y \bigr)
	= 
	\frac{ \Gamma( \gamma ) \, \Gamma(\gamma' ) }{ \Gamma(\alpha) \, \Gamma(\beta) } 
	\sum_{m=0}^{\infty} \sum_{n=0}^{\infty}
	\frac{ \Gamma(\alpha \!+\! m \!+\! n) \, \Gamma(\beta \!+\! m \!+\! n) }{ \Gamma( \gamma \!+\! m) \, \Gamma(\gamma' \!+\! n) }
	\frac{X^m}{m!} \frac{Y^n}{n!}
	\, ,
\label{F4 def}
\end{equation}
and outside of it by its analytical extension. 
This definition implies symmetry properties,
\begin{align}
&
F_4\bigl( \alpha, \beta ; \gamma, \gamma' ; X, Y \bigr) 
	= 
	F_4\bigl( \beta , \alpha ; \gamma, \gamma' ; X , Y \bigr) 
	=
	F_4\bigl( \alpha, \beta ; \gamma', \gamma ; Y, X \bigr)
	\, ,
\label{F4 symmetries}
\end{align}
and also that,
\begin{equation}
F_4 \bigl( 0 , \beta ; \gamma, \gamma' ; X, Y \bigr) \!=\! 1 \, .
\end{equation}
Performing either of the two sums in the definition~(\ref{F4 def}) expresses
it as a single series over hypergeometric functions,
\begin{equation}
F_4 \bigl( \alpha, \beta ; \gamma, \gamma' ; X, Y \bigr)
	\!=\! 
	\frac{ \Gamma( \gamma' ) }{ \Gamma(\alpha) \, \Gamma(\beta) } 
	\! \sum_{n=0}^{\infty} \!
	\frac{ \Gamma( \alpha \!+\! n ) \, \Gamma( \beta \!+\! n ) }
		{ \Gamma( \gamma' \!+\! n) \, n! }
		Y^n
		{}_2F_1 \Bigl( \! \bigl\{ \alpha \!+\! n , \beta \!+\! n \bigr\} , 
		\bigl\{ \gamma \bigr\} , X \Bigr)
	\, ,
\label{F4 hyper series}
\end{equation}
which is particularly useful for defining the analytic continuation to the entire
range of complex coordinates~$X$ and~$Y$. In particular, it makes manifest
the two branch points at~$X\!=\!1$ and~$Y\!=\!1$ that the Appell's fourth 
function possesses. 
We need only two special cases of the result in~(\ref{F4 integral}),
\begin{align}
&
\int_{0}^{\infty} \! dz \, z^{\rho-1} J_{\rho}(cz) J_\lambda(az) J_\lambda(bz) 
\nonumber \\
&	\hspace{2cm}
	=
	\frac{ 2^{\rho-1} \Gamma(\rho\!+\!\lambda) \, c^{-\rho} }
		{ \Gamma^2(1\!+\!\lambda) \, \Gamma(1\!-\!\lambda) }
		\Bigl( \frac{ab}{c^2} \Bigr)^{\!\lambda}
		F_4\biggl( \rho \!+\! \lambda , \lambda ; 1 \!+\! \lambda , 1 \!+\! \lambda ;
			\frac{a^2}{c^2} , \frac{b^2}{c^2} \biggr)
	\, ,
\\
&
\int_{0}^{\infty} \! dz \, z^{\rho-1} J_{\rho}(cz) J_\lambda(az) J_{-\lambda}(bz) 
	=
	\frac{2^{\rho-1} \Gamma(\rho) \, c^{-\rho} }
		{ \Gamma(1\!+\!\lambda) \, \Gamma(1\!-\!\lambda) }
		\Bigl( \frac{a}{b} \Bigr)^{\!\lambda}
	\, ,
\end{align}
so that~(\ref{intermediate inverse}) evaluates to,
\begin{align}
\mathcal{M}_\lambda(x;x')
	={}&
	- \frac{ \bigl[ (1\!-\!\epsilon) H_0 \bigr]^{ D-2 } }
		{ 16 \pi^{ \frac{D+1}{2} } }
	\frac{ \Gamma \bigl( \frac{D - 3}{2} \bigr) \, \Gamma(\lambda) \, \Gamma(-\lambda) }
		{ \bigl[ (1\!-\!\epsilon)^2 \mathcal{H} \mathcal{H}' \| \Delta\vec{x}\|^2 
			\bigr]^{ \frac{D-3}{2} } }
	\biggl[
	e^{i\pi\lambda}
		\Bigl( \frac{\mathcal{H} }{ \mathcal{H}' } \Bigr)^{\!\lambda}
	+
	e^{-i\pi\lambda}
		\Bigl( \frac{\mathcal{H} }{ \mathcal{H}' } \Bigr)^{\!-\lambda}
	\biggr]
\nonumber \\
&	\hspace{-1.7cm}
	+
	\mathscr{F}_\lambda \Bigl( 
		(1\!-\!\epsilon) \mathcal{H} \| \Delta\vec{x} \| , 
		(1\!-\!\epsilon) \mathcal{H}' \| \Delta\vec{x} \|
		\Bigr)
	+
	\mathscr{F}_{-\lambda}\Bigl( 
		(1\!-\!\epsilon) \mathcal{H} \| \Delta\vec{x} \| , 
		(1\!-\!\epsilon) \mathcal{H}' \| \Delta\vec{x} \|
		\Bigr)
	\, ,
\label{M solution 1}
\end{align}
where we have defined a shorthand notation for the two-variable function,
\begin{align}
\mathscr{F}_\lambda(X,Y)
	={}&
	\frac{ \bigl[ (1\!-\!\epsilon) H_0 \bigr]^{ D-2 } }
		{ 16 \pi^{ \frac{D+1}{2} } }
	\frac{ \Gamma(\frac{D-3}{2}\!+\!\lambda) \, \Gamma(-\lambda) }{ \lambda }
	( X Y )^{ - \frac{D-3}{2} - \lambda } \,
\nonumber \\
&	\hspace{2.5cm}
	\times
		F_4\biggl( \frac{D\!-\!3}{2} \!+\! \lambda , \lambda ; 1 \!+\! \lambda , 1 \!+\! \lambda ;
			\frac{1}{X^2}, \frac{1}{Y^2} \biggr)
			\, .
\label{curly F}
\end{align}
This result is obtained for the range of coordinates on which the result~(\ref{F4 integral})
is valid, which according to~(\ref{F4 ranges}) translates into the condition~$y(x;x')\!>\!4$
on the bi-local variable. The result is then extended to
the full range of coordinates by analytic continuation provided by
the implicit~$i\delta$-prescriptions in time coordinates/conformal Hubble 
rates~(\ref{H idelta}). 
It is interesting to note that the bulk solution~(\ref{M solution 1}), when written in
terms of bi-local variables~(\ref{bilocal variables}), depends only on~$y$ and~$v$, while the 
sole dependence on~$u$ is in the overall factor taken out in~(\ref{Xi split}).

The representation~(\ref{M solution 1}) for the integral defined 
in~(\ref{intermediate inverse}) is correct on the entire range of coordinates.
However, in practice this representation is best adapted to 
the super-Hubble regime for which the arguments of Appell's fourth function take
small values. We can derive an alternative representation adapted to the sub-Hubble regime,
by making use of the identity 16.16.10.~from~\cite{Olver:2010,Olver:web},
\begin{align}
\MoveEqLeft[5]
F_4 \bigl( \alpha, \beta ; \gamma, \gamma'; X, Y \bigr)
	= 
	\frac{ \Gamma(\gamma') \, \Gamma(\beta \!-\! \alpha) }{ \Gamma( \gamma' \!-\! \alpha) \, \Gamma(\beta) } \,
		(-Y)^{-\alpha} F_4\biggl( \alpha , \alpha \!-\! \gamma' \!+\! 1 ; \gamma , \alpha \!-\! \beta \!+\! 1 ; \frac{X}{Y} , \frac{1}{Y} \biggr)
\nonumber \\
&
	+
	\frac{ \Gamma(\gamma') \, \Gamma(\alpha\!-\!\beta) }{ \Gamma( \gamma' \!-\! \beta) \, \Gamma(\alpha) } \,
		(-Y)^{-\beta} F_4\biggl( \beta , \beta \!-\! \gamma' \!+\! 1 ; \gamma , \beta \!-\! \alpha \!+\! 1 ; \frac{X}{Y} , \frac{1}{Y} \biggr)
	\, ,
\label{trans1}
\end{align}
that is valid for~$ \bigl| {\rm Arg}(-Y) \bigr| \!<\! \pi$. This turns the 
solution~(\ref{M solution 1}) into,
\begin{align}
\MoveEqLeft[2]
\mathcal{M}_\lambda(x;x') 
	=
	e^{ \frac{i\pi (D-1)}{2} }
	\biggl[
	e^{ i\pi\lambda } \,
	\overline{ \mathscr{F} }_\lambda 
		\Bigl( \frac{\mathcal{H} }{\mathcal{H}' } , 
			(1\!-\!\epsilon) \mathcal{H} \| \Delta\vec{x} \| \Bigr)
	+
	e^{ -i\pi\lambda } \,
	\overline{ \mathscr{F}}_{-\lambda }
		\Bigl( \frac{\mathcal{H} }{\mathcal{H}' } , 
			(1\!-\!\epsilon) \mathcal{H} \| \Delta\vec{x} \| \Bigr)
	\biggr]
\nonumber \\
&
	+
	e^{ - \frac{i\pi( D-1 )}{2} }
	\biggl[
	e^{ - i\pi\lambda }
	\,
	\overline{ \mathscr{F} }_\lambda 
		\Bigl( \frac{\mathcal{H}'}{\mathcal{H}} , 
			(1\!-\!\epsilon) \mathcal{H}' \| \Delta\vec{x} \| \Bigr)
	+
	e^{ i\pi\lambda }
	\,
	\overline{ \mathscr{F} }_{-\lambda }
		\Bigl( \frac{\mathcal{H}'}{\mathcal{H}} , 
			(1\!-\!\epsilon) \mathcal{H}' \| \Delta\vec{x} \| \Bigr)
	\biggr]
	\, ,
\label{M solution 2}
\end{align}
where we define another shorthand notation for a different two-variable function,
\begin{align}
\overline{\mathscr{F}}_\lambda(X,Y)
	={}&
	\frac{ \bigl[ (1\!-\!\epsilon) H_0 \bigr]^{ D-2 } }
		{ 16 \pi^{ \frac{D+1}{2} } }
	\frac{ \Gamma\bigl( \frac{D-3}{2} \!+\! \lambda \bigr) \, \Gamma(-\lambda) }
		{ (D\!-\!3) }  \,
		X^{ \frac{D-3}{2} + \lambda }
\nonumber \\
&
	\hspace{1.5cm}
	\times
	F_4 \biggl(
		\frac{D\!-\!3}{2} \!+\! \lambda , \frac{D\!-\!3}{2} ;
		1 \!+\! \lambda , \frac{ D \!-\! 1 }{2} ;
		X^2, Y^2
		\biggr)
		\, .
\label{Fbar}
\end{align}
Note that we had applied the transformation~(\ref{trans1}) 
to~(\ref{M solution 1}) in a fashion that preserves manifest
invariance under simultaneous complex conjugation and 
interchange~$x \! \leftrightarrow \! x'$ that the two-point function
must respect.

\medskip

It is also possible represent the solution~(\ref{M solution 1}) in 
terms of Appell's first function instead of the fourth one. 
This is accomplished by first applying the argument transformation formula (3.1) 
from~\cite{Brychkov:2017},
\begin{equation}
F_4\bigl( \alpha, \beta ; \gamma, \gamma' ; X, Y \bigr) 
	\!=\!
	F_4\bigl( \alpha, \beta ; \gamma \!-\! 1, \gamma' ; X, Y \bigr) 
	-
	\frac{\alpha\beta X}{ \gamma (\gamma \!-\! 1)}
	F_4\bigl( \alpha+1, \beta+1 ; \gamma+1, \gamma' ; X, Y \bigr) 
	\, ,
\label{F4 contiguous trans}
\end{equation}
followed by the transformation formula~(4.1) 
from~\cite{Bailey:1936} between different Appell's functions, 
\begin{align}
\MoveEqLeft[5]
F_4\biggl( \alpha, \beta ; \gamma, \beta ; \frac{-X}{(1\!-\!X) (1\!-\!Y)}, \frac{-Y}{(1\!-\!X) (1\!-\!Y)} 	\biggr) 
\nonumber \\
&	
	=
	(1\!-\!X)^\alpha (1\!-\!Y)^\alpha
	F_1\bigl( \alpha ; \gamma\!-\!\beta, 1\!+\! \alpha \!-\! \gamma ; \gamma; X, XY \bigr)
	\, .
\label{F4-F1}
\end{align}
While more is known about Appell's first function than the fourth one, 
we do not pursue the representation resulting from~(\ref{F4 contiguous trans}) 
and~(\ref{F4-F1}) above, as we did not find immediate advantages in employing it
compared to the two representations in~(\ref{M solution 1}) and~(\ref{M solution 2}). 
Further representations and transformations of Appell's fourth function
can be derived from transformations of the hypergeometric function~\cite{Brychkov:2017} 
that might be useful when examining different ranges of coordinates.

\subsection{Infrared part}
\label{subapp: Infrared part}

The infrared part of the two-point function in~(\ref{Xi split}) is given by,
\begin{equation}
\mathcal{X}_\lambda(x;x')
	=
	\frac{ (aa')^{-\frac{(D-2)(1-\epsilon)}{2}} (1\!-\!\epsilon)^2 \mathcal{H} \mathcal{H}'}
		{ (2\pi)^{\frac{D-1}{2}} \| \Delta\vec{x} \|^{ \frac{D-3}{2} }  } \!
\int_{0}^{k_0} \!\! dk \, k^{ \frac{D-5}{2} }
	J_{ \frac{D-3}{2} } \bigl( k \| \Delta\vec{x} \| \bigr)
	U_\lambda(\eta,k) U_\lambda^*(\eta',k)
	\, .
\end{equation}
We evaluate this integral by following the procedure of 
Sec.~\ref{subsubsec: IR divergent scalar two-point functions}.
This entails first expanding the CTBD mode functions in the low 
momentum limit~(\ref{more IR}),
keeping only the terms that can generate negative powers of
the IR cutoff~$k_0$,
\begin{align}
\mathcal{X}_\lambda(x;x')
={}&
	\frac{ (aa')^{-\frac{(D-2)(1-\epsilon)}{2}} \,
		(1\!-\!\epsilon) \sqrt{\mathcal{H} \mathcal{H}' }
		\, \Gamma^2(\lambda) \, \Gamma^2(1\!-\!\lambda)  }
		{  2^{1-2\lambda} (2\pi)^{\frac{D+1}{2}} \| \Delta\vec{x} \|^{ \frac{D-3}{2} }  } 
\nonumber \\
&
	\times
	\sum_{N=0}^{\infty} \sum_{n=0}^{N} 
	\frac{ \bigl( -\frac{1}{4} \bigr)^{N} }{ n! ( N \!-\! n )! \, \Gamma(n \!+\! 1 \!-\! \lambda) \, \Gamma(N \!-\! n \!+\! 1 \!-\! \lambda) } 
	\Bigl( \frac{ \mathcal{H} }{\mathcal{H}' } \Bigr)^{\!N-2n }
\nonumber \\
&	\hspace{1cm}
	\times
	\int_{0}^{k_0}\!\! dk \, k^{ \frac{D-5}{2} }
	J_{ \frac{D-3}{2} } \bigl( k \| \Delta\vec{x} \| \bigr)
	\Bigl[ \frac{k^2}{ (1\!-\!\epsilon)^2 \mathcal{H} \mathcal{H}' } \Bigr]^{N -\lambda}
	\, .
\end{align}
Evaluating the resulting integrals according to~(\ref{IR int ref}) then gives,
\begingroup
\allowdisplaybreaks
\begin{align}
\MoveEqLeft[2]
\mathcal{X}_\lambda(x;x')
=
	\frac{ \bigl[ (1 \!-\! \epsilon) H_0 \bigr]^{ D-2 }  }{  (4\pi)^{\frac{D}{2}} } 
	\frac{ \Gamma(\lambda) \Gamma(2\lambda)  }
		{  \Gamma\bigl( \frac{1}{2} \!+\! \lambda \bigr)\, \Gamma\bigr( \frac{D-1}{2} \bigr) } 
	\sum_{N=0}^{ \infty} \sum_{n=0}^{N} 
	\frac{ 1 }{ ( \frac{D-3}{2} \!+\! N \!-\! \lambda  )  }
\nonumber \\
&
	\times
	\frac{ \bigl( -\frac{1}{4} \bigr)^{N} \, \Gamma^2(1\!-\!\lambda) }
		{ n! ( N \!-\! n )! \, \Gamma(n \!+\! 1 \!-\! \lambda) \, \Gamma(N \!-\! n \!+\! 1 \!-\! \lambda) } 
	\Bigl[ \frac{ k_0^2 }{ (1\!-\!\epsilon)^2 \mathcal{H} \mathcal{H}' } 
		\Bigr]^{\frac{D-3}{2} +N -\lambda}
	\Bigl( \frac{ \mathcal{H} }{\mathcal{H}' } \Bigr)^{\! N-2n }
\nonumber \\
&	\hspace{1cm}
	\times
	{}_1F_2\biggl( \Bigl\{ \frac{ D \!-\! 3 }{2} \!+\! N \!-\! \lambda \Bigr\} , 
		\Bigl\{ \frac{ D \!-\! 1}{2} \!+\! N \!-\! \lambda , \frac{D\!-\!1}{2} \Bigr\} , 
		- \frac{ (k_0 \| \Delta\vec{x} \| )^2}{4} \biggr)
		\, ,
\end{align}
\endgroup
where we have used the Legendre duplication formula to simplify the gamma functions
in the overall factor.
Expanding the hypergeometric function for small 
arguments~$k_0 \| \Delta\vec{x} \| \!\ll\! 1$, and reorganizing the series 
produces the final expression,
\begin{align}
\MoveEqLeft[1]
\mathcal{X}_\lambda(x;x')
=
	\frac{ \bigl[ (1 \!-\! \epsilon) H_0 \bigr]^{ D-2 }  }{  (4\pi)^{\frac{D}{2}} } 
	\frac{ \Gamma(\lambda) \Gamma(2\lambda)  }
		{  \Gamma\bigl( \frac{1}{2} \!+\! \lambda \bigr)\, \Gamma\bigr( \frac{D-1}{2} \bigr) } 
	\Bigl[ \frac{ k_0^2 }{ (1\!-\!\epsilon)^2 \mathcal{H} \mathcal{H}' } 
		\Bigr]^{\frac{D-3}{2} -\lambda}
\label{X sol}
\\
&
	\times \!\!\!
	\sum_{N=0}^{ \lfloor \lambda - \frac{D-3}{2} \rfloor  } \!\!
	\sum_{n=0}^{N}
	\sum_{\ell=0}^{N-n}
	\frac{ c_{Nn\ell} }{ \bigl( \frac{D-3}{2} \!+\! N  \!-\! \lambda \bigr) }
	\Bigl[ \frac{ k_0^2 }{ (1\!-\!\epsilon)^2 \mathcal{H} \mathcal{H}' } 
		\Bigr]^{N}
	\Bigl[ (1\!-\!\epsilon)^2 \mathcal{H} \mathcal{H}'\| \Delta\vec{x} \|^2 \Bigr]^{n}
	\Bigl( \frac{ \mathcal{H} }{\mathcal{H}' } \Bigr)^{\! N - n -2\ell }
	\, ,
\nonumber 
\end{align}
where the coefficients~$c_{Nn\ell}$ are given in~(\ref{c coeff def}),
and
where the floor function truncates the external sum at~$N \!\le\! (D\!-\!3)/2$,
thus keeping only negative powers of~$k_0$.

\subsection{Various limits}
\label{subapp: Various limits}

Closed form solutions~(\ref{M solution 1}) and~(\ref{M solution 1})
for the Laplace-inverted two-point function greatly simplify in certain limits.
We derive several of these limits here, including rederiving limits 
from Sec.~\ref{sec: Various limits} starting directly from closed form solutions given 
in terms of Appell's fourth function.

\bigskip

\noindent {\bf de Sitter limit.}
Reproducing the de Sitter limit~(\ref{subsec: De Sitter limit}) 
does not rely on the particular representation of
the inverted Laplacian~(\ref{Xi def}). 
Rather, it hinges on showing that~(\ref{Upsilon dS}) holds,
which is true for any representation.

\bigskip

\noindent {\bf Flat space limit.} When examining particular limits, 
different representations of the Appell's fourth
function are useful. For the flat space limit of Sec.~\ref{subsec: Flat space limit}
we need to derive the asymptotic behaviour 
of~(\ref{Fbar}) around the point~$(X,Y)\!=\!(1,0)$, to be applied in~(\ref{M solution 2}).
This is because the arguments of Appell's functions in~(\ref{M solution 2}) 
in the flat space limit reduce to,
\begin{subequations}
\begin{align}
(1\!-\!\epsilon) \mathcal{H} \| \Delta\vec{x} \|
	\, \overset{H_0 \to 0}{\longsim} \,
	(1\!-\!\epsilon) H_0 \| \Delta\vec{x} \|
	\, ,
\qquad
&
(1\!-\!\epsilon) \mathcal{H}' \| \Delta\vec{x} \|
	\, \overset{H_0 \to 0}{\longsim} \,
	(1\!-\!\epsilon) H_0 \| \Delta\vec{x} \|
	\, ,
\\
\frac{\mathcal{H}^2}{\mathcal{H}'^2}
	\, \overset{H_0 \to 0}{\longsim} \,
	1 + 2(1\!-\!\epsilon) H_0 \Delta\eta 
	\, ,
\qquad
&
\frac{\mathcal{H}'^2}{\mathcal{H}^2}
	\, \overset{H_0 \to 0}{\longsim} \
	1 - 2(1\!-\!\epsilon) H_0 \Delta\eta 
	\, .
\end{align}
\end{subequations}
where the~$i\delta$-prescriptions are implied. Therefore, the limit
$(X,Y)\!\to\!(1,0)$ has to be taken simultaneously because the particular 
ratio of arguments is finite and non-vanishing,
\begin{equation}
X \longrightarrow 1 \, ,\qquad \quad
Y \longrightarrow 0 \, , \qquad \quad
\frac{Y}{(1\!-\!X)^2} \longrightarrow Z \, , \qquad \quad
0<Z<\infty \, .
\label{simultaneous limit}
\end{equation}
This can be accomplished as in~\cite{Exton:1995},
by first applying the transformation formula for the hypergeometric function 9.131.2. from~\cite{Gradshteyn:2007},
\begin{align}
\MoveEqLeft[2.5]
{}_2F_1\Bigl( \! \bigl\{ \alpha , \beta \bigr\} , \bigl\{ \gamma \bigr\} , X \Bigr)
	\!=
	\frac{ \Gamma(\gamma) \, \Gamma (\gamma \!-\! \alpha \!-\! \beta) }
		{ \Gamma(\gamma \!-\! \alpha) \, \Gamma(\gamma \!-\! \beta) } \,
		{}_2F_1 \Bigl( \!\bigl\{ \alpha, \beta \bigr\}, 
			\bigl\{ \alpha \!+\! \beta \!-\! \gamma \!+\! 1 \bigr\} , 1 \!-\!  X \! \Bigr) \ 
\label{HyperOneToZero}
\\
&
	+ 
	\frac{ \Gamma(\gamma) \, \Gamma (\alpha \!+\! \beta \!-\! \gamma) }
		{ \Gamma(\alpha) \, \Gamma(\beta) } \,
	( 1 \!-\! X )^{\gamma - \alpha - \beta} \,
	{}_2F_1\Bigl( \bigl\{ \gamma \!-\! \alpha , \gamma \!-\! \beta \bigr\} , 
		\bigl\{ \gamma \!-\! \alpha \!-\! \beta \!+\! 1 \bigr\} , 1 \!-\! X \Bigr)
	\, ,
\nonumber 
\end{align}
valid for~$ \bigl| {\rm Arg}(1\!-\!X) \bigr| \!<\! \pi$,
to the series representation 
of Appell's fourth function in~(\ref{F4 hyper series}), resulting in,
\begin{align}
\MoveEqLeft[1]
F_4\bigl( \alpha, \beta ; \gamma, \gamma' ; X, Y \bigr) 
	=
	\frac{ \Gamma(\gamma) \, \Gamma(\gamma') }
		{ \Gamma(\alpha) \, \Gamma(\beta) }
	\sum_{n=0}^{\infty}
	\Biggl\{
	\frac{ \Gamma(\alpha\!+\!n) \, \Gamma(\beta \!+\! n) \,
			\Gamma(\gamma\!-\!\alpha\!-\!\beta \!-\! 2n ) }
		{ \Gamma(\gamma \!-\! \alpha \!-\! n) \, \Gamma(\gamma \!-\! \beta \!-\! n) \,
			\Gamma(\gamma' \!+\! n) \, n! } \, Y^n
\nonumber \\
&
	\times\!
	{}_2F_1\Bigl( \bigl\{ \alpha\!+\!n , \beta \!+\! n \bigr\} , 
		\bigl\{ \alpha \!+\! \beta \!-\! \gamma \!+\! 1 \!+\! 2n \bigr\} , 
		1\!-\!X \Bigr)
	+
	\frac{ \Gamma( \alpha \!+\! \beta \!-\! \gamma \!+\! 2n ) }
		{ \Gamma(\gamma' \!+\! n) \, n! }
	(1\!-\!X)^{\gamma-\alpha-\beta}
\nonumber \\
&
	\times\! \biggl[ \frac{Y}{(1\!-\!X)^2} \biggr]^n
	{}_2F_1\Bigl( \bigl\{ \gamma \!-\! \alpha\!-\!n , \gamma \!-\! \beta \!-\! n \bigr\} , 
		\bigl\{ \gamma \!-\! \alpha \!-\! \beta \!+\! 1 \!-\! 2n \bigr\} , 
		1\!-\!X \Bigr)
	\Biggr\}
	\, .
\label{hyper series representation}
\end{align}
Then applying the limit~(\ref{simultaneous limit}) for the variables simplifies this expression,
\begin{align}
\MoveEqLeft[2]
F_4\bigl( \alpha, \beta ; \gamma, \gamma' ; X, Y \bigr) 
	\ \overset{(\ref{simultaneous limit})}{\longsim} \ \
	\frac{ \Gamma(\gamma) \, \Gamma(\gamma \!-\! \alpha \!-\! \beta) }
		{ \Gamma(\gamma \!-\! \alpha) \, \Gamma(\gamma \!-\! \beta) }
\nonumber \\
&	\hspace{1cm}
	+ 
	\frac{ \Gamma(\gamma) \, \Gamma(\gamma') }
		{ \Gamma(\alpha) \, \Gamma(\beta) }
		(1\!-\!X)^{\gamma-\alpha-\beta}
	\sum_{n=0}^{\infty}
	\frac{ \Gamma( \alpha \!+\! \beta \!-\! \gamma \!+\! 2n ) }
		{ \Gamma(\gamma' \!+\! n) \, n! }
\biggl[ \frac{Y}{(1\!-\!X)^2} \biggr]^n
\nonumber \\
={}&
	\frac{ \Gamma(\gamma) \, \Gamma(\gamma \!-\! \alpha \!-\! \beta) }
		{ \Gamma(\gamma \!-\! \alpha) \, \Gamma(\gamma \!-\! \beta) }
	+
	\frac{ \Gamma(\gamma) \, \Gamma(\alpha \!+\! \beta \!-\! \gamma) }
		{ \Gamma(\alpha) \, \Gamma(\beta) }
	(1\!-\!X)^{\gamma-\alpha-\beta}
\nonumber \\
&	\hspace{2cm}
	\times
	{}_2F_1\biggl( \Bigl\{ \frac{\alpha \!+\! \beta \!-\! \gamma \!+\! 1}{2} ,
		\frac{\alpha \!+\! \beta \!-\! \gamma }{2} \Bigr\} ,
		\Bigl\{ \gamma' \Bigr\} , \frac{4Y}{(1\!-\!X)^2} \biggr)
	\, .
\label{F4 flat intermediate}
\end{align}
We apply this result to functions~(\ref{Fbar}),
and plug them into~(\ref{M solution 2}) taking the flat space limit,
\begin{align}
\MoveEqLeft[2]
\mathcal{M}_\lambda(x;x') 
	\ \overset{ H_0 \to 0 }{\longsim} \ \
	\frac{ \bigl[ (1\!-\!\epsilon) H_0 \bigr]^{ 2 } }
		{ (4\pi)^{ \frac{D-1}{2} } }
	\frac{ \Gamma( D\!-\!4 ) }{ \Gamma \bigl( \frac{D-1}{2} \bigr) }
	\times e^{ - \frac{i\pi (D-2)}{2} }
	( \Delta\eta )^{ 4-D }
\nonumber \\
&	\hspace{3.5cm}
	\times
	{}_2F_1\biggl( \Bigl\{ \frac{D\!-\!3}{2} ,
		\frac{ D \!-\! 4 }{2} \Bigr\} ,
		\Bigl\{ \frac{ D \!-\! 1 }{2} \Bigr\} , 
		1 \!+\! \frac{ \Delta x^2}{ \Delta\eta^{2} } \biggr)
		\, .
\label{M flat}
\end{align}
Note that this expression is divergent in~$D\!\to\!4$ by itself, but that it always 
appears in photon two-point function 
with derivatives acting on it that remove the divergence.
The flat space limit we derived in Sec.~\ref{subsec: Flat space limit} relies on 
demonstrating that the limit in~(\ref{check Xi flat limit}) holds. Checking this limit 
here entails taking the derivative of~(\ref{M flat}) with respect to~$y$, which in the 
flat space limit reduces to, 
\begin{equation}
\frac{\partial}{\partial y} \ \overset{ H_0 \to 0 }{\longsim} \
	\frac{1}{ (1\!-\!\epsilon)^2 H_0^2 } \frac{\partial}{\partial (\Delta x^2)}
	\, ,
\end{equation}
and then applying the second transformation formula 
9.131.1 from~\cite{Gradshteyn:2007},
\begin{equation}
{}_2F_1\Bigl( \bigl\{ \alpha, \beta \bigr\} , \bigl\{ \gamma \bigr\} , X \Bigr)
	=
	(1 \!-\! X)^{-\beta}
	{}_2 F_1\biggl( \bigl\{ \gamma \!-\! \alpha , \beta \bigr\} , \bigl\{ \gamma \bigr\} ,
		\frac{X}{X \!-\! 1} \biggr)
		\, ,
\label{mixed id}
\end{equation}
valid for~$\bigl| {\rm Arg}(1\!-\!X) \bigr| \!<\! \pi$.
This finally produces
\begin{equation}
\frac{\partial}{\partial y}
\mathcal{M}_\lambda(x;x') 
	\xrightarrow{ H_0 \to 0 }
	\frac{ \Gamma( D\!-\!2 ) }{ 4 \, \Gamma \bigl( \frac{D+1}{2} \bigr) }
	\frac{ \bigl( \Delta x^2 \bigr)^{\! -\frac{ D - 2 }{2}} }
		{ (4\pi)^{ \frac{D-1}{2} } }
	{}_2 F_1\biggl( \Bigl\{ 1 , \frac{ D \!-\! 2 }{2} \Bigr\} , \Bigl\{ \frac{ D \!+\! 1 }{2} \Bigr\} ,
		1 \!+\! \frac{ \Delta\eta^2 }{ \Delta x^2 } \biggr)
		\, .
\end{equation}
Upon applying some simple identities for gamma function, this is the same as 
expression~(\ref{check Xi flat limit}), up to an overall 
factor~$(aa')^{- \frac{(D-2)\epsilon}{2}}$ that was factored out in~(\ref{Xi split}).

\bigskip

\noindent{\bf Spatial coincidence limit for the bulk part.}
The limit~$\| \Delta \vec{x} \| \!\to\! 0$ is straightforwardly obtained from the second representation
for the bulk part~(\ref{M solution 2}). First we need to consider the appropriate limit
of the function defined in~(\ref{Fbar}), that is obtained by first using
the limit of Appell's fourth function for one vanishing argument,
\begin{equation}
F_4 \bigl( \alpha , \beta ; \gamma, \gamma' ; X, Y \bigr)
	\xrightarrow{Y \to 0}
	{}_2 F_1 \Bigl( \bigl\{ \alpha , \beta \bigr\} , \bigl\{ \gamma \bigr\} , X \Bigr)
	\, ,
\end{equation}
followed by the quadratic transformation formula 15.8.21 from~\cite{Olver:2010,Olver:web},
\begin{align}
\MoveEqLeft[4]
{}_2F_1\Bigl( \bigl\{ \alpha , \beta \bigr\} , 
	\bigl\{ \alpha \!-\! \beta \!+\! 1 \bigr\} , X^2 \Bigr)
\nonumber \\
&
	=
	(1 \!+\! X)^{-2\alpha} \,
	{}_2F_1\biggl( \! \Bigl\{ \alpha , \alpha \!-\! \beta \!+\! \frac{1}{2} \Bigr\} , 
		\Bigl\{ 2 \alpha \!-\! 2\beta \!+\! 1 \Bigr\} , \frac{ 4X }{ (1 \!+\! X)^2 } \biggr)
		\, 
\label{quad hyper}
\end{align}
and the linear transformation 9.132.2 from~\cite{Gradshteyn:2007},
\begin{align}
{}_2F_1\Bigl( \bigl\{ \alpha , \beta \bigr\} , \bigl\{ \gamma \bigr\} , X \Bigr)
	={}&
	\frac{ \Gamma(\gamma) \, \Gamma(\beta \!-\! \alpha) }
		{ \Gamma(\beta) \, \Gamma( \gamma \!-\! \alpha) }
		(-X)^{-\alpha}
		{}_2F_1 \Bigl( \bigl\{ \alpha , \alpha \!+\! 1 \!-\! \gamma \bigr\} , 
			\bigl\{ \alpha \!+\! 1 \!-\! \beta \bigr\} , \frac{1}{X} \Bigr)
\nonumber \\
&	\hspace{-2.5cm}
	+
	\frac{ \Gamma(\gamma) \, \Gamma(\alpha \!-\! \beta) }
		{ \Gamma(\alpha) \, \Gamma(\gamma \!-\! \beta) }
		(-X)^{-\beta}
		{}_2F_1 \Bigl( \bigl\{ \beta , \beta \!+\! 1 \!-\! \gamma \bigr\} , 
			\bigl\{ \beta \!+\! 1 \!-\! \alpha \bigr\} , \frac{1}{X} \Bigr)
	\, ,
\label{hyper inversion}
\end{align}
that is valid for~$\bigl| {\rm Arg}(-X) \bigr| \!<\! \pi$. This sequence transforms the 
function in~(\ref{Fbar}) into,
\begin{align}
&
\overline{\mathscr{F}}_\lambda(X,Y)
	\xrightarrow{Y \to 0}
	\frac{ - \bigl[ (1\!-\!\epsilon) H_0 \bigr]^{ D-2 } }
		{ (4\pi)^{ \frac{D}{2} } (D\!-\!3) \sin(\pi\lambda) }
	\Biggl[
	\frac{ \Gamma( \frac{4-D}{2} ) \, \Gamma\bigl( \frac{D-3}{2} \!+\! \lambda \bigr) }
			{ 2 \, \Gamma( \frac{5-D}{2} \!+\! \lambda ) }
		\bigl[ e^{i {\rm Arg(-X)}} \bigr]^{ - \frac{D-3}{2} - \lambda }
\nonumber \\
&	\hspace{3.5cm}
	\times
		{}_2F_1 \biggl( \Bigl\{ \frac{D\!-\!3}{2} \!+\! \lambda , 
				\frac{D\!-\!3}{2} \!-\! \lambda \Bigr\} , 
			\Bigl\{ \frac{D\!-\!2}{2} \Bigr\} , \frac{ (1 \!+\! X)^2 }{ 4X } \biggr)
\\
&	
	+
	\frac{ \Gamma( \frac{D-4}{2} ) }{ 2 }
		\bigl[ e^{i {\rm Arg(-X)}} \bigr]^{ -\frac{1}{2} - \lambda }
		\biggl[ \frac{ (1\!+\!X)^2 }{4X} \biggr]^{\frac{4 - D}{2} } \!
		{}_2F_1 \biggl( \Bigl\{ \frac{1}{2} \!+\! \lambda , \frac{1}{2} \!-\! \lambda \Bigr\} , 
		\Bigl\{ \frac{6\!-\!D}{2} \Bigr\} , \frac{ (1 \!+\! X)^2 }{ 4X } \biggr)
	\Biggr]
		\, .
\nonumber 
\end{align}
Applying this result to~(\ref{M solution 2}), keeping track of~$i\delta$-prescriptions,
and applying the identity for gamma functions,
\begin{equation}
\frac{\Gamma(\alpha \!+\! \lambda)}{ \Gamma(1 \!-\! \alpha \!+\! \lambda) }
	-
	\frac{ \Gamma(\alpha \!-\! \lambda) }{ \Gamma(1 \!-\! \alpha \!-\! \lambda) }
	=
	- 2 \sin(\pi\lambda) \frac{ \Gamma(\alpha \!+\! \lambda) \, \Gamma(\alpha \!-\! \lambda) }
		{ \Gamma\bigl( \frac{1}{2} \!+\! \alpha \bigr) \, \Gamma\bigl( \frac{1}{2} \!-\! \alpha \bigr) } \, ,
\end{equation}
then yields the sought spatial coincidence limit,
\begin{align}
\mathcal{M}_\lambda(x;x')
={}&
	\frac{-2}{D\!-\!3} \times
\frac{ \bigl[ (1\!-\!\epsilon) H_0 \bigr]^{D-2} }{ (4\pi)^{ \frac{D}{2} } }
	\frac{ \Gamma\bigl( \frac{D-3}{2} \!+\! \lambda \bigr) \, \Gamma\bigl( \frac{D-3}{2} \!-\! \lambda \bigr) }
		{ \Gamma\bigl( \frac{D-2}{2} \bigr) }
\nonumber \\
&
	\times
	{}_2F_1 \biggl( \Bigl\{ \frac{D\!-\!3}{2} \!+\! \lambda , \frac{D\!-\!3}{2} \!-\! \lambda \Bigr\} , 
		\Bigl\{ \frac{D\!-\!2}{2} \Bigr\} , 1 \!+\! \sh^2\Bigl( \frac{v}{2} \Bigr) \biggr)
		\, .
\label{SpatialCoincidenceLimit}
\end{align}
This is precisely the spatial coincidence limit one obtains from the series 
representation~(\ref{Xi solution}) in Sec.~\ref{subsec: Laplace-inverted two-point function},
that receives contribution only from the~$n\!=\!0$ term
since~$\bigl[ y \!+\! 4 \, \sh^2\bigl( \frac{v}{2} \bigr) \bigr] \!=\! (1\!-\!\epsilon)^2 \mathcal{H} \mathcal{H}' \| \Delta \vec{x} \|^2$.

\bigskip

\noindent{\bf Large spatial separations limit for the bulk part.} The limit~$\| \Delta \vec{x} \| \!\to\! \infty$
is best inferred from the first representation for the bulk 
solution~(\ref{M solution 1}). It involves taking the 
large argument limit of the function~(\ref{curly F}) appearing there,
\begin{equation}
\mathscr{F}_\lambda(X,Y)
	\, \overset{ X,Y \! \to \infty }{\longsim} \,
	\frac{ \bigl[ (1\!-\!\epsilon) H_0 \bigr]^{ D-2 } }
		{ 16 \pi^{ \frac{D+1}{2} } }
	\frac{ \Gamma \bigl( \frac{D-3}{2}\!+\!\lambda \bigr) \, \Gamma(-\lambda) }{ \lambda }
	( X Y )^{ - \frac{D-3}{2} - \lambda } \,
	\, ,
\end{equation}
after which the limit of~(\ref{M solution 1}) follows,
\begin{align}
\MoveEqLeft[4]
\mathcal{M}_\lambda(x;x')
	=
	\frac{ \bigl[ (1\!-\!\epsilon) H_0 \bigr]^{ D-2 } }
		{ 16 \pi^{ \frac{D+1}{2} } }
	\Biggl\{
	-
	\frac{ \Gamma \bigl( \frac{D - 3}{2} \bigr) \, \Gamma(\lambda) \, \Gamma(-\lambda) }
		{ \bigl[ (1\!-\!\epsilon)^2 \mathcal{H} \mathcal{H}' \| \Delta\vec{x}\|^2 
			\bigr]^{ \frac{D-3}{2} } }
	\biggl[
	e^{i\pi\lambda}
		\Bigl( \frac{\mathcal{H} }{ \mathcal{H}' } \Bigr)^{\!\lambda}
	+
	e^{-i\pi\lambda}
		\Bigl( \frac{\mathcal{H} }{ \mathcal{H}' } \Bigr)^{\!-\lambda}
	\biggr]
\nonumber \\
&
	+
	\frac{ \Gamma \bigl( \frac{D-3}{2}\!+\!\lambda \bigr) \, \Gamma(-\lambda) }
		{ \lambda \bigl[ (1\!-\!\epsilon)^2 \mathcal{H} \mathcal{H}' \| \Delta\vec{x} \|^2 \bigr]^{ \frac{D-3}{2} + \lambda } }
	-
	\frac{ \Gamma \bigl( \frac{D-3}{2}\!-\!\lambda \bigr) \, \Gamma(\lambda) }
		{ \lambda \bigl[ (1\!-\!\epsilon)^2 \mathcal{H} \mathcal{H}' \| \Delta\vec{x} \|^2 \bigr]^{ \frac{D-3}{2} - \lambda } }
	\Biggr\}
	\, .
\end{align}
Deriving this result from the series representation~(\ref{Xi solution}) might be prohibitively
difficult and whenever large spatial separations are called for it is better to appeal
to the closed form solution~(\ref{M solution 1}). Note that, unlike the spatial coincidence
limit~(\ref{SpatialCoincidenceLimit}), the limit of large spatial separation above 
exhibits a Coulomb-like tail that is usually generated as a response to a point charge.
The same behaviour is exhibited in the Coulomb gauge in flat space~(\ref{FourDimFlat})
where the Coulomb-like contribution is present for large spatial separations (with both
real and imaginary parts), but disappears in the limit of small separations due to the logarithm 
that multiplies it.

\bigskip

\noindent{\bf Temporal coincidence limit for the bulk part.}
The behaviour close to temporal coincidence~$\eta' \! \to\! \eta$ 
of the bulk part of the Laplace-inverted
two point function is best inferred staring from the second representation 
in~(\ref{M solution 2}). However, taking this limit directly is impeded by
the first argument tending to one,~$Y \!\to\! 1$,
where one of the poles of Appell's fourth function is located. 
The location of this pole descends from the corresponding pole of the hypergeometric 
equation, which is revealed by examining the single series
representation~(\ref{F4 hyper series}) applied to Appell's fourth function in~(\ref{Fbar}),
\begin{align}
F_4 \bigl( \alpha, \beta ; \gamma, \gamma' ; X^2 , Y^2 \bigr)
	={}&
	\frac{ \Gamma( \gamma' ) }{ \Gamma(\alpha) \, \Gamma(\beta) } 
	\sum_{n=0}^{\infty}
	\frac{ \Gamma( \alpha \!+\! n ) \, \Gamma( \beta \!+\! n ) }
		{ \Gamma( \gamma' \!+\! n) \, n! }
		Y^{2n}
\nonumber \\
&	\hspace{2.5cm}
	\times
		{}_2F_1 \Bigl( \bigl\{ \alpha \!+\! n , \beta \!+\! n \bigr\} , 
		\bigl\{ \gamma \bigr\} , X^2 \Bigr)
		\, .
\end{align}
The singular behaviour around the pole at~$X\!\to\!1$ is isolated by applying 
formula~(\ref{HyperOneToZero}) for transforming the argument of the hypergeometric 
function,
\begin{align}
\MoveEqLeft[1]
F_4 \bigl( \alpha, \beta ; \gamma, \gamma' ; X^2 , Y^2 \bigr)
	=
	\frac{ \Gamma( \gamma' ) }{ \Gamma(\alpha) \, \Gamma(\beta) } 
	\sum_{n=0}^{\infty}
	\frac{ \Gamma( \alpha \!+\! n ) \, \Gamma( \beta \!+\! n ) }
		{ \Gamma( \gamma' \!+\! n) \, n! }
		Y^{2n}
\nonumber \\
&	
	\times
	\biggl\{
	\frac{ \Gamma(\gamma) \, \Gamma (\gamma \!-\! \alpha \!-\! \beta \!-\! 2n) }
		{ \Gamma(\gamma \!-\! \alpha \!-\! n) \, \Gamma(\gamma \!-\! \beta \!-\! n) } \,
		{}_2F_1 \Bigl( \!\bigl\{ \alpha \!+\! n , \beta \!+\! n \bigr\}, 
			\bigl\{ \alpha \!+\! \beta \!-\! \gamma \!+\! 1 \!+\! 2n \bigr\} , 1 \!-\!  X^2 \! \Bigr) \ 
\nonumber \\
&	\hspace{1cm}
	+ 
	\frac{ \Gamma(\gamma) \, \Gamma (\alpha \!+\! \beta \!-\! \gamma \!+\! 2n) }
		{ \Gamma(\alpha \!+\! n) \, \Gamma(\beta \!+\! n) } \,
		\bigl( 1 \!-\! X^2 \bigr)^{\gamma - \alpha - \beta - 2n} \,
\nonumber \\
&	\hspace{2cm}
	\times
	{}_2F_1\Bigl( \bigl\{ \gamma \!-\! \alpha \!-\! n , \gamma \!-\! \beta \!-\! n \bigr\} , 
		\bigl\{ \gamma \!-\! \alpha \!-\! \beta \!+\! 1 \!-\! 2n \bigr\} , 1 \!-\! X^2 \Bigr)
		\biggr\}
		\, .
\end{align}
Both hypergeometric functions in the resulting expression above are now well
behaved in the limit~$X\!\to\!1$. However, the second one is multiplied by more and 
more negative powers of a small quantity. These powers descend from the behaviour
around the pole. We keep all such leading terms in the second series, and resum the 
series to obtain control over the singular behaviour,
\begin{align}
\MoveEqLeft[0.1]
F_4 \bigl( \alpha, \beta ; \gamma, \gamma' ; X, Y \bigr)
	\ \overset{X\sim1}{\longsim} \
	\frac{ \Gamma(\gamma) \,\Gamma( \gamma' ) }{ \Gamma(\alpha) \, \Gamma(\beta) } 
	\bigl( 1 \!-\! X^2 \bigr)^{\! \gamma - \alpha - \beta }
	\sum_{n=0}^{\infty} \!
	\frac{ \Gamma(\alpha \!+\! \beta \!-\! \gamma \!+\! 2n) }
		{ \Gamma( \gamma' \!+\! n) \, n! }
		\Bigl[ \frac{Y}{1 \!-\! X^2} \Bigr]^{2n}
\nonumber \\
&
	=
	\frac{ \Gamma(\gamma) \, \Gamma(\alpha \!+\! \beta \!-\! \gamma) }
		{ \Gamma(\alpha) \, \Gamma(\beta) } 
	\bigl( 1 \!-\! X^2 \bigr)^{\! \gamma - \alpha - \beta }
	{}_2F_1\biggl( 
	\Bigl\{ \frac{\alpha \!+\! \beta \!-\! \gamma}{2} , 
		\frac{\alpha \!+\! \beta \!-\! \gamma \!+\! 1}{2} \Bigr\} ,
		\Bigl\{ \gamma' \Bigr\} , \Bigl[ \frac{2Y}{1 \!-\! X^2} \Bigr]^{2} \biggr)
		\, .
\end{align}
This result should be followed by another another
transformation formula~(\ref{hyper inversion}) that inverts the argument of the 
hypergeometric function,
\begingroup
\allowdisplaybreaks
\begin{align}
\MoveEqLeft[1]
F_4 \bigl( \alpha, \beta ; \gamma, \gamma' ; X, Y \bigr)
	\ \overset{X\sim1}{\longsim} \
	\frac{ \Gamma(\gamma) \, \Gamma(\gamma') }
		{ 2 \,\Gamma(\alpha) \, \Gamma(\beta) } 
	\bigl( 1 \!-\! X^2 \bigr)^{\! \gamma - \alpha - \beta }
\nonumber \\
&
	\times \!
	\Biggl\{
		\frac{ \Gamma \bigl( \frac{ \alpha+\beta-\gamma }{2} \bigr) }
			{ \Gamma \bigl( \gamma' \!-\! \frac{\alpha + \beta - \gamma}{2} \bigr) }
		\biggl( - \Bigl[ \frac{Y}{1 \!-\! X^2} \Bigr]^{2} 
			\biggr)^{ \!\! \frac{\gamma - \alpha - \beta }{2}}
\nonumber \\
&	\hspace{2cm}
	\times
		{}_2F_1 \biggl( \Bigl\{ \frac{\alpha \!+\! \beta \!-\! \gamma}{2} , 
			\frac{\alpha \!+\! \beta \!-\! \gamma}{2} \!+\! 1 \!-\! \gamma' \Bigr\} , 
		\Bigl\{ \frac{1}{2} \Bigr\} , \Bigl[ \frac{1 \!-\! X^2}{2Y} \Bigr]^{2} \biggr)
\nonumber \\
&	\hspace{0.8cm}
	-
	\frac{ \Gamma \bigl( \frac{\alpha + \beta - \gamma + 1}{2} \bigr) }
		{ \Gamma \bigl( \gamma' \!-\! \frac{\alpha + \beta - \gamma + 1}{2} \bigr) }
		\biggl( - \Bigl[ \frac{Y}{1 \!-\! X^2} \Bigr]^{2} 
			\biggr)^{ \!\! \frac{ \gamma - \alpha - \beta - 1}{2}}
\nonumber \\
&	\hspace{2cm}
	\times
		{}_2F_1 \biggl( \Bigl\{ \frac{\alpha \!+\! \beta \!-\! \gamma \!+\! 1}{2} , \frac{\alpha \!+\! \beta \!-\! \gamma \!+\! 3}{2} \!-\! \gamma' \Bigr\} , 
			\Bigl\{ \frac{3}{2} \Bigr\} , \Bigl[ \frac{1 \!-\! X^2}{2Y} \Bigr]^{2} \biggr)
	\Biggr\}
	\, .
\end{align}
\endgroup
The last step consists of taking the vanishing argument limit of the hypergeometric 
functions above, in which they reduce to unity,
\begin{align}
\MoveEqLeft[1]
F_4 \bigl( \alpha, \beta ; \gamma, \gamma' ; X, Y \bigr)
	\ \overset{X\sim1}{\longsim} \
	\frac{ \Gamma(\gamma) \, \Gamma(\gamma') }
		{ 2 \,\Gamma(\alpha) \, \Gamma(\beta) } 
	\bigl( 1 \!-\! X^2 \bigr)^{\! \gamma - \alpha - \beta }
\\
&
	\times \!
	\Biggl\{
		\frac{ \Gamma \bigl( \frac{ \alpha+\beta-\gamma }{2} \bigr) }
			{ \Gamma \bigl( \gamma' \!-\! \frac{\alpha + \beta - \gamma}{2} \bigr) }
		\biggl( - \Bigl[ \frac{Y}{1 \!-\! X^2} \Bigr]^{2} 
			\biggr)^{ \!\! \frac{\gamma - \alpha - \beta }{2}}
	\!\! -
	\frac{ \Gamma \bigl( \frac{\alpha + \beta - \gamma + 1}{2} \bigr) }
		{ \Gamma \bigl( \gamma' \!-\! \frac{\alpha + \beta - \gamma + 1}{2} \bigr) }
		\biggl( - \Bigl[ \frac{Y}{1 \!-\! X^2} \Bigr]^{2} 
			\biggr)^{ \!\! \frac{ \gamma - \alpha - \beta - 1}{2}}
	\Biggr\}
	\, .
\nonumber 
\end{align}
Applying this limit to~(\ref{Fbar}) now gives,
\begin{align}
\MoveEqLeft[3]
\overline{\mathscr{F}}_\lambda(X,Y)
	\ \overset{X\sim1}{\longsim} \ 
	\frac{ \bigl[ (1\!-\!\epsilon) H_0 \bigr]^{ D-2 } }
		{ 32 \pi^{ \frac{D-1}{2} } \sin(\pi\lambda) }
		X^{ \frac{D-3}{2} + \lambda }
	\bigl( 1 \!-\! X^2 \bigr)^{\! 4-D }
\nonumber \\
&
	\times \!
	\Biggl\{
	\frac{ \Gamma \bigl( \frac{D-3}{2} \bigr) }
		{ 2 }
		\biggl( - \Bigl[ \frac{Y}{1 \!-\! X^2} \Bigr]^{2} 
			\biggr)^{ \!\! \frac{ 3-D }{2}}
		-
		\frac{ \Gamma \bigl( \frac{ D-4 }{2} \bigr) }
			{ \sqrt{\pi} }
		\biggl( - \Bigl[ \frac{Y}{1 \!-\! X^2} \Bigr]^{2} 
			\biggr)^{ \!\! \frac{ 4-D }{2}}
	\Biggr\}
	\, .
\end{align}
Finally, applying this to the representation in~(\ref{M solution 2}) gives
\begin{align}
\MoveEqLeft[2]
\mathcal{M}_\lambda(x;x') 
	\ \overset{\eta'\to\eta}{\longsim} \,
	\frac{ \bigl[ (1\!-\!\epsilon) H_0 \bigr]^{ D-2 } }
		{ 8 \pi^{ \frac{D}{2} } }
	\Biggl[
	\frac{ \sqrt{\pi} \, 
		\Gamma \bigl( \frac{D - 3}{2} \bigr) i v }{ \bigl[ (1\!-\!\epsilon)^2 \mathcal{H} \mathcal{H}' \| \Delta\vec{x} \|^2 \bigr]^{ \! \frac{D-3}{2}} }
	- 
	\frac{ \Gamma \bigl( \frac{D - 4}{2} \bigr) }
		{ \bigl[ (1\!-\!\epsilon)^2 \mathcal{H} \mathcal{H}' \| \Delta\vec{x} \|^2 
			\bigr]^{ \! \frac{ D-4 }{2}} }
	\Biggr]
	.
\label{Mcoincidence}
\end{align}
The imaginary part of this result accounts for the local terms in
equations~(\ref{second Xi eq})--(\ref{Xi spacetime lowering}) for the Laplace-inverted
two-point function in the~$(++)$ prescription. Namely,  two time derivatives
acting close to coincidence produce
\begin{align}
\MoveEqLeft[4]
\partial_0 \partial_0
	\frac{ i \bigl[ \tensor*[^{\scr \!+}]{\Xi}{^{\scr +\!}} \bigr]_\lambda(x;x') }
		{ (1\!-\!\epsilon)^2 \mathcal{H} \mathcal{H}' }
	=
	\partial_0 \partial_0
	\biggl[
		\theta(\eta \!-\! \eta')
		\frac{ i \bigl[ \tensor*[^{\scr \!-}]{\Xi}{^{\scr +\!}} \bigr]_\lambda(x;x') }
			{ (1\!-\!\epsilon)^2 \mathcal{H} \mathcal{H}' }
		+
		\theta(\eta' \!-\! \eta)
		\frac{ i \bigl[ \tensor*[^{\scr \!+}]{\Xi}{^{\scr -\!}} \bigr]_\lambda(x;x') }
			{ (1\!-\!\epsilon)^2 \mathcal{H} \mathcal{H}' }
	\biggr]
\nonumber \\
\ \overset{x' \to x}{\longsim} \ {}&
	\partial_0 \partial_0
	\biggl[
	\frac{ \Gamma \bigl( \frac{D - 3}{2} \bigr)  }
		{ 8 \pi^{ \frac{D-1}{2} } \| \Delta\vec{x} \|^{ D-3 } a^{ D-2}  }
		\frac{ i |v| }{ (1\!-\!\epsilon) \sqrt{ \mathcal{H} \mathcal{H}' } }
	\biggr]
	=
	\frac{ \Gamma \bigl( \frac{D - 3}{2} \bigr) \, i \delta(\eta \!-\! \eta') }
		{ 4 \pi^{ \frac{D-1}{2} } \| \Delta\vec{x}\|^{ D-3 } a^{D-2} }
		\, .
\end{align}
Moreover, the result in~(\ref{Mcoincidence}) also allows us to infer the behaviour 
close to coincidence of the function~$i\Upsilon$ defined in~(\ref{Upsilon}),
\begin{align}
i\Upsilon
	\ \overset{\eta'\to \eta}{\longsim} \,{}&
	-
	\frac{\epsilon}{(1 \!-\! \epsilon) }
		\!\times\!
	\frac{ \bigl[ (1\!-\!\epsilon)^2 HH'\bigr]^{ \frac{D-2}{2} } \, 
		\Gamma \bigl( \frac{D - 2}{2} \bigr) }
		{ 4 \pi^{ \frac{D}{2} } \bigl[ (1\!-\!\epsilon)^2 \mathcal{H} \mathcal{H}' \| \Delta\vec{x} \|^2 
			\bigr]^{ \! \frac{ D-2 }{2}} }
		\, ,
\label{UtimeCoin}
\end{align}
where the real part of~(\ref{Mcoincidence}) contributes,
while the contributions of the imaginary part end up canceling.

\subsection{Series representation}
\label{subapp: Series representation}

Here we derive the series solution~(\ref{check Xi solution}) for the Laplace-inverted
two-point function starting from the closed form solutions~(\ref{M solution 1}) 
and~(\ref{M solution 2}) for the bulk part, and~(\ref{X sol}) for the infrared part.


\bigskip

\noindent {\bf Bulk part.}
The starting point is formula (3.3) from~\cite{Brychkov:2017}
for transforming the parameters of the Appell's fourth function,
\begin{equation}
F_4 \bigl( \alpha, \beta; \gamma, \gamma' ; X , Y \bigr)
	=
	\frac{- \beta}{ \alpha \!-\! \beta} 
	F_4\bigl( \alpha, \beta\!+\!1; \gamma, \gamma' ; X , Y \bigr)
	+
	\frac{\alpha}{ \alpha \!-\! \beta} 
	F_4\bigl( \alpha \!+\! 1 , \beta; \gamma, \gamma' ; X , Y \bigr)
	\, .
\end{equation}
Iterating this identity~$(N\!+\!1)$ times for the special combination of parameters
appearing in~(\ref{curly F}) gives,
\begin{align}
\MoveEqLeft[4]
F_4\bigl( \alpha, \beta \!-\! 1; \beta, \beta ; X , Y \bigr)
	=
	\sum_{n=0}^{N}
	\frac{ ( 1 \!-\! \beta ) \,\Gamma( \alpha \!-\! \beta \!+\! 1 ) \, \Gamma(\alpha \!+\! n) }
		{ \Gamma(\alpha) \, \Gamma( \alpha \!-\! \beta \!+\! 2 \!+\! n ) }
	F_4 \bigl( \alpha \!+\! n, \beta; \beta , \beta ; X , Y \bigr)
\nonumber \\
&
	+
	\frac{ \Gamma( \alpha \!-\! \beta \!+\! 1 ) \, \Gamma(\alpha \!+\! 1 \!+\! N) }
		{ \Gamma(\alpha) \, \Gamma( \alpha \!-\! \beta \!+\! 2 \!+\! N ) }
	F_4 \bigl( \alpha \!+\! 1 \!+\! N , \beta \!-\! 1; \beta, \beta ; X , Y \bigr)
	\, .
\label{iteration identity}
\end{align}
In the end we want to take the limit~$N\!\to\!\infty$. 
Appell's functions in the series part of~(\ref{iteration identity}) are all in a prticular
form that can be reduced to rescale propagator functions~(\ref{F def}) and their derivatives.
We do this by modifying the example from Appendix A of~\cite{Frob:2013qsa}.
The first step is to apply the reduction formula 9.182.7 from~\cite{Gradshteyn:2007},
\begin{align}
\MoveEqLeft[5]
F_4\biggl( \alpha, \beta ; \beta, \beta ; \frac{-X}{(1\!-\!X)(1\!-\!Y)} , \frac{ -Y}{(1\!-\!X)(1\!-\!Y)} \biggr)
\nonumber \\
&
	=
	(1 \!-\! X)^\alpha (1 \!-\! Y)^\alpha \,
	{}_2F_1\Bigl( \bigl\{ \alpha, \alpha \!-\! \beta \!+\!1 \bigr\}, 
		\bigl\{ \beta \bigr\} , XY \Bigr)
		\, ,
\label{Appell reduction}
\end{align}
that reduces them all to  hypergeometric functions. 
Then we apply a sequence of three transformation 
identities for hypergeometric functions;
first the quadratic transformation formula~(\ref{quad hyper}),
then the argument transformation formula~(\ref{mixed id}),
followed by the argument inversion formula~(\ref{hyper inversion}).
This sequence transforms the reduction formula~(\ref{Appell reduction}) into,
\begin{align}
\MoveEqLeft[2]
F_4\bigl( \alpha, \beta ; \beta, \beta ; X , Y \bigr)
	=
	\frac{ \Gamma(2\beta \!-\! 1) }{ \Gamma \bigl( \beta \!-\! \frac{1}{2} \bigr) }
		\bigl( 4 \sqrt{XY} \, \bigr)^{ \! -\alpha}
\nonumber \\
&
	\times
	\Biggl\{
	\frac{ \Gamma \bigl( \beta \!-\! \alpha \!-\! \frac{1}{2} \bigr) }
		{  \Gamma(2 \beta \!-\! \alpha \!-\! 1) } \,
		{}_2F_1\biggl( \Bigl\{ \alpha , \alpha \!-\! 2 \beta \!+\! 2 \Bigr\} , 
			\Bigl\{ \alpha \!-\! \beta \!+\! \frac{3}{2} \Bigr\} , 
			\frac{ \bigl( \sqrt{X} \!+\! \sqrt{Y} \bigr)^2 \!-\! 1 }{ 4\sqrt{XY} }
			\biggr)
\nonumber \\
&	\hspace{1cm}
	+
	\frac{ \Gamma \bigl( \alpha \!-\! \beta \!+\! \frac{1}{2} \bigr) }
		{ \Gamma( \alpha ) }
		\biggl[ \frac{ 1 \!-\! \bigl( \sqrt{X} \!+\! \sqrt{Y} \, \bigr)^2 }{ 4\sqrt{XY} }
			\biggr]^{ \beta-\alpha-\frac{1}{2} }
\nonumber \\
&	\hspace{2.cm}
	\times
		{}_2F_1\biggl( \Bigl\{ \beta \!-\! \frac{1}{2} , \frac{3}{2} \!-\! \beta \Bigr\} , 
			\Bigl\{ \beta\!-\! \alpha \!+\! \frac{1}{2} \Bigr\} , 
			\frac{ \bigl( \sqrt{X} \!+\! \sqrt{Y} \, \bigr)^2 \!-\! 1 }{ 4\sqrt{XY} } \biggr)
	\Biggr\}
		\, .
\label{final reduction form}
\end{align}
Applying this reduction formula to all the terms in the series in
identity~(\ref{iteration identity}), that we apply to Appell's functions in~(\ref{M solution 1})
gives after some tedious algebra,
\begin{align}
\mathcal{M}_\lambda(x;x')
	={}&
	\sum_{n=0}^{N}
	\frac{ ( -1 )^{n} \, \Gamma\bigl( \frac{D-3}{2} \bigr) }
		{ 4 \, \Gamma \bigl( \frac{D-1}{2} \!+\! n \bigr) }
	\Bigl[ y + 4 \, \sh^2 \Bigl( \frac{v}{2} \Bigr) \Bigr]^{ \! n}
	\frac{\partial^{n} }{\partial y^{n} } 
	 I \bigl[ \mathcal{F}_\lambda(y) \bigr]
\nonumber \\
&
	+
	{\rm rem}_\lambda^N \Bigl( (1\!-\!\epsilon) \mathcal{H} \| \Delta\vec{x} \|,
		(1\!-\!\epsilon) \mathcal{H}' \| \Delta\vec{x} \| \Bigr)
	 \, ,
\label{M series inter}
\end{align}
where we have recognized that
\begin{align}
\MoveEqLeft[2]
	\frac{ \bigl[ (1\!-\!\epsilon) H_0 \bigr]^2 }{ (4\pi)^{ \frac{D}{2} } }
	\Bigl( \frac{-1}{4} \Bigr)^{\!n-1} \,
	\frac{ \Gamma\bigl( \frac{D-3}{2} \!+\! \lambda \!+\! n \bigr) \, 
			\Gamma\bigl( \frac{D-3}{2} \!-\! \lambda \!+\! n \bigr) }
		{ \Gamma\bigl( \frac{D-2}{2} \!+\! n \bigr) }
\\
&
	\times
	{}_2F_1\biggl( \Bigl\{ \frac{D\!-\!3}{2} \!+\! \lambda \!+\! n ,
			\frac{D\!-\!3}{2} \!-\! \lambda \!+\! n \Bigr\} , 
		\Bigl\{ \frac{D\!-\!2}{2} \!+\! n \Bigr\} , 1 \!-\! \frac{y}{4} \biggr)
		=
		\frac{\partial^n}{\partial y^n} I\bigl[ \mathcal{F}_\lambda(y) \bigr]
		\, ,
\nonumber 
\end{align}
which follows from the definition of the rescaled propagator function~(\ref{F def}),
and where the remainder in the second line of~(\ref{M series inter}) is
\begin{align}
{\rm rem}_\lambda^N(X,Y)
	={}&
	\frac{ \bigl[ (1\!-\!\epsilon) H_0 \bigr]^{ D-2 } \Gamma \bigl( \frac{D - 3}{2} \bigr) \,
			\Gamma(\lambda) \, \Gamma(-\lambda) }
		{ 16 \pi^{ \frac{D+1}{2} } ( XY )^{ \frac{D-3}{2} } }
	\Biggl[
	- 
	e^{i\pi\lambda}
		\Bigl( \frac{ X }{ Y } \Bigr)^{\!\lambda}
	-
	e^{-i\pi\lambda}
		\Bigl( \frac{ X }{ Y } \Bigr)^{\!-\lambda}
\nonumber \\
&	\hspace{-2.2cm}
	+
	\frac{ \Gamma \bigl( \frac{D-1}{2} \!+\! \lambda \!+\! N \bigr) \, (XY)^{ - \lambda } }
		{ \Gamma \bigl( \frac{D-1}{2} \!+\! N \bigr) \, \Gamma(1 \!+\! \lambda) }
		F_4\biggl( \frac{D\!-\!1}{2} \!+\! \lambda \!+\! N , \lambda ; 1 \!+\! \lambda , 1 \!+\! \lambda ;
			\frac{1}{ X^2}, \frac{1}{ Y^2} \biggr)
\nonumber \\
&	\hspace{-2.2cm}
	+
	\frac{ \Gamma \bigl( \frac{D-1}{2} \!-\! \lambda \!+\! N \bigr) \, (XY)^{ \lambda } }
		{ \Gamma \bigl( \frac{D-1}{2} \!+\! N \bigr) \, \Gamma( 1 \!-\! \lambda ) }
		F_4\biggl( \frac{D\!-\!1}{2} \!-\! \lambda \!+\! N , -\lambda ; 1 \!-\! \lambda , 1 \!-\! \lambda ;
			\frac{1}{ X^2}, \frac{1}{ Y^2} \biggr)
	\Biggr]
	\, .
\label{remainder N}
\end{align}
It is now clear that the first line of~(\ref{M series inter}) reproduces the series
representation~(\ref{Xi solution}) from Sec.~\ref{subsec: Laplace-inverted two-point function}
upon taking the limit~$N \!\to\! \infty$. This would imply that the expression in the second line 
of ~(\ref{M series inter}) has to vanish in the same limit.

Showing the remainder~(\ref{remainder N}) vanishes would be straightforward 
if asymptotic expansions of Appell's fourth functions for large parameters were 
readily available. Unfortunately, that is not the case, and there seem to be no 
readymade results in the literature. Large parameter asymptotic expansions are 
known for some cases for Appell's first function~$F_1$~\cite{Lopez1,Lopez2},
so one might hope that expressing the remainder in terms of~$F_1$ 
using~(\ref{F4 contiguous trans}) and~(\ref{F4-F1}) would help.
Unfortunately, thus obtained representation in terms of~$F_1$
is not covered by the asymptotic expansions in~\cite{Lopez1,Lopez2}.  
The necessary expansions are currently work in progress~\cite{LopezProgress}.


\bigskip

\noindent {\bf IR part.}
Here we show that the IR part~(\ref{X sol}) of the Laplace-inverted two-point function
corresponds to the series representation~(\ref{check Xi solution}) and~(\ref{Xi solution}). 
We start by writing~(\ref{X sol}) in terms of bi-local variables~(\ref{bilocal variables}), and
rewriting it as a primitive function with respect to~$y$,
\begin{align}
\MoveEqLeft[6]
\mathcal{X}_\lambda(x;x')
=
	I \Biggl[ 
	\frac{ \bigl[ (1 \!-\! \epsilon) H_0 \bigr]^{ D-2 } \, \Gamma(\lambda) \Gamma(2\lambda)  }
		{ (4\pi)^{\frac{D}{2}} \, \Gamma\bigr( \frac{D-1}{2} \bigr) \, 
			\Gamma\bigl( \frac{1}{2} \!+\! \lambda \bigr) } 
	\sum_{N=0}^{\infty}
	\sum_{n=0}^{N}
	\sum_{\ell=0}^{N-n}
	\frac{ (n\!+\!1) c_{(N+1)(n+1)\ell} }{ \bigl( \frac{D-1}{2} \!+\! N  \!-\! \lambda \bigr) }
\nonumber 
\\
&
	\times
	\Bigl[ \frac{ k_0^2 e^{-u} }{ (1\!-\!\epsilon)^2 H_0^2 } 
		\Bigr]^{\frac{D-1}{2} -\lambda + N}
	\Bigl[ y + 4 \, \sh^2\Bigl( \frac{v}{2} \Bigr) \Bigr]^{n}
	\,
	\ch\bigl[ ( N \!-\! n \!-\! 2\ell ) v \bigr]
	\Biggr]
	\, .
\end{align}
Next we derive from the definition of the coefficients~(\ref{c coeff def}) that,
\begin{equation}
(n\!+\!1)c_{(N+1)(n+1)\ell}
	=
	- \frac{ c_{Nn\ell} }{2 ( D \!-\! 1 \!+\! 2n )}
	\, .
\end{equation}
If we expand the denominator formally in powers of~$2n$,
\begin{equation}
(n\!+\!1)c_{(N+1)(n+1)\ell}
	=
	- \frac{ 1 }{2(D\!-\!1) }
	\sum_{m=0}^{\infty} \Bigl( \frac{-2}{D\!-\!1 } \Bigr)^{\! m} n^m c_{Nn\ell}
\end{equation}
we can plug it into the triple series, and write the coefficient~$n^m$ in terms of derivatives,
\begin{equation}
n^m \Bigl[ y + 4 \, \sh^2\Bigl( \frac{v}{2} \Bigr) \Bigr]^{n}
	=
	\biggl( \Bigl[ y + 4 \, \sh^2\Bigl( \frac{v}{2} \Bigr) \Bigr]
		\frac{\partial}{\partial y} \biggr)^{\!\!m} \,
	\Bigl[ y + 4 \, \sh^2\Bigl( \frac{v}{2} \Bigr) \Bigr]^{n}
	\, .
\end{equation}
Then, upon recognizing the last two rows as the IR part of the scalar two-point 
function~(\ref{W result}), 
\begin{equation}
\mathcal{X}_\lambda(y,u,v)
=
	I \Biggl[  \,
	\sum_{m=0}^{\infty} \Bigl( \frac{-2}{D\!-\!1 } \Bigr)^{\! m} 
	\biggl( \Bigl[ y + 4 \, \sh^2\Bigl( \frac{v}{2} \Bigr) \Bigr] \frac{\partial}{\partial y} \biggr)^{\!\! m} \,
	\frac{  \mathcal{W}_\lambda(y,u,v) }{2(D\!-\!1) } \Biggr]
	\, .
\end{equation}
This is precisely the IR part written in the intermediate form~(\ref{M series solution}). 
It is written in the final form~(\ref{check Xi solution}) 
following the procedure given in~(\ref{commuting ders})--(\ref{Stirling generatrix}).

\section{Checks for two-point function}
\label{app: Checks for two-point function}

Checking explicitly that our result for the photon two-point function given in the covariant
tensor basis~(\ref{cov rep}) with the structure functions~(\ref{C1 result})--(\ref{C4 result})
satisfies the equations from Sec.~\ref{subsec: Generalities} is important.
This is facilitated by the following expressions for derivatives of the bi-local variables,
%
%
\begin{align}
&
\nabla_\mu \bigl( \partial_\nu y \bigr) = 
	g_{\mu\nu} (1\!-\!\epsilon) H^2 \bigl( 2 \!-\! y \!-\! 2 \epsilon \, e^{-v} \bigr)
	-
	\frac{\epsilon}{ 1\!-\!\epsilon }
	\Bigl[ \bigl( \partial_\mu y \bigr) \bigl( \partial_\nu u \bigr) 
		\!+\! \bigl( \partial_\mu u \bigr) \bigl( \partial_\nu y \bigr) \Bigr]
		\, ,
\\
&
\nabla_\mu \bigl( \partial_\nu u \bigr) = 
	- g_{\mu\nu} (1\!-\!\epsilon) H^2
	- \frac{1\!+\!\epsilon}{1\!-\!\epsilon}
		\bigl( \partial_\mu u \bigr) \bigl( \partial_\nu u \bigr) 
		\, ,
\qquad \quad
\nabla_\mu \bigl( \partial_\nu' u \bigr) = 0
\, ,
\end{align}
%
%
and by the contraction identities for basis tensors given in 
Table~\ref{tensor contractions}. 
%
\begin{table}[h!]
\renewcommand{\arraystretch}{1.3}
\centering
\begin{tabular}{l r}
\hline
$\quad
	g^{\mu\nu} \bigl( \partial_{\mu} y \bigr) \bigl( \partial_{\nu} y \bigr)$
	&
	$(1\!-\!\epsilon)^2 H^2 \bigl( 4y \!-\! y^2 \bigr) \quad$
\\
\hline
$\quad
	g'^{\rho\sigma} \bigl( \partial_{\rho}' y \bigr) \bigl( \partial_{\sigma}' y \bigr)$
	&
	$(1\!-\!\epsilon)^2 H'^2 \bigl( 4y \!-\! y^2 \bigr) \quad$
\\
\hline
$\quad
	g^{\mu\nu} \bigl( \partial_{\mu} y \bigr) \bigl( \partial_{\nu} u \bigr) $
	&
	$(1\!-\!\epsilon)^2 H^2 \bigl( 2 \!-\! y \!-\! 2 e^{-v} \bigr) \quad$
\\
\hline
$\quad
	g'^{\rho\sigma} \bigl( \partial_{\rho}' y \bigr) \bigl( \partial_{\sigma}' u \bigr) $
	&
	$(1\!-\!\epsilon)^2 H'^2 \bigl( 2 \!-\! y \!-\! 2 e^{v} \bigr) \quad$
\\
\hline
$\quad
	g^{\mu\nu} \bigl( \partial_{\mu} u \bigr) \bigl( \partial_{\nu} u \bigr) $
	&
	$- (1\!-\!\epsilon)^2 H^2 \quad$
\\
\hline
$\quad
	g'^{\rho\sigma} \bigl( \partial_{\rho}' u \bigr) \bigl( \partial_{\sigma}' u \bigr) $
	&
	$ - (1\!-\!\epsilon)^2 H'^2 \quad$
\\
\hline
$\quad
	g^{\mu\nu} \bigl( \partial_{\mu} y \bigr) \bigl( \partial_{\nu} \partial_{\!\rho}' y \bigr) $
	&
	$ (1\!-\!\epsilon)^2 H^2 (2\!-\!y) \bigl( \partial_{\rho}' y \bigr) \quad$
\\
\hline
$\quad
	g'^{\rho\sigma} \bigl( \partial_{\mu} \partial_{\rho}' y \bigr) \bigl( \partial_{\sigma}' y \bigr)$
	&
	$ (1\!-\!\epsilon)^2 H'^2 (2\!-\!y) \bigl( \partial_{\mu} y \bigr) \quad$
\\
\hline
$\quad
	g^{\mu\nu} \bigl( \partial_{\mu} u \bigr) \bigl( \partial_{\nu} \partial_{\rho}' y \bigr) $
	&
	$- (1\!-\! \epsilon)^2 H^2 
		\bigl[ \bigl(\partial'_{\rho}y \bigr) + 2 e^{-v} \bigl( \partial'_{ \rho} u \bigr) \bigr]
		\quad$
\\
\hline
$\quad
	g'^{\rho\sigma} \bigl( \partial_{\mu} \partial_{\rho}' y \bigr) \bigl( \partial'_{\sigma} u \bigr) $
	&
	$ - (1\!-\! \epsilon)^2 H'^2 
		\bigl[ \bigl(\partial_{\mu}y \bigr) + 2 e^v \bigl( \partial_{ \mu} u \bigr) \bigr]
		\quad$
\\
\hline
$\quad
	g^{\mu\nu} \bigl( \partial_{ \mu} \partial_{\rho}' y \bigr) 
	\bigl( \partial_{\nu} \partial_{\sigma}' y \bigr) $
	&
	$\qquad (1\!-\!\epsilon)^2 H^2 \bigl[ 4 (1\!-\!\epsilon)^2 H'^2 g'_{\rho\sigma}
			- \bigl( \partial'_\rho y \bigr) \bigl( \partial'_\sigma y \bigr) \bigr] \quad$
\\
\hline
$\quad
	g'^{\rho\sigma} \bigl( \partial_{\mu} \partial_{\rho}' y \bigr) 
	\bigl( \partial_{\nu} \partial_{\sigma}' y \bigr) \qquad$
	&
	$\qquad (1\!-\!\epsilon)^2 H'^2 \bigl[ 4 (1\!-\!\epsilon)^2 H^2 g_{\mu\nu}
			- \bigl( \partial_\mu y \bigr) \bigl( \partial_\nu y \bigr) \bigr] \quad$
\\
\hline
\end{tabular}
\caption{
Contractions of tensor structures (table adopted from~\cite{Glavan:2020zne}).}
\label{tensor contractions}
\end{table}
Additionally, equations from Sec.~\ref{subsubsec: IR divergent scalar two-point functions} 
satisfied by scalar two-point functions,
and equations from Sec.~\ref{subsec: Laplace-inverted two-point function}
satisfied by the Laplace-inverted two-point functions
need to be used.
These checks are greatly facilitated by computer algebra programs such as
{\it Cadabra}~\cite{Peeters:2007wn,Peeters:2006kp,Peeters:2018dyg}, 
and {\it Wolfram Mathematica} that we made 
extensive use of here.

\bigskip

\noindent {\bf Ward-Takahashi identity.}
We can expand the left-hand-side of the Ward-Takahashi identity~(\ref{WT identity})
in the basis of two vectors,
\begin{equation}
\nabla^\mu \, i \bigl[ \tensor*[_\mu^{\tt a\! }]{\Delta}{_\nu^{\tt\!  b}} \bigr](x;x')
	=
	(1\!-\!\epsilon)^2 H^2 
	\Bigl[
	\bigl( \partial'_\nu y \bigr) \mathcal{Z}_1(y,u,v)
	+
	\bigl( \partial'_\nu u \bigr) \mathcal{Z}_2(y,u,v)
	\Bigr]
	\, ,
\end{equation}
where the two structure functions are expressed in terms of the structure functions
of the photon two-point function,
\begin{align}
\mathcal{Z}_1 ={}&
	\biggl[
	(2 \!-\! y) \frac{ \partial }{ \partial y }
	- \frac{ \partial }{ \partial u }
	- \frac{ \partial }{ \partial v}
	- D 
	- \frac{ ( D \!-\! 2 )\epsilon }{ ( 1 \!-\! \epsilon ) }
	\biggr]
	\bigl( \mathcal{C}_1 + \mathcal{C}_3 - \overline{\mathcal{C}}_3 \bigr)	
	-
	2e^{-v} 
	\frac{\partial }{\partial y} \bigl( \mathcal{C}_3 - \overline{\mathcal{C}}_3 \bigr)
\nonumber \\
&	\hspace{-0.9cm}
	+ 
	\biggl[
	\bigl( 4y \!-\!  y^2 \bigr) \frac{\partial }{\partial y}
		+ (D \!+\! 1) (2 \!-\! y) 
		+ \bigl( 2 \!-\! y \!-\! 2 e^{-v} \bigr)
		\biggl( \frac{\partial }{\partial u} \!+\! \frac{\partial }{\partial v} 
		+ \frac{ (D \!-\! 2) \epsilon }{ (1 \!-\! \epsilon) } 
		\biggr)
	\biggr]
	\mathcal{C}_2
	\, ,
\\
\mathcal{Z}_2 ={}&
	- 2 e^{-v}
	\biggl[
	\frac{\partial }{\partial u}
	+ \frac{\partial }{\partial v}
	+ \frac{ (D \!-\! 2) \epsilon }{ ( 1 \!-\! \epsilon) } 
	\biggr] 
	\mathcal{C}_1
	- 
	2 e^{-v} \bigl( \mathcal{C}_3
		- \overline{\mathcal{C}}_3 \bigr)
\nonumber \\
&	\hspace{-0.9cm}
	+
	\biggl[
	\bigl( 4y \!-\! y^2 \bigr) \frac{\partial }{\partial y}
	+ D (2 \!-\! y) 
	+
	\bigl( 2 \!-\! y \!-\! 2 e^{-v} \bigr)
	\biggl( \frac{\partial }{\partial u}
	+ \frac{\partial }{\partial v}
	+ \frac{(D \!-\! 2) \epsilon }{ (1\!-\! \epsilon) } 
	\biggr)
	\biggr]
	\bigl( \mathcal{C}_3 + \overline{\mathcal{C}}_3 \bigr)
\nonumber \\
&
	+ 
	\biggl[
	\bigl( 2 \!-\! y \!-\! 2 e^{-v} \bigr) \frac{\partial }{\partial y}
	- \frac{\partial }{\partial u}
	- \frac{\partial }{\partial v}
	- (D \!-\! 1)
	- \frac{ (D \!-\! 2) \epsilon }{ (1 \!-\! \epsilon) } 
	\biggr]
	\mathcal{C}_4
	\, .
\end{align}
Upon plugging in the photon structure functions~(\ref{C1 result})--(\ref{C4 result}),
and applying the equation~(\ref{Upsilon eom}) satisfied by~$i\Upsilon$,
and the recurrence 
relations~(\ref{generalized recurrence}) for the scalar two-point functions,
and the equations of motion~(\ref{scalar eom even})
and~(\ref{scalar eom odd}) they satisfy,
the result simplifies to,
\begin{equation}
(1\!-\!\epsilon)^2 H^2 \mathcal{Z}_1 
	=
- \xi_s \frac{\partial}{\partial y} i \Delta_{\nu+1}
\, ,
\qquad 
(1\!-\!\epsilon)^2 H^2 \mathcal{Z}_2 =
- \xi_s \Bigl( \frac{\partial}{\partial u} - \frac{\partial}{\partial v} \Bigr) i \Delta_{\nu+1}
\, ,
\end{equation}
which exactly reproduces the right-hand side
of the Ward-Takahashi identity~(\ref{WT identity}).

\bigskip

\noindent {\bf Equation of motion.}
The Ward-Takahashi identity that we just checked also simplifies the equation of 
motion~(\ref{2pt EOM}) satisfied by the two-point functions in the simple covariant gauge,
\begin{equation}
\Bigl(
	\dalembertian \delta_\mu^\rho
	- {R_\mu}^\rho
	\Bigr)
	i \bigl[ \tensor*[_\rho^{\tt a \! }]{\Delta}{_\nu^{\tt \! b}} \bigr](x;x')
	=
	{\tt S}^{\tt ab}
	g_{\mu\nu} \frac{ i \delta^D(x\!-\!x') }{ \sqrt{-g} }
	+
	( 1 \!-\! \xi_s ) \partial_\mu \partial'_\nu
	i \bigl[ \tensor*[^{\tt a \!}]{\Delta}{^{\tt \! b }} \bigr]_{\nu+1}(x;x') \, .
\label{simplified EOM}
\end{equation}
The left-hand-side can be expanded off-coincidence in the appropriate tensor basis,
\begin{align}
\MoveEqLeft[1.5]
\Bigl(
	\dalembertian \delta_\mu^\rho
	- {R_\mu}^\rho
	\Bigr) 
	i \bigl[ \tensor*[_\rho^{\tt a\! }]{\Delta}{_\nu^{\tt\!  b}} \bigr](x;x')
	=
	(1\!-\!\epsilon)^2 H^2 \biggl[ 
	\bigl( \partial_\mu \partial'_\nu y \bigr) \mathcal{E}_1(y,u,v)
	+
	\bigl( \partial_\mu y \bigl) \bigr( \partial'_\nu y \bigr) \mathcal{E}_2(y,u,v)
\nonumber 
\\
&
	+ \Bigl[ \bigl( \partial_\mu y \bigr) \bigl( \partial'_\nu u \bigr) 
		\!+\! \bigl( \partial_\mu u \bigr) \bigl( \partial'_\nu y \bigr) \Bigr] \,
		 \mathcal{E}_3(y,u,v)
	+ \Bigl[ \bigl( \partial_\mu y \bigr) \bigl( \partial'_\nu u \bigr) \!-\! \bigl( \partial_\mu u \bigr) \bigl( \partial'_\nu y \bigr) \Bigr] \, \overline{\mathcal{E}}_3(y,u,v)
\nonumber \\
&
	+ \bigl( \partial_\mu u \bigr) \bigl( \partial'_\nu u \bigr) \, \mathcal{E}_4(y,u,v) 
	\biggr]
	\, ,
\end{align}
where the structure functions are found to be,
\begingroup
\allowdisplaybreaks
\begin{align}
\mathcal{E}_1 ={}&
	\biggl[
	\bigl( 4y \!-\! y^2 \bigr) \frac{\partial^2 }{\partial y^2}
	+
	D (2 \!-\! y) \frac{\partial }{\partial y}
	+ 
	2\bigl( 2 \!-\! y \!-\! 2 e^{-v} \bigr) 
		\biggl( \frac{\partial }{ \partial u} 
		\!+\! \frac{\partial }{ \partial v}
		\!+\! \frac{ ( D \!-\! 4 ) \epsilon }{ 2 ( 1 \!-\! \epsilon ) }  \biggr) 
			\frac{\partial}{\partial y} 
\nonumber \\
&	\hspace{-0.9cm}
	\!-\! 
	\biggl(\! \frac{\partial }{ \partial u} \!+\! \frac{\partial }{ \partial v} 
		\!+\! \frac{ D \!-\! 1 \!-\! 3 \epsilon }{ 1 \!-\! \epsilon }\! \biggr)
		\Bigl( \frac{\partial }{ \partial u} \!+\! \frac{\partial }{ \partial v} \Bigr)
	\!-\! 
	\frac{ D \!-\! 2\epsilon }{ 1 \!-\! \epsilon } 
	\biggr] \mathcal{C}_1
	\!+\!
	\frac{2}{ 1 \!-\! \epsilon }
	\Bigl[
	\bigl( 2 \!-\! y \!-\! 2 \epsilon   e^{-v} \bigr)
	\mathcal{C}_2
	\!-\! 
	\mathcal{C}_3
	\!+\!
	\overline{\mathcal{C}}_3
	\Bigr]
	 ,
\label{structure function E1}\\
\mathcal{E}_2 ={}&
	\biggl[
	\bigl( 4y \!-\! y^2 \bigr) \frac{\partial^2}{\partial y^2}
	+
	(D\!+\! 4) (2 \!-\! y )\frac{\partial}{\partial y}
	+
	2 \bigl( 2 \!-\! y \!-\! 2e^{-v} \bigr) 
		\biggl( \frac{\partial}{\partial u} \!+\! \frac{\partial}{\partial v}
			\!+\! \frac{ (D \!-\! 2) \epsilon }{2(1 \!-\! \epsilon)} \biggr) 
			\frac{\partial}{\partial y}
\nonumber \\
&	\hspace{-0.9cm}
	- 
	\biggl( \frac{\partial}{\partial u} \!+\! \frac{\partial}{\partial v} \!+\!
			\frac{ D \!+\! 1 \!-\! 5\epsilon }{1 \!-\! \epsilon } \biggr)
		\Bigl( \frac{\partial}{\partial u} \!+\! \frac{\partial}{\partial v} \Bigr)
	- 
	\frac{2(D \!-\! 3\epsilon) }{ 1 \!-\! \epsilon } 
	\biggr] \mathcal{C}_2
	-
	\frac{2}{1\!-\!\epsilon} \frac{\partial }{\partial y}
		\bigl( \mathcal{C}_1 \!+\! \mathcal{C}_3 \!-\! \overline{\mathcal{C}}_3 \bigr)
	\, ,
\label{structure function E2}\\
\mathcal{E}_3 ={}&
	- \biggl[
	\frac{ \epsilon }{ 1 \!-\! \epsilon } \bigl( 2 \!-\! y \!+\! 2e^{-v} \bigr) 
		\frac{\partial }{\partial y}
	+ \frac{\partial }{\partial u}
	+ \frac{\partial }{\partial v}
	- \frac{(D \!-\! 2\epsilon)\epsilon}{(1 \!-\! \epsilon)^2} 
	\biggr]
	\mathcal{C}_1
\nonumber \\
&	\hspace{-0.8cm}
	-
	\biggl[
	\frac{ \epsilon }{ 1 \!-\! \epsilon }
	\biggl(\bigl( 4y \!-\! y^2 \bigr) \frac{\partial }{\partial y}\!+\!D(2\!-\!y) \biggr)
	-
	\bigl( 2 \!-\! y \!-\! 2e^{-v} \bigr) 
		\biggl( \frac{\partial}{\partial u} \!+\! \frac{\partial}{\partial v}
			\!-\! \frac{(1\!+\!(D \!-\! 3)\epsilon) \epsilon }{(1 \!-\! \epsilon)^2} \biggr)
	\biggr]
	\mathcal{C}_2
\nonumber \\
&	\hspace{-0.8cm}
	+
	\biggl[
	\bigl( 4y \!-\! y^2 \bigr) \frac{\partial^2}{\partial y^2}
	\!+\!
	(D\!+\!2)(2\!-\!y) \frac{\partial}{\partial y}
	\!+\!
	2\bigl( 2 \!-\! y \!-\! 2e^{-v} \bigr)
		\biggl(
		\frac{\partial}{\partial u}
		\!+\! \frac{\partial}{\partial v}
		\!-\! \frac{1\!-\! (D\!-\!3)\epsilon }{2(1\!-\!\epsilon)}
		\biggr) \frac{\partial}{\partial y}
\nonumber \\
&	\hspace{0.cm}
	-
	\biggl(
	\frac{\partial}{\partial u} 
		\!+\! \frac{\partial}{\partial v} 
		\!+\! \frac{ D\!-\!4\epsilon }{1\!-\!\epsilon}
		\biggr)
		\Bigl(
		\frac{\partial}{\partial u} \!+\! \frac{\partial}{\partial v}
		\Bigr)
	-
	\frac{(D\!-\!1) \!-\!(2D\!+\!1) \epsilon \!+\!4\epsilon^2 }
		{ (1\!-\!\epsilon)^2 }
	\biggr]
	\mathcal{C}_3
\nonumber \\
&	\hspace{-0.85cm}
	+
	\biggl[
	\frac{ 1\!+\!\epsilon }{ 1\!-\!\epsilon } \bigl( 2 \!-\! y \!-\! 2e^{-v} \bigr)
		\frac{ \partial }{ \partial y}
	+
	\frac{\partial}{\partial u}
	+
	\frac{\partial}{\partial v}
	-
	\frac{ ( D \!-\! 1 \!-\! \epsilon ) \epsilon }{(1\!-\!\epsilon)^2}
	-
	\frac{1 \!+\! \epsilon}{1 \!-\! \epsilon}
	\biggr]
	\overline{\mathcal{C}}_3
	-
	\frac{1}{ 1\!-\!\epsilon } \frac{\partial \mathcal{C}_4}{\partial y}
	\, ,
\label{structure function E3}
\\
\overline{\mathcal{E}}_3 ={}&
	\biggl[
	\frac{ \epsilon }{ (1 \!-\! \epsilon) } \bigl( 2 \!-\! y \!-\! 2e^{-v} \bigr) 
		\frac{\partial }{\partial y}
	+ \frac{\partial }{\partial u}
	+ \frac{\partial }{\partial v}
	- \frac{(D \!-\! 2\epsilon)\epsilon}{(1 \!-\! \epsilon)^2} 
	\biggr]
	\mathcal{C}_1
	\!+\! 
	\biggl[
	\frac{\epsilon}{(1\!-\!\epsilon)} (4y \!-\! y^2) \frac{\partial}{\partial y}
\nonumber \\
&	\hspace{-0.cm}
	-
	\bigl( 2 \!-\! y \!+\! 2e^{-v} \bigr)
		\biggl(
		\frac{\partial}{\partial u}
		+
		\frac{\partial}{\partial v}
		+\frac{(D \!-\! 3 \!+\!\epsilon) \epsilon }{(1\!-\!\epsilon)^2}
		\biggr)
	+2(2 \!-\!y)\frac{(D \!-\! 1 \!-\! \epsilon) \epsilon }{(1\!-\!\epsilon)^2}
	\biggr]
	\mathcal{C}_2
\nonumber \\
&	\hspace{-0.9cm}
	+
	\biggl[
	\frac{ 1\!+\!\epsilon }{ 1\!-\!\epsilon } \bigl( 2 \!-\! y \!-\! 2 e^{-v} \bigr)
		\frac{ \partial }{ \partial y}
	+
	\frac{\partial}{\partial u}
	+
	\frac{\partial}{\partial v}
	-
	\frac{ ( D \!-\! 1 \!-\! \epsilon ) \epsilon }{(1\!-\!\epsilon)^2}
	-
	\frac{1 \!+\! \epsilon}{1 \!-\! \epsilon}
	\biggr]
	\mathcal{C}_3
	-
	\frac{1}{ 1\!-\!\epsilon } \frac{\partial \mathcal{C}_4}{\partial y}
\nonumber \\
&	\hspace{-0.9cm}
	+
	\biggl[
	\bigl( 4y\!-\!y^2 \bigr) \frac{\partial^2}{\partial y^2}
		+ 
		(D\!+\!2) (2\!-\!y) \frac{\partial}{\partial y}
		+
	2 \bigl( 2 \!-\! y \!-\! 2 e^{-v} \bigr)
		\biggl(
		\frac{\partial}{\partial u}
		\!+\! \frac{\partial}{\partial v}
		\!-\! \frac{1\!-\!(D\!-\!3)\epsilon }{ 2(1\!-\!\epsilon) } 
		\biggr) \frac{\partial}{\partial y}
\nonumber \\
&	\hspace{-0.cm}
	-
	\biggl( \frac{\partial}{\partial u} 
		\!+\! \frac{\partial}{\partial v} 
		\!+\! \frac{ D\!-\!4\epsilon }{ 1\!-\!\epsilon } \biggr)
		\Bigl( \frac{\partial}{\partial u} \!+\! \frac{\partial}{\partial v} \Bigr)
	-
	\frac{(D \!-\! 1)\!-\!(2D \!+\! 1) \epsilon \!+\! 4\epsilon^2 }{ (1\!-\!\epsilon)^2 }
	\biggr]
	\overline{\mathcal{C}}_3
	\, ,
\label{structure function E3 bar}
\\
\mathcal{E}_4 ={}&
	-
	\biggl[
	\frac{ 2 \epsilon }{ 1 \!-\! \epsilon } \bigl( 4y \!-\! y^2 \bigr) 
		\frac{ \partial }{\partial y}
	-
	2 ( 2 \!-\! y ) 
		\biggl( \frac{\partial }{\partial u}
			\!+\! \frac{\partial}{\partial v} 
			\!-\! \frac{ (D\!-\!2\epsilon) \epsilon }{ (1\!-\!\epsilon)^2 } \biggr)
	\biggr]
	\bigl( \mathcal{C}_3 + \overline{\mathcal{C}}_3  \bigr)
\nonumber \\
&	\hspace{-0.9cm}
	-
	4e^{-v}
		\biggl( \frac{\partial }{\partial u}
			\!+\! \frac{\partial}{\partial v} 
			\!-\! \frac{ 2 \!-\! (D\!-\!4)\epsilon }{ 2(1\!-\!\epsilon) } \biggr)
	\bigl( \mathcal{C}_3 - \overline{\mathcal{C}}_3 \bigr)
	+
	\frac{ 2(D\!-\!2) (1\!+\!\epsilon) \epsilon }{ (1\!-\!\epsilon)^2 } e^{-v} 
		\bigl( \mathcal{C}_1 + \mathcal{C}_3 + \overline{\mathcal{C}}_3 \bigr)
\nonumber\\
&	\hspace{-0.9cm}
	+
	\biggl[
	\bigl( 4y \!-\! y^2 \bigr) \frac{\partial^2}{\partial y^2}
	+D ( 2 \!-\! y ) \frac{\partial}{\partial y}
	+2 \bigl( 2 \!-\! y \!-\! 2 e^{-v} \bigr) 
		\biggl( \frac{ \partial }{ \partial u} \!+\! \frac{ \partial }{ \partial v } 
			- \frac{2\!-\! (D \!-\! 4)\epsilon }{ 2(1\!-\!\epsilon) }  
			\biggr) \frac{ \partial }{ \partial y}
\nonumber \\
&	\hspace{-0.cm}
	- \biggl( \frac{ \partial }{ \partial u } \!+\! \frac{ \partial }{ \partial v } 
			\!+\! \frac{ D\!-\!1\!-\!3\epsilon }{ 1\!-\!\epsilon } \biggr)
		\Bigl( \frac{ \partial }{ \partial u} \!+\! \frac{ \partial }{ \partial v } \Bigr)
	+ \frac{2(D\!-\!1\!-\!\epsilon)\epsilon }{(1\!-\!\epsilon)^2}
	\biggr] \mathcal{C}_4
	\, .
\label{structure function E4}
\end{align}
\endgroup
Upon plugging in the photon structure functions~(\ref{C1 result})--(\ref{C4 result}), 
and applying judiciously 
equations~(\ref{Upsilon eom})--(\ref{Upsilon last eq}) satisfied by the function~$i\Upsilon$
defined in~(\ref{Upsilon}), 
and equations~(\ref{scalar eom even})--(\ref{generalized recurrence}) 
satisfied by the scalar two-point functions
the expressions above evaluate to,
\begingroup
\allowdisplaybreaks
\begin{subequations}
\begin{align}
(1\!-\!\epsilon)^2 H^2 \mathcal{E}_1 ={}&
	(1 \!-\! \xi_s) \frac{\partial}{\partial y} i \Delta_{\nu+1}
	\, ,
\label{Delta E1}\\
(1\!-\!\epsilon)^2 H^2 \mathcal{E}_2 ={}&
	(1 \!-\! \xi_s) \frac{\partial^2}{\partial y^2} i \Delta_{\nu+1}
	\, ,
\label{Delta E2}\\
(1\!-\!\epsilon)^2 H^2 \mathcal{E}_3 ={}&
	(1 \!-\! \xi_s) \frac{\partial^2}{\partial y \partial u} i \Delta_{\nu+1}
	\, ,
\label{Delta E3}\\
(1\!-\!\epsilon)^2 H^2 \overline{\mathcal{E}}_3 ={}&
	- (1 \!-\! \xi_s) \frac{\partial^2}{\partial y \partial v} i \Delta_{\nu+1}
	\, ,
\label{Delta E3 bar}\\
(1\!-\!\epsilon)^2 H^2 \mathcal{E}_4 ={}&
	(1 \!-\! \xi_s) \Bigl( \frac{\partial^2}{\partial u^2} 
		- \frac{\partial^2}{\partial v^2} \Bigr) i \Delta_{\nu+1}
		\, .
\label{Delta E4}
\end{align}
\end{subequations}
\endgroup
These precisely account for the right-hand-side of the equation of 
motion~(\ref{simplified EOM}) off-coincidence.

In order to check that the local terms on the right-hand-side of~(\ref{simplified EOM}) 
are also correctly reproduced for the case of the Feynman (and consequently Dyson)  
propagator it is sufficient to consider only the most singular terms from each 
photon structure function~(\ref{C1 result})--(\ref{C4 result}). For scalar propagators
these are isolated easily using the power-series representation~(\ref{power series})
for the bulk part, as the IR series~(\ref{W result}) does not contribute. From there
we find that the most singular term is just the conformal two-point function,
\begin{equation}
i \Delta_\nu(y,u,v) \ \overset{x'\to x}{\longsim} \ 
	i \Delta_{\scr 1/2}(y,u)
	=
	e^{- \frac{(D-2)\epsilon}{2(1-\epsilon)} u } \mathcal{F}_{\scr 1/2}(y)
	=
	\frac{\bigl[ (1\!-\!\epsilon)^2 HH' \bigr]^ \frac{D-2}{2} \, \Gamma\bigl( \frac{D-2}{2} \bigr) }{ (4\pi)^{ \frac{D}{2} } } 
	\Bigl( \frac{y}{4} \Bigr)^{\! -\frac{D-2}{2} }
	.
\end{equation}
The leading singular behaviour for the function~$i\Upsilon$ the leading singular 
can be inferred from the result in~(\ref{UtimeCoin}) by 
substituting~$(1\!-\!\epsilon)^2 \mathcal{H} \mathcal{H}' \| \Delta\vec{x} \|^2 \! \to \! y$
in the denominator,
\begin{equation}
i \Upsilon(y,u,v) \ \overset{x' \to x}{\longsim} \
	- \frac{ \epsilon}{1 \!-\! \epsilon} i \Delta_{\scr 1/2}(y,u) 
	\, ,
\end{equation}
which can be confirmed by showing that this result satisfies 
equations~(\ref{Upsilon eom})--(\ref{Upsilon last eq}) close to coincidence.
The leading singular behaviour of the photon structure functions is
therefore captured by
\begin{subequations}
\begin{align}
&
\mathcal{C}_1 
	\ \overset{ x' \to x }{\longsim} \ 
	\frac{(D\!-\!2)}{2(1\!-\!\epsilon) \nu_{\scr T}} 
	\mathscr{C}
	\, ,
\qquad \quad
\mathcal{C}_2
	\ \overset{ x' \to x }{\longsim} \ 
	\frac{ ( 1 \!+\! \epsilon ) }{2(1\!-\!\epsilon) \nu_{\scr T}} 
	\frac{\partial}{\partial y} \mathscr{C}
	\, ,
\\
&
\mathcal{C}_3
	\ \overset{ x' \to x }{\longsim} \ 
	\frac{(D\!-\!2) \epsilon }{4(1\!-\!\epsilon) \nu_{\scr T}} 
		\mathscr{C}
	\, ,
\qquad \quad
\overline{\mathcal{C}}_3
	\ \overset{ x' \to x }{\longsim} \ 
	0 \times v \mathscr{C}
	\, ,
\qquad \quad
\mathcal{C}_4
	\ \overset{ x' \to x }{\longsim} \ 
	0 \times \mathscr{C}
	\, ,
\end{align}
\end{subequations}
where,
\begin{equation}
\mathscr{C}(y,u) =
	- \frac{  i \Delta_{\scr 1/2}(y,u)  }{ 2(1\!-\!\epsilon)^2 H H'}
		 \, .
\end{equation}
We can then apply a simple derivative identity,
\begin{align}
\MoveEqLeft[3]
\bigl( \partial_\mu y \bigr) \bigl( \partial_\nu' y \bigr) f(y,u)
	=
	\partial_\mu \partial'_\nu I^2 \bigl[ f(y,u) \bigr]
	-
	\bigl( \partial_\mu \partial'_\nu y \bigr) I \bigl[ f(y,u) \bigr]
\\
&
	-
	\Bigl[ \bigl( \partial_\mu y \bigr) \bigl( \partial'_\nu u \bigr) 
		\!+\! \bigl( \partial_\mu u \bigr) \bigl( \partial'_\nu y \bigr) \Bigr]
		\frac{\partial}{\partial u} I\bigl[f(y,u) \bigr]
	-
	\bigl( \partial_\mu u \bigr) \bigl( \partial'_\nu u \bigr)
	\frac{\partial^2}{\partial u^2} I^2\bigl[ f(y,u) \bigr]
	\, ,
\nonumber 
\end{align}
to the second tensor structure to rewrite the most singular piece 
of the photon propagator in a particularly convenient form,
\begin{align}
i \bigl[ \tensor*[_\mu^{\scr \! + \! }]{\Delta}{_\nu^{\scr \!+\!}} \bigr](x;x')
	\ \overset{ x' \to x }{\longsim} \ {}&
	\frac{ (1 \!+\! \epsilon) }{ 2(1\!-\!\epsilon) \nu_{\scr T} } 
		\partial_\mu \partial'_\nu I[\mathscr{C}]
	+
	\bigl( \partial_\mu \partial'_\nu y \bigr) \mathscr{C}
\nonumber \\
&	\hspace{2cm}
	+
	\Bigl[ \bigl( \partial_\mu y \bigr) \bigl( \partial'_\nu u \bigr) 
		\!+\! \bigl( \partial_\mu u \bigr) \bigl( \partial'_\nu y \bigr) \Bigr]
		\frac{  \epsilon \, \mathscr{C} }{1 \!-\! \epsilon } 
		\, ,
\label{singularFeynman}
\end{align}
where we have neglected an irrelevant piece where the fourth tensor structure
is multiplied by $y^{- \frac{D-4}{2}}$, which is not 
sufficiently singular to produce a local contribution in the equaton of motion.
We then use the identities which generate the temporal delta function,
\begin{subequations}
\begin{align}
\partial_0 \partial_0 y_{\scr ++}
	={}&
	\partial_0 \partial_0 y
	+
	4 (1\!-\!\epsilon)^2 \mathcal{H}^2 \! \times \! i\delta \!\times\! \delta(\eta\!-\!\eta')
	\, ,
\\
\partial_0 \partial_0' y_{\scr ++}
	={}&
	\partial_0 \partial_0' y
	-
	4 (1\!-\!\epsilon)^2 \mathcal{H}^2 \! \times \!i\delta \!\times\! \delta(\eta\!-\!\eta')
	\, ,
\end{align}
\end{subequations}
and the covariant~$D$-dimensional delta function,
\begin{equation}
- 4(1\!-\!\epsilon)^2 H^2 \!\times\!  i\delta \!\times\! \delta(\eta\!-\!\eta') \times  \frac{\partial}{\partial y}
	i \Delta_{\scr 1/2}\bigl(y_{\scr ++},u_{\scr ++} \bigr)
	=
	\frac{i \delta^{D}(x\!-\!x') }{ \sqrt{-g} }
	\, ,
\end{equation}
to compute the action of the photon kinetic operator from~(\ref{D operator}) onto the
most singular part of the propagator~(\ref{singularFeynman}),
\begin{align}
&
{\mathcal{D}_\mu}^\rho \biggl\{ \partial_\rho \partial'_\nu I[ \mathscr{C} ]
 \bigl(y_{\scr ++},u_{\scr ++} \bigr)\biggr\}
	\ \overset{ x' \to x }{\longsim} \ 
	-  \frac{ 2 }{ \xi_s } 
	 	\bigl( a^2 \delta_\mu^0\delta_\nu^0 \bigr)
	\frac{i \delta^{D}(x\!-\!x') }{ \sqrt{-g} }
	\, ,
\\
&
{\mathcal{D}_\mu}^\rho \biggl\{ \bigl( \partial_\rho \partial'_\nu y \bigr) \mathscr{C}
 \bigl(y_{\scr ++},u_{\scr ++} \bigr)\biggr\}
 	\ \overset{ x' \to x }{\longsim} \ 
 	\biggl[ g_{\mu\nu} + \Bigl( 1 \!-\! \frac{1}{\xi_s} \Bigr) 
 		\bigl( a^2 \delta_\mu^0 \delta_\nu^0 \bigr) \biggr]
 		\frac{i \delta^{D}(x\!-\!x') }{ \sqrt{-g} }
 		\,  ,
\\
&
{\mathcal{D}_\mu}^\rho \biggl\{ \Bigl[
 \bigl( \partial_\rho y_{\scr ++} \bigr) \bigl( \partial'_\nu u_{\scr ++} \bigr)
		\!+\! \bigl( \partial_\rho u_{\scr ++} \bigr) \bigl( \partial'_\nu y_{\scr ++} \bigr) \Bigr]
		\mathscr{C} \bigl(y_{\scr ++},u_{\scr ++} \bigr)\biggr\}
	\ \overset{ x' \to x }{\longsim} \ 
	0 \, .
\end{align}
When combined, these relations correctly reproduce the local terms in the Feynman 
propagator equation of motion~(\ref{2pt EOM}).


\end{document}